\documentclass[12pt,a4paper,fleqn,twoside]{report}

\usepackage{amsmath}                               
\usepackage{amssymb}                               
\usepackage{bbm}                                   
\usepackage{a4}                                    
\usepackage{graphicx}                              
\DeclareMathOperator{\tr}{tr}                      
\newcommand{\LRa}{\Leftrightarrow}                 
\newcommand{\eps}{\varepsilon}                     
\renewcommand{\d}{\partial}                        
\renewcommand{\vec}{\boldsymbol}                   
\newcommand{\abs}[1]{\left| #1 \right|}            
\newcommand{\et}[3]{\eta^{#1}_{#2 #3}}             
\newcommand{\etb}[3]{\bar\eta^{#1}_{#2 #3}}        
\newcommand{\PO}{\mathcal P}                       
\newcommand{\lam}{\lambda}                         
\newcommand{\Lam}{\Lambda}                         
\newcommand{\al}{\alpha}                           
\newcommand{\eq}[1]{Eq.\,\eqref{#1}}               
\newcommand{\eqs}[1]{Eqs.\,\eqref{#1}}             
\newcommand{\fig}[1]{Fig.\,\ref{#1}}               
\newcommand{\UM}{\mathbbm 1}                       
\renewcommand{\o}[1]{{\mathcal O}(#1)}             
\newcommand{\absatz}[1]{\noindent {\it #1.}}

\begin{document}

\pagenumbering{roman}

\begin{titlepage}
\centering
\renewcommand{\baselinestretch}{1.5}
\sc 
\huge Faculty of Physics and Astronomy \\
\LARGE University of Heidelberg \\
\vfill
\rm
\large
{\bf
Diploma thesis \\
in Physics \\
submitted by \\
Ulrich Herbst \\
born in Landau in der Pfalz, Germany \\
2005
} 
\end{titlepage}

\thispagestyle{empty}
\cleardoublepage

\begin{titlepage}
\centering
\renewcommand{\baselinestretch}{1.5}
\vspace*{3cm}
\sc
\huge 
The deconfining phase of SU(2) Yang-Mills thermodynamics \\
\vfill
\rm
\large
{\bf
This diploma thesis has been carried out by 
Ulrich Herbst
at the \\
Institut f\"ur Theoretische Physik \\
under the supervision of \\
Priv.-Doz. Dr. habil. Ralf Hofmann 
}
\end{titlepage}

\thispagestyle{empty}
\cleardoublepage

\thispagestyle{empty}

\vspace*{3.5cm}
\begin{center}
\bf Abstract
\end{center}
\begin{quote}
Phase and modulus of an energy- and pressure-free, composite and adjoint field 
in an SU(2) Yang-Mills theory are computed.
This field is generated by trivial holonomy calorons of topological charge one.
It possesses nontrivial $S_1$-winding on the group manifold.
The two-loop contribution to the thermodynamical pressure of an SU(2) Yang-Mills theory
in the electric (deconfining) phase is computed in the real time formalism of finite temperature field theory.
The result supports the picture of only very weakly interacting quasiparticles.
\end{quote}

\vspace{3.5cm}
\begin{center}
\bf Zusammenfassung
\end{center}
\begin{quote}
Phase und Betrag eines energie- und druckfreien, zusammengesetzten und adjungierten Feldes
in SU(2) Yang-Mills Theorie werden berechnet.
Dieses Feld wird von Kaloronen trivialer Holonomie und topologischer Ladung eins erzeugt.
Es besitzt nichttriviales Verhalten auf der Gruppenmannigfaltigkeit.
Der Zwei-Schleifen Beitrag zum thermodynamischen Druck einer SU(2) Yang-Mills Theorie
in der elektrischen (deconfinierten) Phase wird im Real-Zeit-Formalismus berechnet.
Das Ergebnis st\"utzt das Bild nur sehr schwach wechselwirkender Quasiteilchen.
\end{quote}

\clearpage
\thispagestyle{empty}
\cleardoublepage

\tableofcontents
\cleardoublepage

\pagenumbering{arabic}

\chapter{Introduction}

Gauge theories such as the Standard Model of particle physics
have been investigated mainly in the framework of perturbation theory.
Due to the tremendous complexity of (especially nonabelian) gauge theories, 
this is the only feasible approach for problems concerning single fundamental processes:
It is not possible to solve even the simplest QED processes exactly.
Concerning the macroscopic properties (e.\,g. thermodynamics) of a system,
chances for a nonperturbative treatment in form of an effective theory are much better.
Superconductivity in metals, for example, was described phenomenologically 
by the Landau-Ginzburg theory \cite{Ginzburg Landau, Abrikosov}.
Subsequently, a quantum theory of superconductivity was developed on the microscopic level \cite{Bardeen Cooper Schrieffer}.

In the Standard Model,
a number of striking results has been obtained using perturbation theory,
among them the anomalous magnetic moment of the electron, 
asymptotic freedom in QCD and many others.
In spite of these great successes, perturbation theory has some intrinsic problems:

Perturbation theory is an expansion in powers of the gauge coupling.
This is by definition only applicable to cases of small coupling.
It is completely impossible to address strongly coupled problems in perturbation theory.
Moreover, a perturbation expansion is an expansion about the trivial vacuum of the theory.
But there exist objects in Yang-Mills theory, e.\,g. instantons, that are topologically distinct from the trivial vacuum.
Such objects can by no means be included in an expansion about the trivial vacuum.
This can also be seen from the partition function:
The instanton enters the partition function with the measure $e^{-S} = e^{-8\pi^2/g^2}$, 
which has an essential singularity at zero coupling such that
the Taylor series of this function about $g=0$ vanishes identically.
Therefore, excitations of nontrivial topology are completely ignored in perturbation theory.

For SU(N) Yang-Mills theory at finite temperature, further problems are known.
It was shown that a perturbative calculation of the thermodynamical pressure can not be driven beyond order $g^5$ \cite{Linde}.
This is essentially due to the existence of weakly screened, soft magnetic modes
which cause infrared instabilities.

There is also a number of experimental and observational results in particle physics and cosmology
the explanation of which is still open or being disputed.
Examples are the nondetection of the Higgs particle at LEP,
the existence of dark matter and dark energy in the universe,
and a quadrupole signal in one of the power spectra of the cosmic microwave background.
A nonperturbative description of gauge theories may be helpful or even necessary in 
understanding these aspects.

At present a large number of theories and approaches 
which try to include nonperturbative aspects in field theories is being considered.
An analytical and nonperturbative approach to SU(2) and SU(3) Yang-Mills thermodynamics is presented in \cite{Hofmann}.
In the spirit of the Ginzburg-Landau theory of superconductivity, a macroscopic field is used
to account for microscopic processes in an effective theory.
More precisely, a composite adjoint Higgs field $\phi$,
which describes the BPS saturated and topologically nontrivial part of the ground state, is introduced.
The Higgs field $\phi$ is generated at an asymptotically high temperature by noninteracting
calorons of topological charge one and trivial holonomy.
The field $\phi$ is quantum mechanically and 
thermodynamically stabilized and can thus be used as a background
for the topologically trivial sector of the theory.
Interactions between trivial-holonomy calorons are included via a macroscopic pure-gauge configuration $a_\mu^{bg}$
which is a solution to the equation of motion for the topologically trivial sector
in the presence of the background $\phi$.

As the modulus of the Higgs field decreases with temperature as $|\phi| \sim \sqrt{ \Lam^3/T }$, 
where $\Lam$ is the Yang-Mills scale,
the effects of topological defects die off at large temperature in a power-like fashion.
Asymptotic freedom and infrared-ultraviolet decoupling of the fundamental theory,
which are results obtained in perturbation theory at $T=0$, are preserved.

Thermodynamical quantities are in this framework calculated as loop expansions about the nontrivial ground state
consisting of the Higgs field $\phi$ and the pure-gauge configuration $a_\mu^{bg}$.
On tree level the excitations in the high temperature (or electric) phase 
are either massive thermal quasiparticles or massless 'photons'.
The interactions between these quasiparticles appear to be very weak.
The effective theory is both infrared- and ultraviolet-finite.
The former is due to the existence of caloron-induced gauge boson masses (IR cutoff),
the latter to constraints on loop momenta arising from the existence of the compositeness scale $|\phi|$ (UV cutoff).

The purpose of this thesis is 
to compute the dynamical generation of the macroscopic, composite field $\phi$
and to calculate the pressure of the Yang-Mills gas on two-loop level for the SU(2) case.
The thesis is organized as follows:
Chapter~2 reviews basic properties of Lie groups and pure Yang-Mills theory. 
The physics of some topological objects in gauge theories, 
namely the Abrikosov-Nielsen-Olesen vortex, the 't~Hooft-Polyakov monopole, and instantons is sketched. 
The focus is on the latter.
Chapter~3 first presents a brief outline of the physics of the electric phase according to the approach in \cite{Hofmann}.
The definition of the phase of the composite Higgs field in terms of 
a spatial and scale parameter average over an adjointly transforming two-point function is given and discussed.
The average has to be evaluated on trivial holonomy caloron and anticaloron configurations.
We discuss the uniqueness of the given definition and perform the evaluation.
We show how under the assumption of an externally given scale the modulus of the field can be determined.
The potential is deduced from the BPS equation.
Chapter~4 contains the calculation of the two-loop contributions 
to the thermodynamical pressure of the SU(2) Yang Mills gas.
We determine the contributing diagrams and state the Feynman rules.
The computation of the diagrams is performed in the real time formalism of finite temperature field theory.
The resulting integrals have to be evaluated numerically.
We compare the two-loop contribution to the one-loop contribution and give an interpretation.
Chapter~5 gives a short summary and an outlook on further research. 
The appendices contain technical details concerning the calculations in Chapters~3 and 4.

\chapter{Basics of SU(N) Yang-Mills Theory and Solitonic Configurations}

\section{Lie groups and Lie algebras}
For a continuous group $G$, the group elements may be parameterized by a set of continuously varying real parameters, 
$\al^a$ ($a=1,2,\dots,K$).
The total number of parameters $K$ is called the order of the group.
There are groups with compact parameter space such as SU(N), where $K = \rm N^2 - 1$, and SO(N), where $K=\frac{\rm N(N-1)}{2}$.
There are also groups with non-compact parameter space, like the Poincar\'e group in four dimensions, where $K=10$.

\absatz{Lie group}
A Lie group is a continuous group $G$ where the set of parameters represents a differentiable manifold.
The latter is referred to as group manifold.
The multiplication map 
\begin{equation}
G \times G \to G: \, (g_1,g_2) \mapsto g_1 \cdot g_2
\end{equation}
and the inverse map 
\begin{equation}
G \to G: \, g \mapsto g^{-1}
\end{equation}
are differentiable.
In terms of the parameters $\al$ and $\al'$ this means
\begin{equation}
g(\al) \cdot g(\al') = g(\al'')
\,,
\end{equation}
where $\al''$ is an analytic function of $\al$ and $\al'$ and similarly for the inverse map.
As the dependence of the group elements on the parameters $\al^a$ is analytic,
any infinitesimal element $g\in G$ can be power expanded about the unit element of the group:
\begin{equation}
g(\al) = 1 + i \al^a T^a + O(\al^2)
\,.
\end{equation}
The objects 
\begin{equation}
T^a = \left( \frac{ \d g(\al) }{ \d \al^a } \right)_{\al=0}
\end{equation} 
are called infinitesimal generators of the group $G$.
For convenience, the unit element $e$ has parameters $\al^a=0$, $g(0)=e$.

\absatz{Lie algebra}
A vector space $V$ together with a bilinear operation, the Lie-Bracket, given as
\begin{equation}
V \times V \to V : \, (X,Y) \mapsto [X,Y]
\end{equation}
and satisfying
\begin{equation} 											\label{prop1}
[X,X] = 0  
\end{equation}
and the Jacobi identity
\begin{equation}  											\label{jacobi}
[X,[Y,Z]] + [Y,[Z,X]] + [Z,[X,Y]] = 0
\end{equation}
is called a Lie algebra.
The property \eq{prop1} implies the antisymmetry of the Lie bracket,
\begin{equation}
[X,Y] = - [Y,X]
\,.
\end{equation}

\noindent
The generators of a Lie group $G$ always form a Lie algebra $\mathfrak g$.
The dimension of the vector space $V$ is equal to the number of generators (that is the order) of $G$.
If $G$ is a matrix group, then the Lie bracket is the usual matrix commutator 
and the Jacobi identity \eq{jacobi} is trivially fulfilled.

Since the generators $T^a$ of the group provide a basis of the Lie algebra, 
the Lie bracket of two generators must again be a linear combination of generators,
\begin{equation} 											\label{LieAlg}
[T^a,T^b] = i f^{abc} T^c
\,.
\end{equation}
The symbols $f^{abc}$ are called structure constants. 
They can be chosen to be completely antisymmetric and real.
Using the structure constants, the Jacobi identity for the generators
\begin{equation}
[T^a,[T^b,T^c]] + [T^b,[T^c,T^a]] + [T^c,[T^a,T^b]] = 0
\end{equation}
can be phrased as
\begin{equation}
f^{ade} f^{bcd} + f^{bde} f^{cad} + f^{cde} f^{abd} = 0
\,.
\end{equation}

\noindent
Many properties of a Lie group can be derived from the Lie algebra;
e.\,g. the commutation relations \eq{LieAlg} of a Lie algebra 
(which themselves often are called the Lie algebra)
completely determine the multiplication law of the associated Lie group in the vicinity of the unit element. 
The connection between Lie algebra and Lie group is established through the exponential map,
\begin{equation}
t \in \mathfrak g \quad \Rightarrow \quad \exp t \in G
\,,
\end{equation}
which, in case of a matrix group, is the usual exponential map defined by the power series
\begin{equation}
\exp A = \sum_{n=0}^{\infty} \frac{A^n}{n!}
\,.
\end{equation}
There are Lie groups that have the same Lie algebra
but group manifolds of different global structure and topology.

A Lie group $G$ that contains no invariant sub-group (ideal) other than $\{e\}$ and $G$ itself is called simple.
If $G$ does not contain any Abelian invariant sub-group other than $\{e\}$, it is called semi-simple.
The corresponding definitions apply to the Lie algebras.

For a Lie algebra, it is possible that some of the generators $T^a$ commute (i.\,e. their Lie-bracket is zero). 
These generators form the so-called Cartan sub-algebra, their number is known as the rank of the Lie algebra.
For example SU(N) has rank $\rm N-1$; SO(2) and SO(3) both have rank 1.

\absatz{Representations}
A linear representation $R$ of a group $G$ is a map
which maps every group element $g \in G$ onto a linear transformation $R(g)$ of a vector space $V$.
This map has to respect the group multiplication, i.\,e.
\begin{equation}
R(g_1)R(g_2)=R(g_1g_2)
\end{equation}
and to map the unit element $e$ onto the unit matrix,
\begin{equation}
R(e) = \UM
\,.
\end{equation}
In other words, this map is a homomorphism.
When a basis for $V$ is chosen, $R(g)$ are matrices.
The dimension of the representation is by definition the dimension of the vector space $V$.
In quantum mechanics, one is usually interested in finite dimensional unitary representations
because unitarity is closely connected to probability density conservation.

A representation is called reducible if it is possible to find a basis
in which all representation matrices have the form
\begin{equation}
R(g) = \begin{pmatrix} R^1(g) & A(g) \\ 0 & R^2(g) \end{pmatrix} 
\,,
\end{equation}
otherwise it is called irreducible.
If additionally $A(g) = 0$ for all $g\in G$, the representation is called fully reducible.
For a semi-simple group, all reducible representations are fully reducible (Weyl's theorem).

Two representations $R$, $R'$ of a group are said to be equivalent, 
if they only differ by a similarity transformation
\begin{equation}
R'(g) = S^{-1} R(g) S \qquad \forall g \in G
\end{equation}
with a nonsingular matrix $S$.
It can be shown that every representation of a compact group is equivalent to a unitary representation.

To form a representation of the Lie algebra, 
the representation matrices have to respect the commutation relations \eq{LieAlg} of the Lie algebra.
The elements of a matrix group can be viewed as linear transformations of $\mathbb R^n$. 
Thus the group elements themselves form a linear representation of the group, 
the fundamental representation.
The $r$ generators of the Lie group $G$ carry a representation of dimension $r$, the adjoint representation.
The generator $T^a$ is mapped onto the mapping (denoted by the same symbol $T^a$)
\begin{equation}
T^a \colon \mathfrak g \to \mathfrak g; \quad T^b \mapsto [ T^a , T^b ]
\,.
\end{equation}
The representation matrices of the adjoint representation are given by the structure constants,
\begin{equation}
R^{\text{adj}}_{ac}(T^b) = i f^{abc}
\,.
\end{equation}
As a direct consequence of the Jacobi identity \eq{jacobi},
the adjoint representation matrices fulfill the Lie algebra commutation relations \eq{LieAlg}, as requested.
The adjoint representation is always a real representation because the structure constants are real and antisymmetric.
The adjoint representation of a simple Lie group is always irreducible.

\absatz{The group SU(N)}
SU(N) is the group of unitary $\rm N \times N$ complex matrices with unit determinant,
\begin{equation}
U^\dagger U = \UM \,, \qquad \det U = 1
\,.
\end{equation}
It is a compact simple Lie group. Its Lie algebra has rank $\rm N-1$.
The generators of SU(N) are the Hermitian and traceless $\rm N \times N$ matrices  
\begin{equation}
H^\dagger = H \,, \qquad \tr H = 0
\,.
\end{equation}
The generators of SU(2) and SU(3) are usually taken to be the Pauli and Gell-Mann matrices, respectively.
The adjoint representation of SU(N) has dimension $\rm N^2 - 1$.

\section{SU(N) Yang-Mills theory}
We are considering a four-dimensional Minkowskian spacetime.
An SU(N) Yang-Mills theory is governed by a Lagrangian which is invariant under any local SU(N) transformation.
Every field in the theory has to transform under a unitary and finite dimensional representation of SU(N).
A Yang-Mills theory containing only gauge fields but no matter fields is often called pure.

The ordinary derivative $\d_\mu$ cannot be used to construct gauge invariant quantities.
So, in order to be able to include derivative terms (i.\,e. kinetic terms for the fields) in the Lagrangian, 
the covariant derivative $D_\mu$ and a gauge field $A_\mu$ have to be introduced\footnote{
One can conventionally absorb the gauge coupling $e$ in the definition of the gauge field 
and write $D_\mu = \d_\mu - i A_\mu$ and 
$F_{\mu\nu} = \d_{\mu} A_{\nu} - \d_{\nu} A_{\mu} - i \left[ A_{\mu},A_{\nu} \right]$ etc.
This notation is convenient for considering nonperturbative aspects.
We will use it for working with instantons in Secs. \ref{inst0}, \ref{instfin} and in Chapter 3.},
\begin{equation}  
D_{\mu} = \d_{\mu} - i e A_{\mu} 
\,.
\end{equation}
The gauge field $A_\mu$ can be expanded in terms of the Hermitian generators of SU(N),
\begin{equation}
A_{\mu} = A^a_{\mu} \frac{\lam^a}{2}
\,,
\end{equation}
where the SU(N) generators $\lam^a$ are normalized such that 
$\tr \lam^a \lam^b = 2 \delta_{ab}$.
When applying the covariant derivative to a matter field, 
the gauge field $A_\mu$ is understood to act in the representation of the matter field.
That is for a fundamental field $\varphi$
\begin{equation}
D_\mu \varphi = \d_\mu \varphi - i e A^a_\mu \frac{\lam^a}{2} \varphi
\end{equation}
and for an adjoint field $\phi$
\begin{equation}
(D_\mu \phi)_a = \d_\mu \phi_a + e f^{abc} A^b_\mu \phi_c
\,.
\end{equation}

\noindent
The covariant derivative $D_\mu$ is constructed such that
the covariant derivative of a field has exactly the same transformation law as the field itself.
To satisfy this request, the gauge field has to transform under an SU(N) gauge transformation $U(x)$ according to
\begin{equation}
A_\mu (x) \to U(x) \, A_\mu (x) \, U^\dagger (x) + \frac{i}{e} \, U(x) \, \d_\mu \, U^\dagger (x)
\,.
\end{equation}
Viewing the gauge group (being a Lie group) as an analytic manifold, 
the gauge field $A_\mu$ is a connection on this manifold.
The curvature of the manifold (in differential geometry) 
corresponds to the field strength tensor (in field theory), namely
\begin{equation}  											\label{fieldstrength}
F_{\mu\nu} 
= \frac ie \left[ D_{\mu},D_{\nu} \right] 
= \d_{\mu} A_{\nu} - \d_{\nu} A_{\mu} - i e \left[ A_{\mu},A_{\nu} \right]
\,. 
\end{equation}
In contrast to Abelian theories, 
the field strength in nonabelian theories is not a gauge invariant quantity
but transforms under the adjoint representation as
\begin{equation}
F_{\mu\nu}(x) \to U(x) \, F_{\mu\nu} (x) \, U^\dagger(x)
\,.
\end{equation}
The field strength can also be written in matrix notation, 
\begin{equation}
F_{\mu\nu} = F^a_{\mu\nu} \frac{\lam^a}{2}  
\,,
\end{equation}
with the components
\begin{equation}
F^a_{\mu\nu} = \d_\mu A^a_\nu - \d_\nu A^a_\mu + e f^{abc} A^b_\mu A^c_\nu
\,, 
\end{equation}
where $f^{abc}$ are the structure constants of the gauge group.
The kinetic term for the gauge field $A_\mu$ in the Lagrangian is
\begin{equation} 											\label{L}
\mathcal L 
= - \frac12 \tr F_{\mu\nu} F^{\mu\nu}
= - \frac14 F^a_{\mu\nu} F^{\mu\nu a}
\,.
\end{equation}
If one demands Lorentz invariance, gauge symmetry, renormalizability and CP-invariance, 
no further terms are allowed in pure Yang-Mills theory.
Relaxing the demand for CP-invariance, an additional term proportional to
$F^a_{\mu\nu} \tilde F^{\mu\nu a}$
may be added to the Lagrangian in \eq{L}.
Here
$\tilde F_{\mu\nu} = \frac12 \eps_{\mu\nu\alpha\beta} F^{\alpha\beta}$
denotes the dual field strength.

In particular a mass term for the gauge field is forbidden because it has the gauge variant form $m^2 \tr A_\mu A_\mu$.
Nevertheless, gauge bosons can acquire mass by dynamical symmetry breaking which is manifested by the Higgs mechanism.

The commutator term in \eq{fieldstrength} vanishes for Abelian gauge groups,
so the Lagrangian in this case is quadratic in the gauge field, 
the equations of motion are linear in the gauge field,
and hence there is no self interaction.
If, in contrast, the gauge group is nonabelian (as is SU(N)), this is no longer true.
The Lagrangian contains terms cubic and quartic in the gauge field (besides quadratic terms)
and thus allows for three- and four-gauge boson vertices.
Because of this a pure SU(N) Yang-Mills theory is interacting. 

The equations of motion for the field $A_\mu$ are derived via the minimal action principle.
For SU(N) pure Yang-Mills theory, they are
\begin{equation}
D_\mu F^{\mu\nu} = 0
\end{equation}
or in components
\begin{equation}
\d^\mu F^a_{\mu\nu} + e f^{abc} A^{b\mu} F^c_{\mu\nu} = 0
\,.
\end{equation}
The right hand side of the equation of motion is zero because of the absence of external sources, i.\,e. charged matter. 
Nevertheless, this is an interacting theory since the gauge field couples to itself due to the nonlinearity in the field tensor.

On the classical level the Lagrangian \eq{L} does not contain any dimensionful parameter:
The gauge coupling $e$ is dimensionless, and gauge boson masses are forbidden.
This is the reason for the invariance of the action under a rescaling of fields and spacetime arguments.
On the quantum level, however, 
a mass scale comes into existence due to the mechanism of dimensional transmutation \cite{'t Hooft Veltman}.

\section{The Abrikosov-Nielsen-Olesen vortex line}
Consider a theory with a U(1) gauge field $A_\mu$ and a scalar Higgs field $\phi$ defined by the Lagrangian
\begin{equation}											\label{HiggsLagrangian}
\mathcal L = - \frac14 F_{\mu\nu} F^{\mu\nu} + (D_\mu \phi) (D^\mu \phi) - m^2 \phi^* \phi - \lam (\phi^* \phi)^2
\,.
\end{equation}
If $m^2 < 0$, the gauge symmetry is spontaneously broken, and the Higgs field acquires a vacuum expectation value 
\begin{equation}
|\phi|_{\text{vac}} = \sqrt{ \frac{-m^2}{2\lam} }
\,.
\end{equation}
The Higgs Lagrangian \eq{HiggsLagrangian} exhibits static string-like solitonic solutions, so-called vortices.
They were discovered by Nielsen and Olesen in 1973 \cite{Nielsen Olesen}.
Consider a field configuration with cylindrical symmetry and the asymptotic behavior 
\begin{equation}											\label{b1}
\phi = |\phi|_{\text{vac}} \, e^{i n \theta}  \qquad\text{($r\to\infty$)}
\end{equation}
for the Higgs field, and 
\begin{equation}											\label{b2}
A_\mu = \frac1e \d_\mu (n \theta) \qquad\text{($r\to\infty$)}
\end{equation}
for the gauge field.
In cylindrical coordinates $z$, $r \equiv |\vec x|$ and $\theta$, the latter reads
\begin{equation}
\begin{split}
A_z &= 0  \\
A_r &= 0  \\
A_\theta &= - \frac{n}{er}   \qquad\text{($r\to\infty$)}
\,.
\end{split}
\end{equation}
Because of the demand for single valued fields, $n$ has to be an integer.
In the asymptotic, $A_\mu$ is a pure gauge and thus the field strength vanishes,
\begin{equation}
F_{\mu\nu} = 0   \qquad\text{($r\to\infty$)}
\,.
\end{equation}
The Higgs field $\phi$ is covariantly constant, 
\begin{equation}
D_r \phi = 0 \qquad\text{and}\qquad D_\theta \phi = 0   \qquad\text{($r\to\infty$)}
\,,
\end{equation}
and the potential $V = m^2 \phi^* \phi + \lam (\phi^* \phi)^2$ evaluates to zero.
So the energy density at $r\to\infty$ is $\mathcal H = - \mathcal L = 0$, 
and static solutions with finite energy and the above boundary conditions can exist in principal.
The equations of motion obtained from the Langrangian \eq{HiggsLagrangian} are
\begin{equation}
\begin{split}
& D^\mu (D_\mu \phi) = - m^2 - 2 \lam \phi |\phi|^2  \\
& ie ( \phi \, \d_\mu \phi^* - \phi^* \, \d_\mu \phi ) + 2e^2 A_\mu |\phi|^2 = \d^\nu F_{\mu\nu}
\,.
\end{split}
\end{equation}
One can check that the equations of motion allow for configurations with the asymptotic behavior \eqs{b1} and \eqref{b2}.
To find a solution to the equations of motion which satisfies the above boundary conditions,
one makes the ansatz
\begin{equation}
\begin{split}
A_z (r) &= 0   \\
A_r (r) &= 0   \\
A_\theta (r) &\equiv A(r)
\end{split}
\end{equation}
for the gauge field, and
\begin{equation}												\label{vortex ansatz}
\phi = \chi(r) e^{in\theta}
\end{equation}
with
\begin{equation}
\begin{split}
\chi(r) \xrightarrow[r\to0]{} 0   
\qquad\text{and}\qquad   
\chi(r) \xrightarrow[r\to\infty]{} |\phi|_{\text{vac}}
\end{split}
\end{equation}
for the Higgs field.
The magnetic field $\vec B$ will have only a $z$-component,
\begin{equation}
B_z = \frac1r \frac{d}{dr} [r A(r)]
\,.
\end{equation}
Inserting the above ansatz into the equations of motion yields a system of differential equations for the functions $A(r)$ and $\chi(r)$,
\begin{equation}
\begin{split}
& \frac1r \frac{d}{dr} \left( r \frac{d\chi}{dr} \right) - \left[ \left( \frac nr -eA \right)^2 + m^2 + 2\lam\chi^2 \right] \chi = 0
\\
& \frac{d}{dr} \left( \frac1r \frac{d}{dr} (rA) \right) - 2e \left( \frac ne + eA \right) \chi^2 = 0 
\,.
\end{split}
\end{equation}
No exact solutions to these equations are known.
The asymptotic behavior of the gauge field and the magnetic field has been deduced by Nielsen and Olesen as
\begin{equation}
A = - \frac{n}{er} - \frac ce K_1 \big( |e| |\phi|_{\text{vac}} \, r \big) 
\xrightarrow[r\to\infty]{} 
- \frac{n}{er} - \frac ce \left( \frac{\pi}{2|e| |\phi|_{\text{vac}} \, r} \right)^{1/2} e^{-|e| |\phi|_{\text{vac}} \, r} + \cdots
\end{equation}
and
\begin{equation}
B_z = c \chi K_0 \big( |e| |\phi|_{\text{vac}} \, r \big) 
\xrightarrow[r\to\infty]{} 
\frac ce \left( \frac{\pi  |\phi|_{\text{vac}} }{2|e|r} \right)^{1/2} e^{-|e| |\phi|_{\text{vac}} \, r} + \cdots
\,,
\end{equation}
where $K_0$ and $K_1$ denote modified Bessel functions, and $c$ is a constant of integration.

The line integral over $A_\mu$ around a circle $S_1$ at infinity yields the magnetic flux through the surface enclosed,
\begin{equation}
\Phi = \int \vec B \cdot d \vec \sigma = \oint A_\mu dx^\mu = \oint A_\theta \, r \, d \theta = - \frac{2\pi}{e} n
\,,
\end{equation}
The magnetic flux is quantized: it appears only in multiples of the flux quantum $\frac{2\pi}{e}$.

The vortex line owes its existence to the topological structure of the gauge manifold.
The boundary condition \eq{b1} defines a mapping of the boundary $S_1$ in physical space onto the group manifold of U(1), 
which again is $S_1$.
There are infinitely many classes of such mappings
which can not be continuously deformed into one another,
\begin{equation}
\pi_1(U(1)) = \pi_1(S_1) = \mathbb Z
\,.
\end{equation}
A gauge theory, where the gauge group $G$ has $\pi_1(G) = 0$, does not exhibit vortex lines.
This is the case for SU(2), for example.

The Lagrangian \eq{HiggsLagrangian} is the 
relativistic generalization of the Landau-Ginzburg free energy of a type II superconductor.
The electromagnetic field $A_\mu$ interacts with the bosonic field $\phi$, which describes the Cooper pairs.
In the superconducting phase the photon becomes massive, 
and if an external magnetic field can enter the superconducting material (i.\,e. a type II superconductor),
it does so only in the form of (quantized) Abrikosov flux tubes.

\section{The 't~Hooft-Polyakov monopole}
Magnetic monopoles in gauge theories were first considered in 1974 
by 't~Hooft \cite{'t Hooft 1974} and Polyakov \cite{Polyakov 74}
in the context of an SU(2) Yang-Mills theory with an isovector Higgs field $\phi^a$. 
The Langrangian of such a theory is
\begin{equation}
\mathcal L = -\frac14 F^a_{\mu\nu} F^{\mu\nu a} + \frac12 (D_\mu \phi^a) (D^\mu \phi^a)
- \frac{m^2}{2} \phi^a \phi^a - \lambda (\phi^a \phi^a)^2
\end{equation}
with the parameter $m^2$ chosen negative such that the Higgs field has a non-zero vacuum expectation value $F$ with
\begin{equation}
F^2 = - \frac{m^2}{4\lam}
\,.
\end{equation}
Now, consider static solutions with the asymptotic behavior
\begin{equation} 											\label{asymp1}
\begin{split}
A^a_i &= -\eps_{iab} \frac{x_b}{e r^2} \qquad \text{($r\to\infty$)}  \\
A^a_0 &= 0
\end{split}
\end{equation}
for the SU(2) gauge field and
\begin{equation}  											\label{asymp2}
\phi^a = F \frac{x^a}{r} \qquad \text{($r\to\infty$)}
\end{equation}
for the Higgs field.
Here $r$ is the norm of the spatial vector, $r = |\vec x|$.
Note that in \eqs{asymp1} and \eqref{asymp2} space and isospace indices are mixed.
Fields of the form \eq{asymp2} are known as "hedgehogs".

By the presence of a nonvanishing vacuum expectation value of the Higgs triplet
the SU(2) gauge symmetry is broken.
The field strength corresponding to the unbroken U(1) subgroup is the 't~Hooft tensor
\begin{equation}
F_{\mu\nu} = \frac{1}{|\phi|} \phi^a F^a_{\mu\nu} - \frac{1}{e|\phi|^3} \eps_{abc} \phi^a (D_\mu \phi^b) (D_\nu \phi^c)
\,.
\end{equation}
Upon inserting
\begin{equation}
\begin{split}
A_\mu^{1,2} = 0 \qquad\qquad & A_\mu^3 = A_\mu  \\
\phi^{1,2} = 0  \qquad\qquad & \phi^3 = F
\,,
\end{split}
\end{equation}
it reduces to the usual definition of the electro-magnetic field tensor. 
Inserting the nontrivial asymptotic conditions \eqs{asymp1} and \eqref{asymp2} the 't~Hooft tensor evaluates to
\begin{equation}
F_{\mu\nu} = - \frac{1}{er^3} \eps_{\mu\nu a} x^a \qquad \text{($r\to\infty$)}
\,.
\end{equation}
This corresponds to the radial magnetic field of magnetic point charge,
\begin{equation}
B_a = \frac{x^a}{er^3} \qquad \text{($r\to\infty$)}
\end{equation}
with a total flux or magnetic charge
\begin{equation}
q_{mag} = \Phi = \oint_{S_2} d \sigma_a B_a
= \frac{4\pi}{e}
\,.
\end{equation}
In \cite{'t Hooft 1974} it is shown that there exist configurations with the requested asymptotic behavior
that are smooth and hence have finite energy for all $\lam$ and $m^2<0$.
In \cite{Bogomolnyi, Prasad Sommerfield} they are explicitly given for the so-called BPS limit $\lam \to 0$.
In the BPS limit, the mass of the monopole is 
\begin{equation}
M_{\text{m}} = \frac{4\pi}{e^2} M_{\text{W}}
\,,
\end{equation}
where $M_W = e F = \frac{em}{2 \sqrt\lambda}$ is the vector boson mass. 
For general $\lam$, the mass of the monopole is larger but still of the same order.

The magnetic charge can also be expressed as
\begin{equation} 											\label{mag charge}
q_{mag} = \frac{1}{4\pi} \int d^3x \, K^0
\end{equation}
with the current
\begin{equation}   											\label{current}
\begin{split}
K^\mu &= - \frac{1}{2e} \, \eps^{\mu\nu\rho\sigma} \, \eps_{abc} \,
         \d_\nu \hat\phi^a \, \d_\rho \hat\phi^b \, \d_\sigma \hat\phi^c  \\
      &= \d_\nu \tilde F_{\mu\nu}
\,,      
\end{split}       
\end{equation}
where $\hat\phi^a = \frac{\phi^a}{|\phi|}$.
This current is identically conserved, $\d_\mu K^\mu = 0$;
it is not a Noether current corresponding to some symmetry of the Lagrangian, but of a topological nature.
Inserting \eq{current} into \eq{mag charge} and applying Gauss' theorem yields
\begin{equation}  											\label{mag charge 2}
q_m = - \frac{1}{8\pi e} \int_{S_2} d\sigma_i \, \eps_{ijk} \, \eps_{abc} \,
      \hat\phi^a \, \d_j \hat\phi^b \, \d_k \hat\phi^c
\,,
\end{equation}
where the integral has to be performed over an $S_2$ with infinite radius.
From \eq{mag charge 2} it can be seen that the total magnetic flux is completely carried by the Higgs field.
Moreover, \eq{mag charge 2} states that the magnetic charge does only depend on the asymptotic behavior of the fields.
The existence of configurations with non-zero magnetic charge is due to the possibility
to demand nontrivial boundary conditions for the fields, that means:
There are maps from the surface of space ($S_2$) onto the manifold $S_2$ (corresponding to rotations in isospace) 
which can not be continuously deformed into the constant map.
Mathematically speaking, the second homotopy group of $S_2$ is nontrivial,
\begin{equation}
\pi_2(S_2) = \mathbb Z
\,.
\end{equation}
Conservation of magnetic monopole charge is due to this topological argument, 
namely the transition between nonhomotopic gauge configurations needs infinite energy.

\section{Instantons at zero temperature} 								\label{inst0}
Introductory material on instantons and other topological objects in gauge theories is presented in \cite{Ryder}.
Reviews on instanton physics are \cite{Schaefer Shuryak, Gross Pisarski Yaffe}.

\absatz{SU(N) vacuum}
The (Minkowskian) vacuum of an SU(N) Yang-Mills theory is infinitely degenerate.
There is not only the trivial vacuum $A_\mu (x) \equiv 0$, but every field configuration of the form
\begin{equation}  											\label{pure gauge}
A_\mu(x) = i \, U(x) \, \d_\mu \, U^\dagger(x)
\,,
\end{equation}
where $U(x)$ is an SU(N) matrix, 
differs from $A_\mu=0$ only by a gauge transformation 
and hence has zero field strength and energy density, as well.
Configurations as in \eq{pure gauge} are referred to as pure gauge.
These pure gauge configurations fall into topologically distinct classes, 
and hence they cannot be smoothly connected.
The configurations \eq{pure gauge} are classified by a winding number, the Pontryagin index,
defined as
\begin{equation}
n_W = \frac{1}{24\pi^2} \int d^3x \, \eps^{ijk} \tr (U^\dagger \d_i U)(U^\dagger \d_j U)(U^\dagger \d_k U) 
\,.
\end{equation}
The quantity $n_W$ is an invariant under smooth deformations. 
It can also be expressed in terms of the gauge field as
\begin{equation} 											\label{ChernSimons0}
n_W = \frac{1}{16\pi^2} \int d^3x \, \eps^{ijk} \left( A_i^a \d_j A_k^a + \frac13 f^{abc} A_i^a A_j^b A_k^c \right)
\,.
\end{equation}
It is not possible to find a solution to the equation of motion with finite energy
that connects two vacua with different winding number.
Configurations corresponding to a transition between different vacua of the theory 
via a tunneling process are called instantons.
They have to be considered in the framework of Euclidean field theory.

\absatz{Tunneling solutions}
Tunneling configurations are solutions to the Euclidean equations of motion
(i.\,e. they minimize the action of the theory, formulated in a Euclidean spacetime).
This can be motivated by an example from the quantum mechanics of a point particle.

\begin{figure}[t]
\includegraphics[width=7cm]{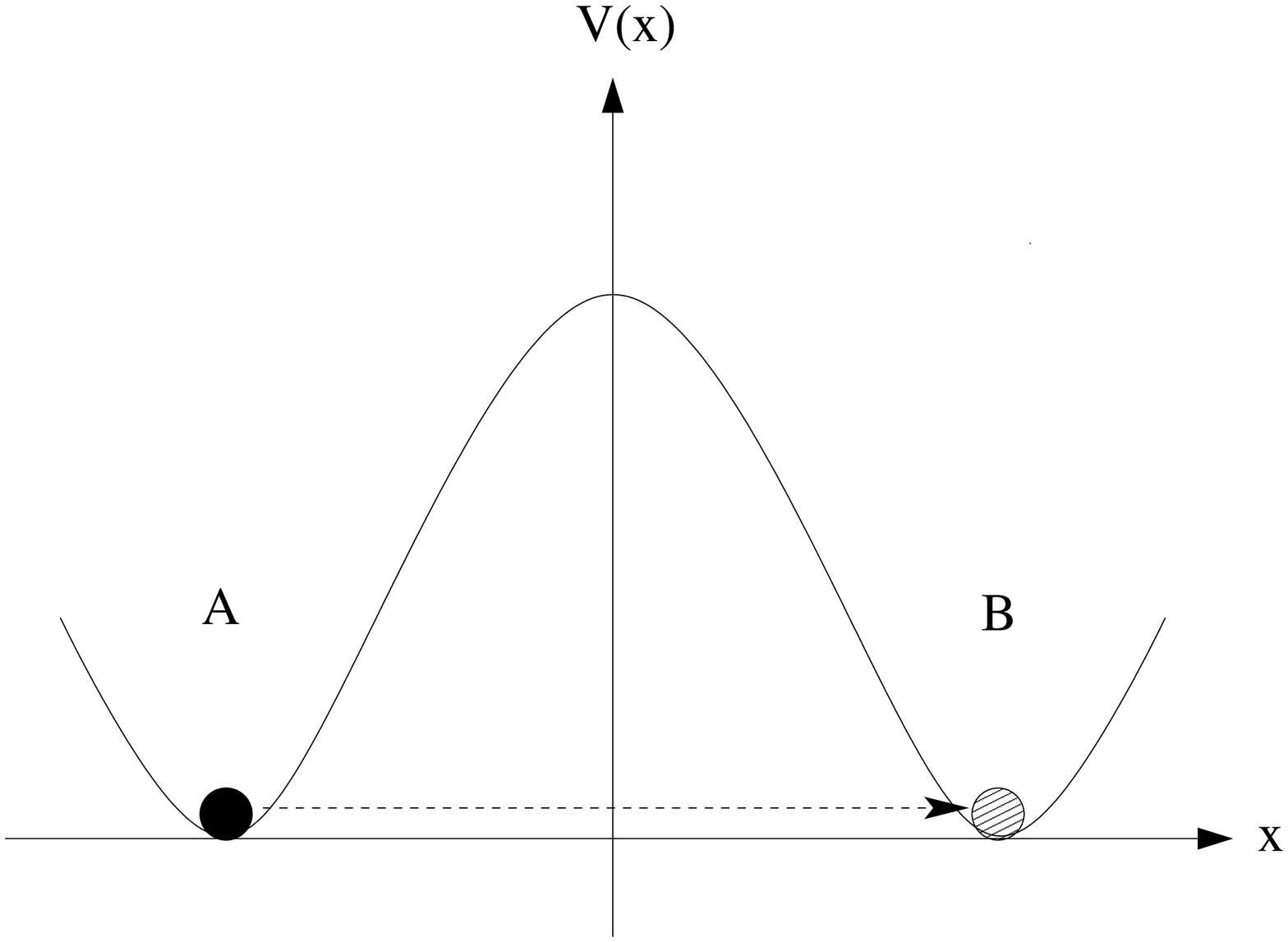}
\hfill
\includegraphics[width=7cm]{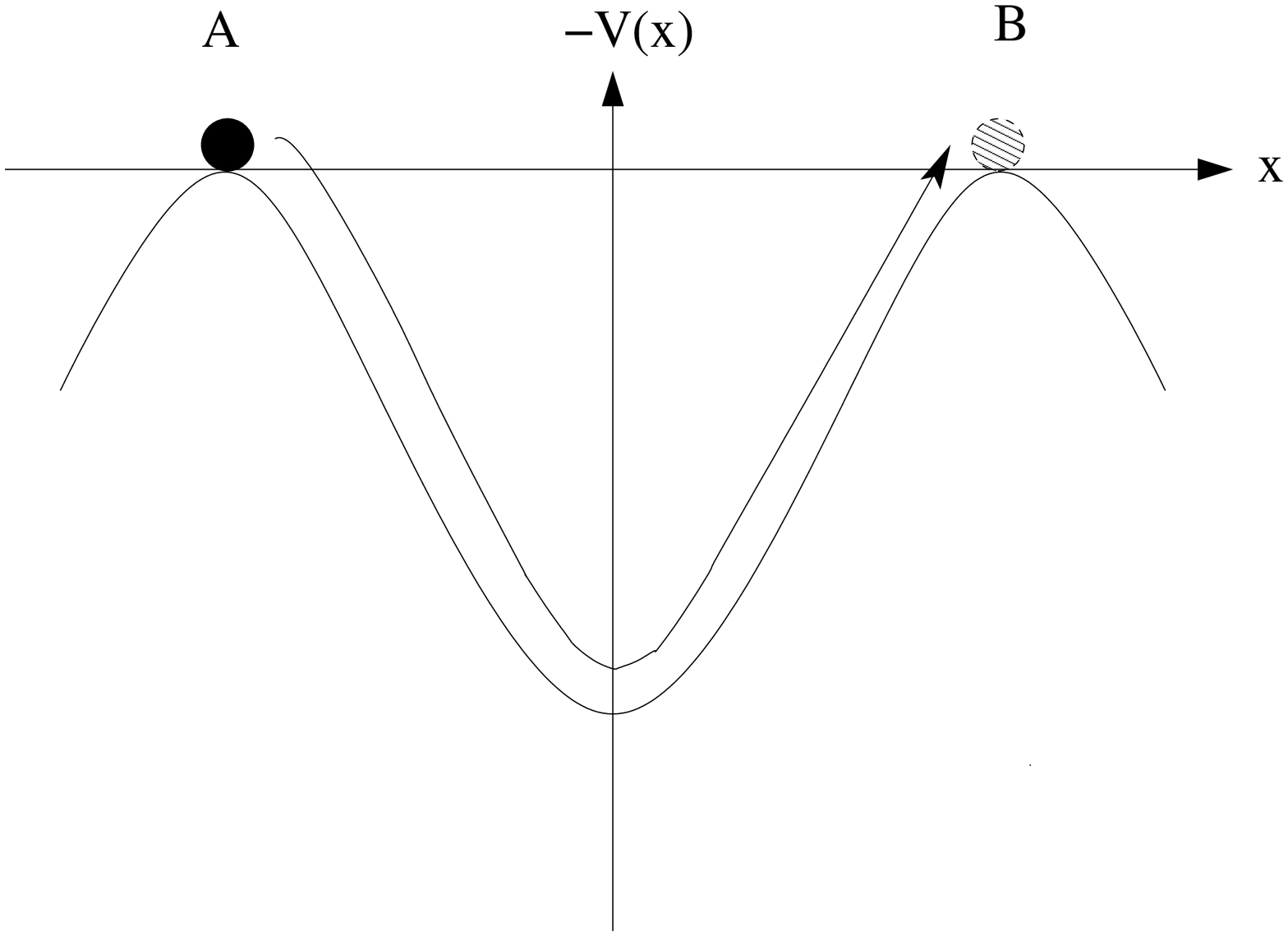}
\caption{  												\label{tunnel}
Left-hand side: In the potential $V(x)$ the transition from A to B is a tunneling process
and hence classically impossible. 
Right-hand side: In the reversed potential $-V(x)$ the transition from A to B is classically possible 
along the indicated trajectory because allowed and forbidden region have been interchanged.}
\end{figure}

Imagine a point particle with kinetic energy $T$ moving in a time independent potential $V$.
A semiclassical approximation (that is an expansion around the classical path)
will not include any tunneling process because there are no classical tunneling paths.
Now letting
\begin{equation}  											\label{sign}
E-V \to V-E  
\end{equation}
clearly interchanges allowed and forbidden region, so that a classical path exists, see \fig{tunnel}.
An expansion around this classical path yields the WKB approximation for the tunneling process.
But $E-V=T$, and $T$ is the square of a time derivative.
Hence the sign change \eq{sign} is equivalent to the Euclidean continuation,
\begin{equation}
t \to i \tau \qquad\qquad \text{($\tau$ real)}
\,.
\end{equation}
So tunneling processes are solutions to the Euclidean equations of motion.
This is true in quantum field theory as well.

\absatz{Finite action and boundary conditions}
In order to get finite action configurations, 
one has to demand that the field strength $F_{\mu\nu}$ approaches zero at the boundary of spacetime faster than $1/x^2$.
This can not only be fulfilled trivially (by approaching $A_\mu=0$), 
but also nontrivially by demanding 
that the gauge field $A_{\mu}$ is pure gauge on the boundary of Euclidean spacetime (which is $S_3$),
\begin{equation}
A_{\mu} = i \, U(x) \d_{\mu} U^{\dagger}(x)
\qquad \text{($\abs{x}\to\infty$)}
\,.
\end{equation}
This condition defines a mapping from the surface $S_3$ to the group space of SU(N).
The equivalence classes of homotopic mappings from $S_3$ into a manifold $X$, $f:S_3\to X$, 
form the so-called third homotopy group of the manifold, $\pi_3(X)$.
So the finite action field configurations fall into topologically distinguished classes
if and only if the third homotopy group of the gauge group manifold is nontrivial.
The homotopy group $\pi_3$ of most gauge groups is known, see for example \cite{WeinbergII}.
The case of basic interest is SU(2), where
\begin{equation}  											\label{pi3}
\pi_3(\text{SU(2)}) = \pi_3(S_3) = \mathbb Z
\,.
\end{equation}
The integer attached to a given mapping $f\colon S_3 \to S_3$ indicates how often the gauge group manifold $S_3$ is "wrapped" 
around the spacetime-$S_3$ and hence is referred to as winding number.
It is also known as the Brouwer degree of the mapping $f$.

\absatz{Pontryagin index and topological charge}
For a given field configuration $F_{\mu\nu}$ the four dimensional Pontryagin index is defined as
\begin{equation}    											\label{top charge}
Q = \frac1{32\pi^2} 
\int d^4x \, F^a_{\mu\nu} \tilde F^a_{\mu\nu}
\,.
\end{equation}
The integrand of \eq{top charge}, the Pontryagin density, can also be written as a total divergence,
\begin{equation}                                            
\frac1{32\pi^2} F^a_{\mu\nu} \tilde F^a_{\mu\nu}
= 
\d_\mu K_\mu
\,,
\end{equation}
where $K_\mu$ is the Chern-Simons current,
\begin{equation}   											\label{ChernSimons}
K_\mu = \frac{1}{16 \pi^2} \, \eps_{\mu\al\beta\gamma}
\left( A^a_\al \d_\beta A^a_\gamma + \frac13 f^{abc} A^a_\al A^b_\beta A^c_\gamma \right)
\,.
\end{equation}
In contrast to the Pontryagin density, the Chern-Simons current is a gauge variant quantity.
If the integrand is nonsingular, the volume integral in \eq{top charge}
can be converted to a surface integral by means of Gauss' theorem,
\begin{equation}   											\label{top charge surface}
Q = \int d^4x \, \d_\mu K_\mu = \int d\sigma_\mu \, K_\mu 
\,.
\end{equation}
Although this surface integral has to be evaluated on a sphere $S_3$ with infinite radius,
it is not necessarily zero:
If the gauge field $A_\mu$ falls off less than $1/|x|$, then the Pontryagin index will be non-zero.

From \eq{top charge surface} it is obvious that (for regular gauge fields) the Pontryagin index
is solely determined by the asymptotic behavior of the gauge field 
or in other words by the boundary conditions.
Thus the existence of fields with non-zero Pontryagin index is due to the possibility of
demanding nontrivial boundary conditions, as already indicated in the context of \eq{pi3}.
The Pontryagin index is a topological charge because its conservation 
does not follow from a continuous symmetry of the Lagrangian
but is solely due to the stability of the boundary conditions under continuous perturbations.

\absatz{Bogomolnyi bound and self-duality}
For the construction of a configuration of minimal Euclidean action
connecting two vacua with different winding (Chern-Simons) number, the Yang-Mills action 
is written in form of a Bogomolnyi decomposition,
\begin{equation}   											\label{bogomolnyi}
\begin{split}
S &= \frac{1}{4g^2} \int d^4x \, F^a_{\mu\nu} F^a_{\mu\nu}  \\
&= \frac{1}{4g^2} \int d^4x \left( \pm F^a_{\mu\nu} \tilde F^a_{\mu\nu}
+ \frac12 ( F^a_{\mu\nu} \mp \tilde F^a_{\mu\nu} )^2 \right)
\end{split}
\end{equation}
where
\begin{equation}
\tilde F_{\mu\nu} = \frac12 \, \eps_{\mu\nu\al\beta} F_{\al\beta}
\end{equation}
is the dual field strength tensor. 
The second term in \eq{bogomolnyi} is a square and hence always positive;
the first term is the Pontryagin index $Q$ defined as in \eq{top charge}, a topological invariant.
Hence any self-dual (or anti self-dual) field configuration, i.\,e.
\begin{equation} 											\label{SD}
F_{\mu\nu} = \pm \tilde F_{\mu\nu}
\,,
\end{equation}
is a minimum of the Euclidean action (in a given topological sector).
It has the action
\begin{equation}
S = \frac{8\pi^2}{g^2} |Q|
\,.
\end{equation}

\noindent
In the dual field strength $\tilde F_{\mu\nu}$
the roles of electric and magnetic fields are interchanged as opposed to $F_{\mu\nu}$.
So, (anti) self-duality can (at least in temporal gauge) also be characterized by 
\begin{equation}
\vec E^a = \pm \vec B^a
\,,
\end{equation}
where $\vec E^a$ and $\vec B^a$ are the color electric and magnetic fields respectively.
From the decomposition \eq{bogomolnyi} it is clear that a self-dual gauge field is a (local) minimum
of the Yang-Mills action, and indeed self-duality and the Bianchi identity imply that the equations of motion are satisfied,
\begin{equation}   											\label{eom}
D_\mu F_{\mu\nu} = \pm D_\mu \tilde F_{\mu\nu} = 0
\,.
\end{equation}
In contrast to the equation of motion \eqref{eom}, which is a second-order differential equation,
the self-duality equation \eqref{SD} is first order.
The Bogomolnyi decomposition \eq{bogomolnyi} gives a lower bound for the action in a given topological sector.
Configurations saturating this bound (i.\,e. self-dual configurations) are also referred to as BPS-saturated,
\cite{Bogomolnyi, Prasad Sommerfield}.
It can be shown that the energy-momentum tensor vanishes identically on BPS-saturated fields.

\absatz{BPST instanton}
Here we consider the gauge group SU(2).
In the boundary condition
\begin{equation}
A_\mu \to i U \d_\mu U^\dagger
\end{equation}
which is necessary to achieve a configuration of finite Euclidean action (see above), one chooses
\begin{equation}
U(x) = \frac{ x_4 + i x_i \lambda_i }{ |x| }   \qquad\text{($|x| \to \infty$)}
\,,
\end{equation}
where $|x| = \sqrt{ x_\mu x_\mu }$, and $\lam_i$ the are Pauli matrices.
Therefore the gauge potential has to have the asymptotic form
\begin{equation}                                        
A^a_\mu \to 2 \et a\mu\nu \frac{x_\nu}{x^2}  \qquad\text{($|x| \to \infty$)}
\,,
\end{equation}
where the 't~Hooft symbols $\et a\mu\nu$ and $\etb a\mu\nu$ are defined as
\begin{equation}
\begin{split}
\et a\mu\nu &= \eps_{a\mu\nu} + \delta_{a\mu} \delta_{\nu4} - \delta_{a\nu} \delta_{\mu4}
\\
\etb a\mu\nu &= \eps_{a\mu\nu} - \delta_{a\mu} \delta_{\nu4} + \delta_{a\nu} \delta_{\mu4}
\,.
\end{split}
\end{equation}
This leads to the ansatz
\begin{equation}  											\label{ansatz}
A^a_\mu = 2 \et a\mu\nu \frac{ x_\nu f(x^2) }{x^2}
\,,
\end{equation}
where the scalar function $f$ has to satisfy the boundary condition $f \to 1 \text{ for } |x| \to \infty$.
Inserting this ansatz into the self-duality equation \eqref{SD} yields the differential equation
\begin{equation}
f (1-f) - x^2 f' = 0
\,,
\end{equation}
which is solved by
\begin{equation}
f = \frac{x^2}{x^2+\rho^2}
\,,
\end{equation}
where $\rho$ is a constant of integration.
This gauge field configuration is the Belavin-Polyakov-Schwartz-Tyupkin instanton \cite{BPST}
\begin{equation} 											\label{BPST}
A^a_\mu (x) = 2 \et a\mu\nu \frac{ x_\nu }{ x^2 + \rho^2 }
\,.
\end{equation}
For this instanton solution, the Chern-Simons current \eq{ChernSimons} has the asymptotic behavior
\begin{equation}
K_\mu \to \frac{1}{2\pi^2} \frac{x_\mu}{|x|^4} \qquad\text{($|x|\to\infty$)}
\,,
\end{equation}
and inserting this into \eq{top charge surface} yields $Q=1$ for the topological charge of the BPST instanton. 
Replacing $\et a\mu\nu$ with $\etb a\mu\nu$ yields a solution with $Q=-1$.
The corresponding field strength is
\begin{equation}
F^a_{\mu\nu} = - 4 \et a\mu\nu \frac{ \rho^2 }{ (x^2+\rho^2)^2 }
\,.
\end{equation}
Although the instanton gauge potential only falls of as $1/x$,
the field strength $F_{\mu\nu}^a$ falls of as $1/x^4$.
So instantons are well localized in (Euclidean) time and in space.

The solution \eq{BPST} can be generalized by shifting the instanton center to an arbitrary point $z_\mu$,
\begin{equation} 											\label{BPST1}
A^a_\mu (x) = 2 \et a\mu\nu \frac{ (x-z)_\nu }{ (x-z)^2 + \rho^2 }
\,,
\end{equation}
and changing the color orientation of the instanton by a global gauge rotation (containing three parameters).
Furthermore the BPST instanton contains an arbitrary parameter $\rho$, 
which gives the size of the instanton.
These make up eight parameters in the generalized BPST instanton.
Since the action of the configuration \eq{BPST1}, namely
\begin{equation}
S = \frac{8\pi^2}{g^2} \abs{Q} = \frac{8\pi^2}{g^2} 
\,,
\end{equation}
is independent of these eight parameters, they correspond to zero modes.
The gauge in which the BPST instanton has the form \eqref{BPST} is referred to as regular gauge.

\absatz{Singular gauge}
The BPST instanton \eq{BPST} has a singularity at infinity.
This singularity can by means of the singular gauge transformation
\begin{equation}
U(x) = \frac{ x_4 + i x_i \lambda_i }{ |x| } 
\end{equation}
be shifted from infinity to the origin, such that the instanton 
in so-called singular gauge has the gauge potential 
\begin{equation} 											\label{singular}
A^a_\mu (x) = 2 \etb a\mu\nu \frac{x_\nu}{x^2} \frac{ \rho^2 }{ x^2 + \rho^2 }
= - \etb a\mu\nu \, \d_\nu \ln \left( 1+ \frac{\rho^2}{(x-z)^2} \right)
\,.
\end{equation}
The singularity of the gauge field $A_\mu$ at $x = 0$ is not physical;
field strength, action density and topological charge density are smooth.
The field strength of an instanton in singular gauge is 
\begin{equation}
F^a_{\mu\nu} (x) 
= 
- \frac{4\rho^2}{(x^2+\rho^2)^2} 
\left( \etb a\mu\nu - 2 \etb a\mu\kappa \frac{x_\kappa x_\nu}{x^2} - 2 \etb a\kappa\nu \frac{x_\mu x_\kappa}{x^2} \right)
\,.
\end{equation}
In regular gauge the integral over the topological charge density 
picked up only contributions at $|x|\to\infty$, cf. \eq{top charge surface}.
In contrast to that, in singular gauge not only the action density, 
but also the topological charge density is localized at the center of the instanton.
It is this property that allows for the construction of multi instantons in singular gauge.

\absatz{Multi instanton solutions}
The 't~Hooft ansatz for the gauge potential
\begin{equation}   											\label{thooftansatz}
A^a_{\mu} (x) = - \etb a\mu\nu \, \d_{\nu} \ln \Pi(x)
\end{equation}
is a generalization of the BPST instanton in singular gauge, \eq{singular}.
The demand for self-duality translates into a condition for the scalar pre-potential $\Pi(x)$, 
it has to fulfill the Laplace equation
\begin{equation}
\Pi^{-1}(x) \cdot \d_\mu \d_\mu \Pi(x) = 0
\,.
\end{equation}
Hence there is (only in singular gauge) some kind of superposition principle:
Superposing the pre-potentials of single instantons with different centers and scale parameters
again yields a self-dual solution to the equations of motion, a multi instanton.
Therefore the pre-potential $\Pi(x)$ has the general form
\begin{equation} 											\label{prepot K}
\Pi (x) = 1 + \sum_{n=1}^K \frac{\rho_n^2}{(x-z_n)^2}
\,.
\end{equation}
The configuration \eq{thooftansatz} with the pre-potential \eq{prepot K} inserted has Pontryagin index $Q = K$. 
Replacing $\et a\mu\nu$ in the 't~Hooft ansatz with $\etb a\mu\nu$ yields a solution with $Q = -K$.
The 't~Hooft ansatz in particular allows for the construction of periodic instantons.

\absatz{Collective coordinates}
The collective coordinates of the SU(2) instanton with topological charge $|Q|=1$ are 
the instanton size $\rho$, the instanton position $z$
and three parameters determining the color orientation, making up 8 parameters in total.
The action on an instanton is independent of all theses parameters; 
i.\,e. they are  moduli of the instanton solution and correspond to zero modes.

A multi instanton of topological charge $K$ is the superposition of $K$ single instantons 
and accordingly has $8K$ parameters, \cite{Gross Pisarski Yaffe}.  
Note that the multi instanton solution as displayed in \eqs{thooftansatz} and \eqref{prepot K}
does not cover all of this moduli space:
the $K$ instantons are chosen to have the same color orientation, which must not necessarily be the case.
The general SU(2) multi instanton solution has been constructed by Atiyah et.~al. \cite{ADHM}.

\absatz{Gauge groups other than SU(2)}
For the case of a simple gauge group $G$, basically no instantons other than in SU(2) arise, 
but the various possible embeddings of SU(2) in $G$ have to be taken into account.
The general SU(N) multi-instanton with Pontryagin index $Q$ has $4{\rm N}Q$ parameters,
\cite{Schaefer Shuryak, Gross Pisarski Yaffe}.

\absatz{Tunneling interpretation}
Assuming that the (regular) gauge potential falls of rapidly at spatial infinity,
one can write the topological charge $Q$ from \eq{top charge surface} as
\begin{equation}
\begin{split}
Q 
&= \int d^4x \, \d_\mu K_\mu  \\
&= \int d\tau \, \frac{d}{d\tau} \int d^3x \, K_0 + \int d\tau \int d\sigma_i K_i  \\
&= \int_{\tau=\infty} d^3x \, K_0 + \int_{\tau=-\infty} d^3x \, K_0  \\
&= n_{W} (\tau=\infty) - n_{W} (\tau=-\infty)
\,,
\end{split}
\end{equation}
where the fact that the zero component of the Chern-Simons current \eq{ChernSimons}
indeed is equal to the integrand of \eq{ChernSimons0} was used.
That means an instanton with Pontryagin index $Q \neq 0$ connects topologically different vacua
with the according difference in winding number via tunneling.

\section{Instantons at finite temperature}							\label{instfin}
\absatz{Finite temperature}
The equilibrium thermodynamics of a quantum field theory at finite temperature 
usually is considered in the framework of Euclidean field theory.
The finite temperature $T$ corresponds to imaginary time being compactified 
on a circle with circumference $\beta = \frac1T$.
The physical fields have to be periodic in Euclidean time.
For a gauge field the periodicity condition is less strict;
only periodicity up to a gauge transformation is demanded, 
\begin{equation}
A_\mu (\beta,\vec x) = \Omega(0,\vec x) \, A_\mu (0,\vec x) \, \Omega^\dagger(0,\vec x) 
			+ \frac ig \Omega(0,\vec x) \, \d_\mu \, \Omega^\dagger(0,\vec x)   
\end{equation}
with $\Omega \in \text{SU(N)}$.
We now have to discern time and space and write $x_\mu = (\tau,\vec x)$.

\absatz{Periodic instanton}
To find an instanton solution which is periodic in Euclidean time, 
one starts from the 't~Hooft ansatz \eq{thooftansatz} with the pre-potential \eq{prepot K}
describing a multi instanton built out of $K$ instantons with scales $\rho_n$ and centered at $z_n$ respectively.
All of them have the same color orientation.
Taking the sum over infinitely many instanton pre-potentials centered at $(n\beta,0)$, $n\in \mathbb Z$
with the same scale parameter $\rho$ yields a periodic instanton (or caloron) centered at $(0,0)$,
\begin{equation}
\Pi(\tau, \vec x) = 1 + \sum_{n=-\infty}^{\infty} \frac{\rho^2}{(\tau-n\beta,\vec x)^2}
\,.
\end{equation}
The sum has been performed by Harrington and Shepard, see \cite{Harrington Shepard},
\begin{equation} 											\label{prepotential}
\Pi (\tau,\vec x) = 1 + \frac{\pi \rho^2}{\beta r} 
\frac{ \sinh \frac{2\pi}{\beta} r }
{ \cosh \frac{2\pi}{\beta} r - \cos \frac{2\pi}{\beta} \tau }
\,,
\end{equation}
where $r = \abs{\vec x}$.
The caloron of \eq{prepotential} still has the action 
$S = \frac{8\pi^2}{g^2}$;
the action is, on the classical level, independent of temperature.
By superposing (the pre-potentials of) single-calorons a caloron solution of higher topological charge can be constructed.
As in the case of the multi-instanton, this is only possible in singular gauge.
Calorons of higher topological charge also possess more moduli,
some of which (namely the core sizes and the core distances) are dimensionful.

\absatz{Polyakov loop}
The Polyakov loop is a time-like Wilson loop 
\begin{equation}
\mathbf P (\vec x) = \PO \exp \left[ ig \int_0^{1/T} d\tau \, A_4(\tau,\vec x) \right]
\,,
\end{equation}
where $\PO$ denotes the path ordering operation.
It has to be evaluated in periodic gauge, $A_\mu(\tau+\beta,\vec x) = A_\mu(\tau,\vec x)$.
The Polyakov loop is only defined at finite temperature.
At spatial infinity, the Polyakov loop does not depend on $\vec x / r$ any longer.
Its value $\mathbf P(|\vec x| \to \infty)$ is a topological invariant.

Calorons are classified according to the eigenvalues of their Polyakov loop at spatial infinity.
By definition, trivial holonomy means for SU(2) calorons that
$\mathbf P (|\vec x| \to \infty) = \pm \UM$.
Calorons with non-unity eigenvalue of $\mathbf P$ at spatial infinity are said 
to have nontrivial holonomy, accordingly.
The caloron solution from \eq{prepotential} has Polyakov loop $\mathbf P = \UM$.
More general solutions have been constructed by Nahm \cite{Nahm 1984}, Lee, Lu \cite{Lee Lu},
and Kraan, van Baal \cite{Kraan van Baal 1}

\absatz{Monopole constituents}
In 1998 it has been shown independently by Lee and Lu \cite{Lee Lu} 
and Kraan and van Baal \cite{Kraan van Baal 1, Kraan van Baal 2}
that SU(N) calorons of nontrivial holonomy contain constituent BPS magnetic monopoles.
These monopoles are subject to an attractive interaction in the case of small holonomy (i.\,e. close to trivial holonomy),
or to a repulsive potential for large holonomy (i.\,e. far from trivial holonomy), see \cite{Diakonov et. al.}.
The latter can lead to dissociation of the caloron into a monopole-antimonopole pair.

\chapter{Composite Adjoint Higgs Field in SU(2) Yang-Mills Theory} 
In \cite{Hofmann} an analytical and nonperturbative 
approach to the thermodynamics of SU(2) and SU(3) Yang-Mills theory in four dimensions is developed.
An essential ingredient is the existence of an adjoint Higgs field $\phi$ composed of trivial holonomy calorons.

Sec.~\ref{outline} gives a brief outline of this approach as presented in \cite{Hofmann}.
We confine ourselves to the case of SU(2) and its so-called electric phase.
We will give and evaluate a microscopic definition of the composite field $\phi$ in Sec.~\ref{phi}, 
and briefly comment on the nonexistence of alternative possibilities for its definition.

\section{An outline of the approach in \cite{Hofmann}}             					\label{outline}
\absatz{Decomposition of gauge fields}
Each SU(2) gauge field $A_\mu$ appearing in the partition function of the fundamental theory
is uniquely decomposed into  
a topologically nontrivial and BPS saturated part $A_\mu^{top}$ represented by calorons, 
and a topologically trivial remainder $a_\mu$,
\begin{equation}
A_\mu = A_\mu^{top} + a_\mu
\,.
\end{equation}
At a large temperature, calorons with topological charge one and trivial-holonomy 
are assumed to generate a macroscopic adjoint scalar field $\phi$.
Interactions between the trivial holonomy calorons via topologically trivial fluctuations are not included in the field $\phi$ 
but are accounted for at a later stage by means of a pure-gauge background $a_\mu^{bg}$.
The field $\phi$ has to have the following properties:

1. It describes (part of) the ground state of a thermal system, 
so its (gauge invariant) modulus $|\phi|$ must be independent of space and time.

2. The action of a classical caloron is independent of temperature.
So no explicit $T$ dependence may arise in $\phi$'s definition.

3. Calorons are BPS saturated (or self-dual) solutions to the Yang-Mills equations of motion 
in four-dimensional Euclidean spacetime (with time $\tau$ compactified on a circle).
In particular, their energy-momentum tensor is precisely zero.
The adjoint scalar field $\phi$, being composed of noninteracting calorons,
inherits this property, which in turn is expressed through a BPS equation for $\phi$'s time dependence,
\begin{equation}
\d_\tau\phi = V^{(1/2)}
\,.
\end{equation}
Here,  $V^{(1/2)}$ is a 'square-root' of the potential $V (\phi) = \tr (V^{(1/2)})^\dagger V^{(1/2)}$, 
which governs the dynamics of $\phi$.
The potential $V(\phi)$ is determined by the demand for BPS saturation.

In Sec.~\ref{phi} we will give a microscopic definition for $\phi$ and after evaluation see 
that this field is (up to a global gauge rotation) of the form
\begin{equation}
\phi(\tau) = \sqrt{ \frac{\Lambda^3}{2\pi T} } \; \lam_1 \exp (-2\pi i T \lam_3 \tau )
\,.
\end{equation}
The potential is found to be
\begin{equation}
V(\phi) = \Lambda^6 \tr \phi^{-2} = 4 \pi T \Lambda^3
\,.
\end{equation}
Here, $\Lam$ is a fixed mass scale generated by dimensional transmutation.
The modulus $|\phi|$ falls off as $\frac1{\sqrt T}$. 
This dependence shows that the nontrivial-topology sector is strongly suppressed at large temperature.

\absatz{Effective theory}
Minimally coupling of $\phi$ to the (up to now not included) topologically trivial fluctuations $a_\mu$
results in the effective action for the electric phase,
\begin{equation}  											\label{Seff}
S_E 
= \int_0^{1/T} d\tau \int d^3x \left( \frac12 \tr G_{\mu\nu} G_{\mu\nu} 
+ \tr \mathcal D_{\mu} \phi \mathcal D_{\mu} \phi + \Lam^6 \phi^{-2} \right)
\,,
\end{equation}
where 
\begin{equation}
G^a_{\mu\nu} = \d_{\mu} a^a_{\nu} - \d_{\nu} a^a_{\mu} - e \eps^{abc} a^b_{\mu} a^c_{\nu}
\end{equation}
is the field strength of topologically trivial fluctuations $a_\mu$,
\begin{equation}
\mathcal D_{\mu} \phi = \d_{\mu} + i e \left[ \phi , a_{\mu} \right]
\end{equation}
is the covariant derivative, and $e$ denotes the effective gauge coupling.
There is no reason why the effective gauge coupling constant 
should be equal to the coupling constant of the fundamental theory.

The field $\phi$ is seen to be quantum mechanically and statistically inert:
The mass associated with its excitations is much larger than
both temperature and the scale of admissible quantum fluctuations of $|\phi|$ itself,
\begin{equation}    
\frac{ \d_{|\phi|}^2 V }{ T^2 } = 12 \pi^2
\qquad \text{and} \qquad 
\frac{ \d_{|\phi|}^2 V }{ |\phi|^2 } = 3\lam^3
\,,
\end{equation}
where $\lam = \frac{2\pi T}{\Lam}$ is the dimensionless temperature.
In the electric phase of the theory, $\lam > 11.65$ (see below), so that $3\lam^3$ indeed is a large number.
The quantum mechanically and thermodynamically stabilized field $\phi$
can now be taken as a background for the equation of motion governing topologically trivial fluctuations,
\begin{equation}													\label{EoM}
\mathcal D_\mu G_{\mu\nu} = 2ie [\phi,\mathcal D_\nu \phi]
\,.
\end{equation}
This solution is to be part of the ground-state description 
and hence must not break the rotational invariance of the system. 
So it has to be a pure gauge. 
A pure-gauge solution to \eq{EoM} reads 
\begin{equation}
a_\mu^{bg} = \delta_{\mu4} \frac{\pi}{e} T \lambda_3  
\,.
\end{equation}
It takes into account holonomy changing interactions between calorons mediated by topologically trivial fluctuations.
Moreover, we have 
\begin{equation}
\mathcal D_\mu \phi = 0
\end{equation}
on the ground-state configuration $\phi$, $a_\mu^{bg}$.
As a consequence, the action density in \eq{Seff} evaluates to $V(\phi)$ on the ground-state. 
This corresponds to a ground-state energy density $\rho_{g.s.} = V(\phi) = 4\pi\Lam^3T$ 
and a ground-state pressure $P_{g.s.} = -V(\phi) = -4\pi\Lam^3T$.

The inert scalar $\phi$ and the pure-gauge solution $a_\mu^{bg}$ form the ground state 
about which loop expansions are performed.
Therefore, the topologically trivial sector $a_\mu$ is split into the pure-gauge ground-state part $a_\mu^{bg}$
and fluctuations $\delta a_\mu$ about this background. 
The adjoint scalar $\phi$ renders some of the fluctuations $\delta a_\mu$ massive 
through the adjoint Higgs mechanism.
In the case of SU(2), the gauge symmetry is dynamically broken\footnote{
With the introduction of a composite field which breaks the gauge symmetry partially 
this approach conceptually resembles the Landau-Ginzburg-Abrikosov theory of superconductivity \cite{Ginzburg Landau, Abrikosov}.} 
to U(1).
Two of the gauge bosons acquire the temperature dependent mass
\begin{equation}  											\label{TLH mass}
m^2 = 4e^2 |\phi|^2 = 4 e^2 \frac{\Lam^3}{2\pi T}
\,,
\end{equation}
while the third gauge boson is left massless\footnote{
The calculation of the mass spectrum as well as other 
explicit calculations (such as the thermodynamical pressure in Chapter~\ref{pressure})
are carried out after a transformation to unitary gauge, where $a_\mu^{bg} \equiv 0$ and hence
\begin{equation*}
G^a_{\mu\nu} [a_\mu] = G^a_{\mu\nu} [\delta a_\mu]
\,.
\end{equation*}
The scalar field in this gauge takes the form
\begin{equation*}
\phi(\tau) = \sqrt{ \frac{\Lam^3}{2\pi T} } \; \lam_1 
\,.
\end{equation*}
It can be shown that the employed gauge transformation is admissible, see \cite{Hofmann}.
}.
The former are referred to as tree-level heavy (TLH) modes, the latter as tree-level massless (TLM) modes.
Since the masses of TLH excitations are temperature dependent, they are thermal quasiparticle fluctuations.
The masses induced by the Higgs mechanism provide infrared cutoffs, so that no infrared divergences occur
in loop expansions of thermodynamical quantities.

\absatz{Compositeness constraints}
The existence of the compositeness scale $|\phi|$
present in the vacuum additionally implies cutoffs for the momenta propagating as vacuum fluctuations:

1. The offshellness of a quantum fluctuation must not exceed the scale $|\phi|$,
\begin{equation}												\label{no1}
| p^2 - m^2 | \le | \phi |^2
\end{equation}
where $p$ is the four-momentum of the quantum fluctuation and $m$ its mass.

2. The center-of-mass energy flowing into or out of a four-vertex must not exceed the scale $|\phi|$,
\begin{equation}                          							\label{no2}
\abs{ (p_1 + p_2)^2 } < |\phi|^2
\,,
\end{equation}
where $p_i$ are the ingoing momenta.

The violation of any of these two conditions would immediately imply
that the fluctuations generated out of the vacuum are able to destroy this vacuum, which is impossible.
Both conditions \eqs{no1} and \eqref{no2} have been formulated for a Minkowskian signature here, 
but can be rotated to the Euclidean.

By virtue of the cutoffs as they emerge above, the phase space for vacuum diagrams is strongly restricted.
As a consequence of this, the interactions between quasiparticle fluctuations are very weak 
and (for sufficiently large temperatures) generate only tiny higher loop corrections to thermodynamical quantities. 
Moreover, ultraviolet divergences do not occur
and the usual renormalization procedure is not needed in the effective theory.

\absatz{Thermodynamical self-consistency}
The thermodynamical quantities pressure, energy density, entropy density etc.
as obtained from the fundamental SU(N) Yang-Mills theory are related by Legendre transforms.
These relations hold in general, so their validity in the effective theory described by the action \eq{Seff}
has to be arranged for by imposing a condition of thermodynamical self-consistency. 
This condition has to assure that the $T$-derivatives of the TLH masses and the ground-state pressure cancel one another,
so that only the explicit $T$-dependence arising from the Boltzmann weight enters $T$-derivatives of thermodynamical quantities.
In particular, the relation
\begin{equation}                                  							\label{legendre}
\rho = T \frac{dP}{dT} - P
\end{equation}
translates into 
\begin{equation} 											\label{selfcons}
\frac{\d P}{\d a} = 0
\,,
\end{equation}
where
\begin{equation}
a = 2 \pi e \, \lam^{-3/2} = e \, \sqrt{ \frac{\Lam^3}{2\pi T^3} } = \frac{m}{2T}
\end{equation}
is a dimensionless measure for the quasiparticle mass.
\eq{selfcons} results into an evolution equation $\lambda(a)$ 
for temperature as a function of the tree-level gauge boson mass.
The evolution has two fixed points, namely $a=0$ and $a=\infty$.
They signal the existence of both a lowest and a highest attainable temperature in the electric phase.
These are denoted as $\lambda_c = \lambda(a=\infty)$ and $\lam_P = \lam(a=0)$ respectively.
The evolution is subject to the initial condition $ \lam (a=0) \equiv \lam_P$.
The low-temperature behavior of $\lam(a)$ is practically independent of $\lam_P$ as long as $\lam_P$ is sufficiently large:
The temperature at which the field $\phi$ emerges does not influence the low-temperature physics.
This is a signal of ultraviolet-infrared decoupling. 

\begin{figure}[t]
\centering
\includegraphics[width=8cm]{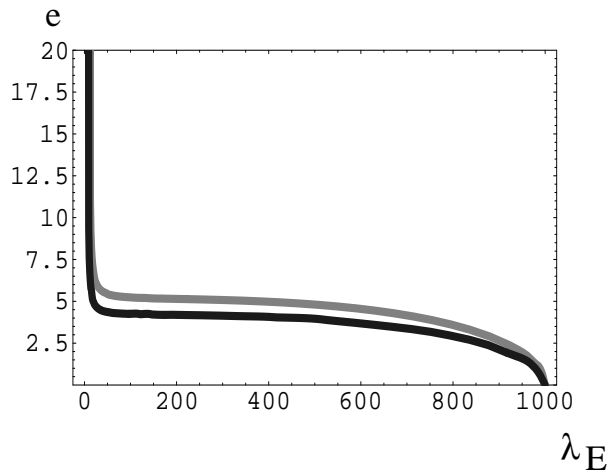}
\caption{                                         							\label{coupling}
Evolution of the effective gauge coupling $e$ in the electric phase 
for SU(2) (grey line) and SU(3) (black line).
The gauge coupling diverges logarithmically at $\lam_c = 11.65$ (SU(2)) and $\lam_c = 8.08$ (SU(3)).
The plateau values are $e = 5.1$ (SU(2)) and $e = 4.2$ (SU(3)).
The graph is taken from \cite{Hofmann}.}
\end{figure}

The evolution equation $\lam(a)$ can be inverted to yield an evolution $e(\lambda)$
for the effective gauge coupling as a function of temperature.
For SU(2) and SU(3) this is shown in \fig{coupling}:
The critical temperature is $\lam_c = 11.65$ for SU(2), 
and the evolution exhibits a plateau, 
where the effective gauge coupling has the value $e = 5.1$.

\absatz{Phase transition to magnetic phase}
At the critical temperature $\lam_c$, the theory undergoes a second order phase transition 
from the electric to the so-called magnetic phase.
The effective gauge coupling shows a logarithmic pole of the form
\begin{equation}
e(\lam) \sim - \log (\lam-\lam_c)  
\,.
\end{equation}
Hence the mass of the constituent BPS monopoles liberated by dissociating nontrivial holonomy calorons, given as
\begin{equation}
M_{\text{monopole}} \sim \frac{4\pi}{e} T  
\,,
\end{equation}
approaches zero with the consequence that magnetic monopoles condense.

In total, the theory is seen to have three phases:
the electric phase at high temperatures,
the magnetic phase for a small range of temperatures comparable to the scale $\Lam$, 
and a center phase for low temperatures.
The electric phase is deconfining, the magnetic phase is preconfining, and the center phases completely confining.
Here, we are only addressing the physics in the electric phase.

\section{The composite adjoint Higgs field $\phi$}     							\label{phi}
In this section, the composite adjoint scalar field $\phi \equiv \phi^a \lam^a$ 
($\lam^a$ denote the Pauli matrices)
which represents the topologically nontrivial, 
BPS saturated and trivial-holonomy part of the ground state is to be computed.

The field $\phi$ can be written as a product of modulus and phase.
As we are aiming for a description of a thermodynamical ground state, 
and the modulus $|\phi|$ is a (gauge invariant and hence) physical quantity, 
it has to be homogenous both in space and time.
To be able to calculate the modulus $|\phi|$, the Yang-Mills scale must be known.
The phase $\frac{\phi^a}{|\phi|}$ is a dimensionless quantity, 
so that for its calculation no information about the Yang-Mills scale $\Lam$ is necessary. 
The classical caloron action $S = \frac{8\pi^2}{g^2}$ is independent of temperature;
this excludes explicit $\beta$-dependences of $\phi$'s phase.
The phase may depend on the temperature only via the periodicity of the caloron.
Briefly, we can write
\begin{equation}                            								\label{decomp1}
\phi^a = |\phi|(\Lam,\beta) \cdot \frac{\phi^a}{|\phi|} \left( \frac{\tau}{\beta} \right)
\,.
\end{equation}
The calorons generating the field $\phi$ are BPS saturated;
as an immediate consequence, the energy-momentum tensor vanishes identically on a caloron. 
As already mentioned, the field $\phi$ will not include interactions,
so the energy-momentum tensor has to vanish on $\phi$ as well.
This in turn can be described by a BPS equation for $\phi$,
\begin{equation}
\d_\tau \phi = V^{(1/2)}
\,,
\end{equation}
where $V^{(1/2)}$ is the 'square-root' of the potential $V \equiv \tr ( V^{(1/2)} )^\dagger \, V^{(1/2)}$.

In sections \ref{abschnitt definition} and \ref{abschnitt calculation},
we will define an adjointly transforming integral over a two-point function and
demand that $\phi$'s phase obeys the same equation of motion.
In section \ref{abschnitt BPS} the BPS equation for $\phi$ will be determined.
In section \ref{abschnitt modulus} we will see that introducing the Yang-Mills scale $\Lam$ externally
allows us to obtain $\phi$'s modulus in terms of $\Lam$ and $\beta$.
With phase and modulus known, the BPS equation is used to determine the potential governing the dynamics of $\phi$. 
The potential is unique, if analytical dependence of the right-hand side of the BPS equation on the field $\phi$ is demanded.

\subsection{Definition}                             							\label{abschnitt definition}

We define the phase of the field $\phi$ as proposed in \cite{Hofmann} by
\begin{equation}                             \label{definition}
\frac{\phi^a}{|\phi|} \left( \frac{\tau}{\beta} \right) 
\sim
\int d^3x \int d\rho \, \tr 
\lambda^a \, 
F_{\mu\nu} (\tau,0) \, \left\{ (\tau,0),(\tau,\vec x) \right\} \,
F_{\mu\nu} (\tau,\vec x) \, \left\{ (\tau,\vec x),(\tau,0) \right\}    
\,,
\end{equation}
where
\begin{equation}  											\label{abk}
\begin{split}
|\phi|^2 &\equiv \frac12 \tr \phi^2  \\
\left\{(\tau,0),(\tau,\vec x)\right\} 
&\equiv 
\PO \exp \left[ i \int_{(\tau,0)}^{(\tau,\vec x)} dz_{\mu} \, A_{\mu}(z) \right]  \\
\left\{(\tau,\vec x),(\tau,0)\right\} 
&\equiv 
\PO \exp \left[ -i \int_{(\tau,0)}^{(\tau,\vec x)} dz_{\mu} \, A_{\mu}(z) \right] 
\,.
\end{split}
\end{equation}
The Wilson lines in \eq{abk} are to be calculated along 
the straight line connecting the points $(\tau,0)$ and $(\tau,\vec x)$. 
$\PO$ denotes the path-ordering symbol, and $\tr$ the SU(2) trace.
In \eqref{definition}, the dependences of the integrand on the
caloron scale parameter $\rho$ and inverse temperature $\beta$ suppressed. 

\eq{definition} 
is a definition of the phase $\frac{\phi^a}{|\phi|}$ in the following sense: 
We demand that $\frac{\phi^a}{|\phi|}$ obeys the same homogenous evolution equation in $\tau$ 
as the right hand side of \eqref{definition} does,
\begin{equation}     											\label{evolequ}
\mathcal D \, \frac{\phi}{|\phi|} = 0  
\,,
\end{equation}
where $\mathcal D$ is a differential operator in $\tau$.
Thus \eqref{evolequ} is an equation of motion for $\phi$'s phase. 

The right-hand side of \eqref{definition} is to be evaluated both on the caloron-field and the anticaloron-field
and afterwards the sum is to be taken.
We will see in the course of the calculation, 
that the definition \eqref{definition} contains quite a number of ambiguities 
which span the solution space of the differential operator $\mathcal D$.

Under a gauge transformation $\Omega(\tau,\vec x)$, the involved objects transform as follows:
\begin{equation}
\begin{split}
\{(\tau,0),(\tau,\vec x)\} &\to \Omega^\dagger (\tau,0) \, \{(\tau,0),(\tau,\vec x)\} \, \Omega (\tau,\vec x)  \\
\{(\tau,\vec x),(\tau,0)\} &\to \Omega^\dagger (\tau,\vec x) \, \{(\tau,\vec x),(\tau,0)\} \, \Omega (\tau,0)  \\
F_{\mu\nu}(\tau,\vec x) &\to \Omega^\dagger (\tau,\vec x) \, F_{\mu\nu} (\tau,\vec x) \, \Omega (\tau,\vec x)  
\,.
\end{split}
\end{equation}
Hence the right-hand side of \eqref{definition} indeed transforms like an adjoint scalar, namely
\begin{equation}
\frac{\phi^a}{|\phi|}(\tau) \to R^{ab}(\tau) \frac{\phi^b}{|\phi|}(\tau)
\,,
\end{equation}
where $R^{ab}$ is the SO(3) matrix 
\begin{equation}
R^{ab}(\tau) \lambda^b = \Omega(\tau,0) \lam^a \Omega^\dagger (\tau,0)
\,.
\end{equation}
The field $\phi$ only transforms under the time dependent part of the gauge transformation, 
the spatial dependence of the gauge transformation is lost in the macroscopic description.

The construction in \eqref{definition} is invariant under spatial translations 
and hence the integration over spatial translations is trivial.
For a gauge variant density as the integrand in \eqref{definition} is,
averaging over global color rotations yields zero and thus is forbidden.
The same applies to time translations.
The only modulus of the caloron solution that is integrated over is the scale parameter $\rho$ with flat measure.

\subsection{Are there alternative possibilities for the definition of $\phi$'s phase?} 						\label{alternative}
The definition \eqref{definition} for $\phi$'s phase is not at all arbitrarily chosen.
Indeed, trying to generalize the right-hand side of \eqref{definition} 
always requires the introduction of 
either explicit temperature dependences (which are not allowed)
or additional scales (which on the classical level do not exist):
\begin{itemize}
\item
Every local definition including the field strength $F_{\mu\nu}$ only, such as
\begin{equation}
\tr \lam^a F_{\mu\nu} F_{\nu\kappa} F_{\kappa\mu}
\,, \qquad
\tr \lam^a F_{\mu\nu} F_{\nu\kappa} F_{\kappa\rho} F_{\rho\mu}
\quad \text{etc.}
\end{equation}
or
\begin{equation}
\tr \eps_{abc} F_{\mu\nu} \lam^b F_{\nu\kappa} \lam^c F_{\kappa\mu}
\quad \text{etc.}
\end{equation}
yields zero when evaluated on the (anti)caloron field.
This is due to the (anti)self-duality of the (anti)caloron field \eq{thooftansatz}.
\item
One could be tempted to consider higher $n$-point functions of the type employed in \eq{definition}, 
such as
\begin{equation} 											\label{3pkt}
\begin{split}
\beta^{-1}
\int d^3x \int d^3y \int d\rho \, \tr 
\lambda^a \, 
F_{\mu\nu} (\tau,0) \, \left\{ (\tau,0),(\tau,\vec x) \right\} \,
F_{\nu\kappa} (\tau,\vec x) \, 
\\ \cdot
\left\{(\tau,\vec x),(\tau,\vec y)\right\} \,
F_{\kappa\mu} (\tau,\vec y) \, \left\{ (\tau,\vec y),(\tau,0) \right\}  \\
\end{split}
\end{equation}
In such an $n$-point function, one has to introduce an additional factor $\beta^{2-n}$ to get a dimensionless object.
$\phi$'s phase is supposed to depend on temperature only via the temperature dependence of the caloron field, 
but not explicitly. 
(This is due to the temperature-independence of the caloron action.)
Therefore the $n$-point function \eqref{3pkt} and its generalization for $n\ge3$ are forbidden.
\item
One could think of shifting the spatial part of the starting point of the construction in \eqref{definition}
from 0 to an arbitrary point $\vec z$. 
Any given value of $|\vec z| \neq 0$ would introduce an additional scale;
but, on the classical level, such a scale does not exist.
The same argument applies to replacing the straight Wilson lines with curved arcs.
Their curvature again corresponds to an additional scale which is physically not present.
\item
The definition \eqref{definition} does not include caloron solutions with topological charge $|Q| >1$.
This has the following reason:
A caloron of topological charge $|Q|>1$ has $m>1$ dimensionful moduli. 
For example, a caloron with $|Q|=2$ has three dimensionful moduli, 
namely two scale parameters and the distance between its two centers.
An $n$-point function of the type displayed in \eq{3pkt} contains $n$ field strength tensors (mass dimension 2)
and $n-1$ integrations over 3-space (mass dimension $-3$). 
The $m$ dimensionful moduli of the caloron have to be integrated as well (mass dimension $-1$ each).
These combine to mass dimension $2n - 3(n-1) - m = 3 - n - m$.
We are looking for a dimensionless object without any explicit dependence on $\beta$.
But this is not possible with $n>2$ and $m>1$.
Therefore, calorons of higher topological charge are excluded.
\end{itemize}

\subsection{Calculation}                            							\label{abschnitt calculation}
In this section, we want to evaluate the right-hand side of \eqref{definition}, 
\begin{equation}                                                   				\label{twopoint}
\int d^3x \int d\rho \, \tr \lambda^a \, F_{\mu\nu} (\tau,0) \, \left\{ (\tau,0),(\tau,\vec x) \right\} \,
F_{\mu\nu} (\tau,\vec x) \, \left\{ (\tau,\vec x),(\tau,0) \right\}
\,,
\end{equation}
on the single caloron solution (cf. Sec.~2)
\begin{equation}          										\label{caloron}
A_{\mu} (\tau,\vec x) = - \etb a\mu\nu \, \frac{\lambda^a}2 \, \d_\nu \ln \Pi (\tau,r)  
\end{equation}
with the pre-potential $\Pi(\tau,r)$ given as
\begin{equation} 											\label{prepotential2}   
\Pi(\tau,r)
= 
1 + \frac{\pi\rho^2}{\beta r} 
\frac{\sinh \frac{2\pi r}{\beta} }
{ \cosh \frac{2\pi r}{\beta} - \cos \frac{2\pi\tau}{\beta} }
\,,
\end{equation}
and $r \equiv | \vec x |$.

\absatz{The Wilson lines}
The single caloron solution has a hedgehog like behavior in the sense that the spatial part of 
the scalar product $x_\mu A_\mu(\tau,\vec x)$ has the same orientation both in 3-space and color-space,
\begin{equation}
\begin{split}
x_i \, A_i^a (\tau,\vec x)
&= - x_i \, \etb ai\nu \, \d_\nu \ln \Pi (\tau,r)  \\
&= - x_i \left( \eps_{ai\nu} - \delta_{\nu4}\delta_{ai} \right) \d_\nu \ln\Pi(\tau,r)  \\
&= - \eps_{ain} \frac{x_i x_n}{r} \, \d_r \ln \Pi(\tau,r) + x_i \, \delta_{ai} \, \d_4 \ln\Pi(\tau,r)  \\
&=  x_a \, \d_4 \ln\Pi(\tau,r)  \,.
\end{split}
\end{equation}
This property allows to discard the path ordering operation in the calculation 
of the Wilson line $\{(\tau,0),(\tau,\vec x)\}$ defined in \eq{abk}.
We parameterize the path as $z_\mu (s) = (\tau,s\vec x)$ with $0 \le s \le 1$.
Hence we have
\begin{equation}
\begin{split}
\left\{(\tau,0),(\tau,\vec x)\right\}
&= \PO \exp \left[ i \int_{(\tau,0)}^{(\tau,\vec x)} dz_{\mu} \, A_{\mu}(z) \right]  \\
&= \PO \exp \left[ i \int_0^1 ds \, x_i \, A_i (\tau,s\vec x) \right]  \\
&= \PO \exp \left[ i \int_0^1 ds \, \frac{x_i}{2} \, \lambda_i \, \d_4 \ln\Pi(\tau,sr) \right]  \\
&= \exp \left[ i \lambda^i \, \frac{x_i}{2} \, \int_0^1 ds \, \d_4 \ln\Pi(\tau,sr) \right]  \\
&= \cos \left( g(\tau,r) \right) 
+ i \, \lambda^i \, \frac{x_i}{r} \, \sin \left( g(\tau,r) \right)
\,,
\end{split}
\end{equation}
where we define
\begin{equation}  											\label{g(r)}
g(\tau,r) 
\equiv 
\int_0^1 ds \, \frac r2 \, \d_4 \ln \Pi(\tau,sr) 
\,.
\end{equation}
The evaluation of the Wilson line on the single anticaloron solution yields just a change of sign
in the argument of the exponential function.
So, for caloron and anticaloron we have
\begin{equation} 											\label{wline}
\left\{ (\tau,0),(\tau,\vec x) \right\}_{C,A} 
= \cos \left( g(\tau,r) \right) 
\pm i \lambda_i \frac{x_i}{r} \sin \left( g(\tau,r) \right)
\end{equation}
respectively.
The following relations hold:
\begin{equation}
\begin{split}
  \left\{ (\tau,0) , (\tau,\vec x)  \right\}_C
&= \left\{ (\tau,\vec x) , (\tau,0)  \right\}_A
= \left\{ (\tau,0) , (\tau,-\vec x) \right\}_A
= \left\{ (\tau,0) , (\tau,-\vec x) \right\}_C^\dagger 
\\
&= \left\{ (\tau,\vec x) , (\tau,0)  \right\}_C^\dagger 
= \left\{ (\tau,0) , (\tau,\vec x)  \right\}_A^\dagger 
= \left\{ (\tau,-\vec x) , (\tau,0) \right\}_C  \,.
\end{split}
\end{equation}
The integrand of \eq{g(r)} for large $r$ behaves like a $\delta$-function in $s$,
\begin{equation}    											\label{delta}
\lim_{r\to\infty} \frac r2 \, \d_4 \ln \Pi(\tau,sr) 
=
\delta(s) \cdot f(\tau)
\,.
\end{equation}
This property has been established numerically.

\absatz{Caloron field strength}
The field strength $F_{\mu\nu}$ on a caloron can be calculated as
\begin{align}
	F_{\mu\nu}^a 
&= 
	\d_\mu A_\nu^a - \d_\nu A_\mu^a - i \left[ A_\mu,A_\nu \right]^a   
\\ \notag 
&= 
	\etb a\mu\nu \frac{ (\d_\kappa \Pi)(\d_\kappa \Pi) }{\Pi^2} 
	- \etb a\nu\kappa \frac{ \Pi (\d_\mu \d_\kappa \Pi ) - 2 (\d_\mu\Pi)(\d_\kappa\Pi) }{\Pi^2}
	+ \etb a\mu\kappa \frac{ \Pi (\d_\nu \d_\kappa \Pi ) - 2 (\d_\nu\Pi)(\d_\kappa\Pi) }{\Pi^2}  
\,.
\end{align}
The calculation is performed in Appendix~\ref{P&Pmunu}.
To obtain the field strength of an anticaloron, $\bar\eta$ has to be replaced with $\eta$.
In particular, the field strength at $\vec x=0$ is
\begin{equation}  											\label{bei null}
F_{\mu\nu}^a (\tau,0) 
= 
\et a\mu\nu \left( \frac{ [\d_4\Pi(\tau,0)]^2 }{\Pi^2(\tau,0)} 
- \frac23 \frac{\d_4^2\Pi(\tau,0)}{\Pi(\tau,0)} \right)  
\,.
\end{equation}

\absatz{The integrand of \eq{twopoint}}
Inserting the result for the Wilson lines \eq{wline} into the expression \eqref{twopoint} 
and writing the field strength in components, $F_{\mu\nu} = F_{\mu\nu}^a \frac{\lam^a}{2}$, we see that 
\begin{equation} 											\label{4terme 1}
\begin{split}
& 
\tr \lambda^a \, F_{\mu\nu} (\tau,0) \, \left\{ (\tau,0),(\tau,\vec x) \right\} \,
F_{\mu\nu} (\tau,\vec x) \, \left\{ (\tau,\vec x),(\tau,0) \right\}  
\bigg|_{\text{Caloron}}
\\ & =
\frac12 \tr 
\Bigg[
\lam^a \lam^b \left( \cos g(\tau,r) + i \lam^c \frac{x^c}{r} \sin g(\tau,r) \right)
\lam^d \left( \cos g(\tau,r) - i \lam^e \frac{x^e}{r} \sin g(\tau,r) \right)  
\Bigg]
\\ & \quad
\cdot F_{\mu\nu}^b (\tau,0) F_{\mu\nu}^d (\tau,\vec x) 
\end{split}
\end{equation}
is to be computed.
Performing the trace and contracting Lorentz and color indices is a straightforward 
but somewhat lengthy calculation; for details see Appendix \ref{details}.
The result is\footnote{
Note that $\d_r$ denotes the derivative with respect to the radial coordinate $r \equiv |\vec x|$. 
For more notations and conventions, see Appendix~\ref{not&conv}.}
\begin{equation}  											\label{kontrahiert}
\begin{split}
& 
	\tr \lambda^a \, F_{\mu\nu} (\tau,0) \, \left\{ (\tau,0),(\tau,\vec x) \right\} \,
	F_{\mu\nu} (\tau,\vec x) \, \left\{ (\tau,\vec x),(\tau,0) \right\}  
	\bigg|_{\text{Caloron}}
\\ & =
	2 i \frac{x^a}{r}
	\left( \frac{[\d_4\Pi(\tau,0)]^2}{\Pi^2(\tau,0)} - \frac23 \frac{\d_4^2\Pi(\tau,0)}{\Pi(\tau,0)} \right)
\\ & \quad \cdot
	\Bigg\{
		2 \cos (2 g(\tau,r))
		\left( 2 \frac{[\d_4\Pi(\tau,r)][\d_r\Pi(\tau,r)]}{\Pi^2(\tau,r)} - \frac{\d_4\d_r\Pi(\tau,r)}{\Pi(\tau,r)} \right)
\\ & \quad 
		+ \sin (2 g(\tau,r))
		\left( 2 \frac{[\d_r\Pi(\tau,r)]^2}{\Pi^2(\tau,r)} - 2 \frac{[\d_4\Pi(\tau,r)]^2}{\Pi^2(\tau,r)} 
		+ \frac{\d_4^2\Pi(\tau,r)}{\Pi(\tau,r)} - \frac{\d_r^2\Pi(\tau,r)}{\Pi(\tau,r)} \right)
	\Bigg\}
\,.
\end{split}
\end{equation}
The factor
\begin{equation}   											\label{beinull}
\frac{[\d_4\Pi(\tau,0)]^2}{\Pi^2(\tau,0)} - \frac23 \frac{\d_4^2\Pi(\tau,0)}{\Pi(\tau,0)} 
=
- \frac{16 \pi^4}{3} \frac{\rho^2}{\beta^2}
\frac{ \pi^2\rho^2 + \beta^2 \left( 2 + \cos \frac{2\pi\tau}{\beta} \right) }
{ \left( 2\pi^2\rho^2 + \beta^2 \left( 1 - \cos \frac{2\pi\tau}{\beta} \right) \right)^2 }
\end{equation}
arises from the field strength at the point $(\tau,0)$, cf. \eq{bei null},
and contains no dependence on $r$.
Note that the expression \eqref{kontrahiert} is proportional to the (spatial) unit vector $\frac{x^a}{r}$.

\absatz{Integration over position space}
The integration over position space demanded in \eqref{twopoint} is performed in polar coordinates.
On the one hand, the angular integration over the unit vector $\frac{x^a}{r}$ yields zero.
But on the other hand, the radial integral is infinite.
This can be seen as follows:
The pre-potential \eqref{prepotential2} for large $r$ behaves as
\begin{equation}
\Pi(\tau,r) \to 1 + \frac{ \pi \rho^2 }{ \beta r }
\qquad \text{($r \gg \beta / 2\pi$)}
\,,
\end{equation}
and its second spatial derivative approaches
\begin{equation}
\d_r^2 \Pi(\tau,r) \to \d_r^2 \left( 1 + \frac{ \pi \rho^2 }{ \beta r } \right) = \frac{2\pi\rho^2}{\beta r^3}
\qquad \text{($r \gg \beta / 2\pi$)}
\,.
\end{equation}
Hence the very last term in \eq{kontrahiert} has the asymptotic behavior
\begin{equation}
\frac{\d_r^2\Pi(\tau,r)}{\Pi(\tau,r)} \to \frac{2\pi\rho^2}{\beta r^3} 
\qquad \text{($r \gg \beta / 2\pi$)}
\end{equation}
and gives rise to the logarithmically divergent integral\footnote{
Note that the integral is convergent at $r=0$, since we have
\begin{equation*}
\lim_{r\to0} \frac{\d_r^2\Pi(\tau,r)}{\Pi(\tau,r)}
= 
- \frac{4\pi^4\rho^2}{3\beta^2} 
\frac{1}{ \sin^2 \left( \frac{\pi\tau}{\beta} \right) }
\frac{ \left( 2 + \cos \left( \frac{2\pi\tau}{\beta} \right) \right) }
{ 2 \pi^2 \rho^2 + \beta^2 \left( 1 - \cos \left( \frac{2\pi\tau}{\beta} \right) \right) }
\,.
\end{equation*}
}
\begin{equation}  											\label{logdiv}
\begin{split}
\int_0^{\infty} dr \, r^2 \sin (2g(\tau,r)) \frac{\d_r^2\Pi(\tau,r)}{\Pi(\tau,r)} 
&= \text{finite} + \int_R^{\infty} dr \sin (2g(\tau,r)) \frac{2\pi}{\beta r} 
\\
&= \text{finite} + \frac{2\pi\rho^2}{\beta} \left( \lim_{r\to\infty} \sin (2g(\tau,r)) \right) \int_R^{\infty} \frac{dr}{r}  
\,,
\end{split}
\end{equation}
where $R \gg \beta / 2\pi$.
The function $g(r)$ defined in \eq{g(r)} has a finite limit for $r\to\infty$, 
which is reached very rapidly. It has been taken out of the integral.

All the other terms in \eq{kontrahiert} give rise to finite integrals in $r$, and hence do not contribute: 
For the square of the first spatial derivative, we have
\begin{equation}
\frac{ [\d_r\Pi(\tau,r)]^2 }{ \Pi^2(\tau,r) } \to 
\frac{\pi^2\rho^4}{\beta^2 r^4}
\qquad \text{($r \gg \beta / 2\pi$)}
\,,
\end{equation}
which is convergent.
All the other terms contain at least one time derivative and thus vanish exponentially, e.\,g.
\begin{equation}
\d_\tau \Pi(\tau,r) 
\to \frac{2\pi\rho^2}{\beta r} \sin \left( \frac{2\pi\tau}{\beta} \right) \exp \left( -\frac{2\pi r}{\beta} \right)
\qquad \text{($r \gg \beta / 2\pi$)}  
\,.
\end{equation}
Thus the only nonvanishing contribution to the expression \eqref{twopoint} is
\begin{align}        											\label{nonvanishing}
& 
- 2 i 
\int d\rho 
\left( \frac{[\d_4\Pi(\tau,0)]^2}{\Pi^2(\tau,0)} - \frac23 \frac{\d_4^2\Pi(\tau,0)}{\Pi(\tau,0)} \right)
\int_{S_2} d\Omega \, \frac{x^a}{r}
\int_0^\infty dr \, r^2 \,
\sin (2 g(\tau,r)) \,
\frac{\d_r^2\Pi(\tau,r)}{\Pi(\tau,r)} 
\notag 
\\ & 
= 
\frac{64 i \pi^5}{3} 
\int d\rho \,
\frac{\rho^4}{\beta^3}
\frac{ \pi^2\rho^2 + \beta^2 \left( 2 + \cos \frac{2\pi\tau}{\beta} \right) }
{ \left( 2\pi^2\rho^2 + \beta^2 \left( 1 - \cos \frac{2\pi\tau}{\beta} \right) \right)^2 }
\int_{S_2} d\Omega \, \frac{x^a}{r}
\int_R^\infty \frac{dr}{r} \,
\sin (2 g(\tau,r)) 
\,.
\end{align}

\absatz{Regularization}
As already pointed out, this expression contains a product zero $\times$ infinity 
and thus needs to be given a finite value  by prescribing a regularization procedure.
The radial integration in \eq{nonvanishing} is regularized according to
\begin{equation}
\int_R^{\infty} \frac{dr}{r} \to \beta^{\eps} \int_R^{\infty} \frac{dr}{r^{1+\eps}}
\end{equation}
with $\eps>0$. 
This integral is
\begin{equation}
\begin{split}
\beta^{\eps} \int_R^{\infty} \frac{dr}{r^{1+\eps}}
& = \beta^{\eps} \int_0^{\infty} \frac{dr}{(r+R)^{1+\eps}}
= \frac{1}{\eps} \left( \frac{\beta}{R} \right)^\eps
\\ &
= \frac{1}{\eps} - \log \left( \frac{R}{\beta} \right) + \frac12 \eps \log^2 \left( \frac{R}{\beta} \right) + 
\o{\eps^2}
\,.
\end{split}
\end{equation}
The right-hand side is a regular expression for $\eps \neq 0$. 
It can be regarded as the analytical continuation of the integral on the left-hand side for $|\eps| \ll 1$.
When smearing the regulator $\eps$ over a small interval $[-\eta,\eta]$ with $0<\eta\ll1$ as
\begin{equation} 											\label{rad reg}
\begin{split}
& 
\frac{1}{2\eta} \int_{-\eta}^{\eta} d\eps
\left( \frac{1}{\eps \pm i0} - \log \left(\frac R\beta\right) 
+ \frac12 \eps \log^2 \left(\frac R\beta \right) + 
\o{\eps^2} 
\right)
=
\mp \frac{\pi i}{2\eta} - \log \left( \frac R\beta \right) 
+ \o{\eta^2} 
\,,
\end{split}
\end{equation}
an ambiguity appears: There are two possibilities how to circumvent the pole.

The angular integration is regularized via the introduction 
of a defect (or surplus) angle in the azimuthal integration, 
\begin{equation} 											\label{ang reg1}
\int_{0}^{\pi} d\theta \sin\theta \int_0^{2\pi} d\varphi
\to
\int_{0}^{\pi} d\theta \sin\theta \int_{\al_C \pm \eta'}^{\al_C + 2\pi \mp \eta'} d\varphi
\end{equation}
with $0<\eta'\ll1$ and an arbitrary angle $0\le\al_C<2\pi$.
The regulated radial integral yields
\begin{equation} 											\label{ang reg2}
\begin{split}
\int_{0}^{\pi} d\theta \sin\theta \int_{\al_C \pm \eta'}^{\al_C + 2\pi \mp \eta'} d\varphi \, \frac{x^a}{r}
&=
\mp \pi \sin \eta' ( \delta_{a1} \cos \al_C + \delta_{a2} \sin \al_C )  \\
&= 
\mp \pi \eta' ( \delta_{a1} \cos \al_C + \delta_{a2} \sin \al_C ) + \o{\eta'^2}
\,.
\end{split}
\end{equation}
The regularization prescription \eq{ang reg1} clearly singles out 
the axis with unit vector \linebreak $(\cos\al_C,\sin\al_C,0)$. 
Later this will be seen to relate to a global gauge choice. 

At the end of the calculation, both regularization parameters $\eta$ and $\eta'$ have to be sent to zero. 
Their ratio in this limit will be denoted by $\Xi_C$,
\begin{equation}
\lim_{\eta,\eta' \to 0} \frac{ \eta' }{ \eta } = \Xi_C
\,.
\end{equation}
This is a positive but otherwise unknown number.
Inserting the regularized expressions \eqs{rad reg} and \eqref{ang reg2} and into \eqref{nonvanishing}, 
we get
\begin{equation}
\begin{split}
& 
\left.
\int d^3x \int d\rho \, \tr \lambda^a \, F_{\mu\nu} (\tau,0) \, \left\{ (\tau,0),(\tau,\vec x) \right\} \,
F_{\mu\nu} (\tau,\vec x) \, \left\{ (\tau,\vec x),(\tau,0) \right\}
\right|_{\text{Caloron}}
\\ & = 
\pm \Xi_C ( \delta_{a1} \cos\al_C + \delta_{a2} \sin\al_C ) \mathcal A \left( \frac{2\pi\tau}{\beta} \right)
\end{split}
\end{equation}
where the dimensionless function $\mathcal A$ is defined as
\begin{equation}    											\label{mathcalA}
\mathcal A \left( \frac{2\pi\tau}{\beta} \right)
\equiv
\frac{32}{3} \frac{\pi^7}{\beta^3}
\int d\rho 
\left( \lim_{r\to\infty} \sin 2g(\tau,r) \right)
\rho^4
\frac{ \pi^2\rho^2 + \beta^2 \left( 2 + \cos \frac{2\pi\tau}{\beta} \right) }
{ \left( 2\pi^2\rho^2 + \beta^2 \left( 1 - \cos \frac{2\pi\tau}{\beta} \right) \right)^2 }
\,.
\end{equation}

\absatz{Integration over scale parameter}
Up to now, we did not specify the range for the integration over the scale parameter $\rho$.
As the integrand in \eq{mathcalA} asymptotically behaves like $\sim \rho^2$, 
integrating $\rho$ from zero to infinity yields an infinite expression.
To see what is going on, a cutoff is introduced in units of inverse temperature $\beta$, 
\begin{equation}
\int d\rho \to \int_0^{\zeta\beta} d\rho \qquad (\zeta>0)
\,.
\end{equation}
This generates an additional dependence of $\mathcal A$ on $\zeta$.
For $\zeta \gg 1$, we will have $\mathcal A \propto \zeta^3$.
The integral in \eq{mathcalA} has to be evaluated numerically.
Due to the property \eq{delta}, the limit $\lim_{r\to\infty} \sin 2g(\tau,r)$ is reached very fast; 
for our purposes, putting $r=10$ in numerical calculations is fully sufficient.
\begin{figure}[tb]
\centering
\includegraphics[width=14cm]{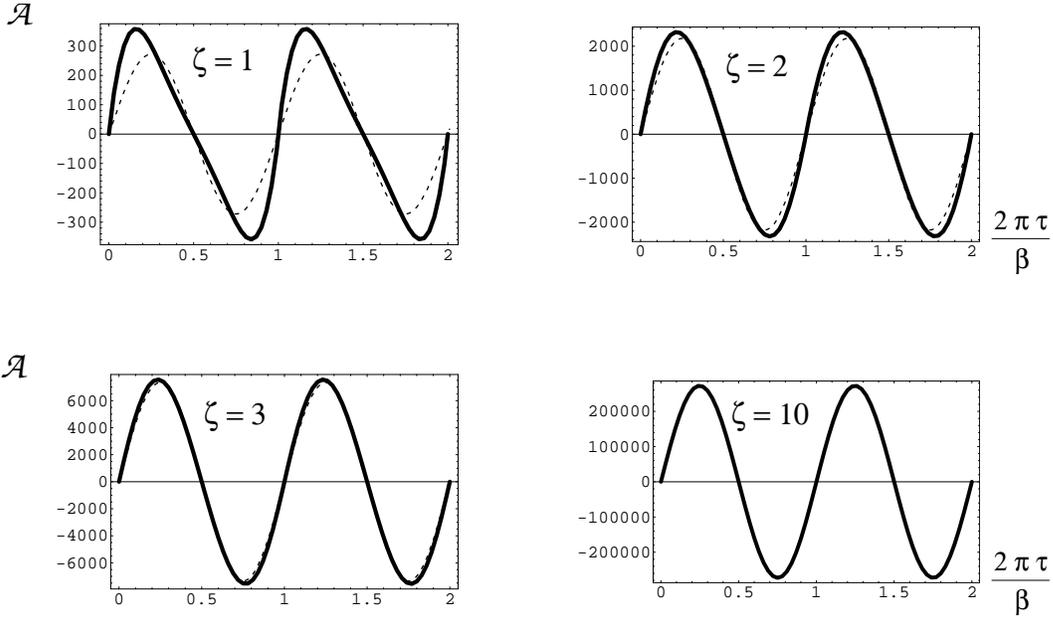}
\caption{   												\label{bild mathcal A}
The function $\mathcal A ( \frac{2\pi\tau}{\beta} )$ 
plotted over two periods with different values of $\zeta$.
For comparison the function $272 \zeta^3 \sin ( \frac{2\pi\tau}{\beta} )$ is plotted as a dashed line. 
Already for $\zeta = 10$ the difference cannot be resolved any more.}
\end{figure}
\fig{bild mathcal A} shows the $\tau$-dependence of $\mathcal A$ for various values of $\zeta$.
With growing $\zeta$, the function $\mathcal A$ rapidly approaches a sine curve, 
\begin{equation}
\mathcal A \left( \frac{2\pi\tau}{\beta} , \zeta \right)
\to
272 \zeta^3 \sin \left( \frac{2\pi\tau}{\beta} \right)
\qquad \text{($\zeta \to \infty$)}
\,.
\end{equation}
Already for $\zeta=10$, the difference between the two curves could not be resolved any more in the figure.
The prefactor 272 has been fitted numerically, see Table~\ref{fit}.
\begin{table}[t]
\centering
\begin{tabular}{r|r p{2mm} r|r p{2mm} r|r}
$\zeta$ & $\mathcal A ( \frac\pi2 ) / \zeta^3$ &&
$\zeta$ & $\mathcal A ( \frac\pi2 ) / \zeta^3$ &&
$\zeta$ & $\mathcal A ( \frac\pi2 ) / \zeta^3$ \\
\cline{1-2}\cline{4-5}\cline{7-8}
1 & 301.295 && 10 & 272.776 &&  100 & 272.026 \\
2 & 285.012 && 20 & 272.216 &&  200 & 272.020 \\
3 & 278.828 && 30 & 272.107 &&  300 & 272.018 \\
4 & 276.161 && 40 & 272.068 &&  400 & 272.018 \\
5 & 274.794 && 50 & 272.050 &&  500 & 272.018 \\
\multicolumn{6}{r}{}         & 1000 & 272.018
\end{tabular}
\caption{ 												\label{fit}
Value of the function $\mathcal A$ at $\frac\pi2$ for several values of the cutoff $\zeta$.
The cutoff dependence $\zeta^3$ has been divided out.}
\end{table}
Therefore, the evaluation of the expression \eqref{twopoint} on the single caloron yields
\begin{equation}                             \label{sinus}
\begin{split}
& 
\left.
\int d^3x \int d\rho \, \tr \lambda^a \, F_{\mu\nu} (\tau,0) \, \left\{ (\tau,0),(\tau,\vec x) \right\} \,
F_{\mu\nu} (\tau,\vec x) \, \left\{ (\tau,\vec x),(\tau,0) \right\}
\right|_{\text{Caloron}}
\\ & =
\pm 272 \zeta^3 \Xi_C ( \delta_{a1} \cos\al_C + \delta_{a2} \sin\al_C ) 
\sin \left( \frac{2\pi\tau}{\beta} \right)
\,.
\end{split}
\end{equation}

\subsection{Ambiguities and BPS saturation}     							\label{abschnitt BPS}
The expression \eqref{twopoint} has been evaluated on the caloron in the previous section.
Evaluation on the anticaloron yields the same result, 
if we agree upon circumventing the pole appearing in \eq{rad reg} in the opposite way as compared to the caloron.
According to the definition, both contributions are to be added.
In this process some ambiguities occur:
\begin{enumerate}
\item 
The undefined number $\lim_{\eta,\eta'\to0} \frac{\eta'}{\eta}$ 
needs not necessarily be the same in the caloron and anticaloron case,
we chose to call it $\Xi_A$ for the anticaloron (instead of $\Xi_C$ for the caloron).
\item 
The same applies to the arbitrarily chosen axis singled out by angular regularization.
It is no restriction of generality to assume that both axes lie in the $x_1x_2$-plane,
but with different azimuthal angles $\al_C$ and $\al_A$.
\item 
As \eqref{definition} is supposed to define only the equation of motion for $\phi$,
we may as well introduce a shift in time; this does not change the operator $\mathcal D$.
Again, this shift may be different for the two contributions,
namely $\tau \to \tau + \tau_C$ for the caloron and $\tau \to \tau + \tau_A$ for the anticaloron.
\end{enumerate}
The requested sum of caloron and anticaloron contribution (including all the above ambiguities) is then
\begin{equation}                                             						\label{result}
\begin{split}
\frac{ \phi^a }{ |\phi| }
\sim
272 \zeta^3 
\Bigg\{
& 
	\pm \Xi_C ( \delta_{a1} \cos\al_C + \delta_{a2} \sin\al_C ) 
	\sin \left( \frac{2\pi(\tau+\tau_C)}{\beta} \right)
\\ & 
	\pm \Xi_A ( \delta_{a1} \cos\al_A + \delta_{a2} \sin\al_A ) 
	\sin \left( \frac{2\pi(\tau+\tau_A)}{\beta} \right)
\Bigg\}
\,.
\end{split}
\end{equation}

\noindent
The caloron (or anticaloron) contribution in \eq{result} alone 
represents a linearly polarized harmonic oscillation in adjoint color space.
The three ambiguities given above can be viewed as the free parameters of such an oscillation, 
namely modulus, phase-shift and polarization axis.
The sum of caloron and anticaloron contribution \eq{result} taking into account all the above ambiguities
is an elliptically polarized oscillation in adjoint color space.
Here, the polarization plane is the $x_1x_2$-plane; 
this is only due to our choice of the angular regularization and no physical property.

The right-hand side of \eq{result} obviously is annihilated by the second order differential operator
\begin{equation}                                                 					\label{D}
\mathcal D = \d_\tau^2 + \left( \frac{2\pi}{\beta} \right)^2
\,.
\end{equation}
Note that the ambiguities inherent to the definition \eqref{definition} span the solution space of $\mathcal D$.

\absatz{Imposing BPS saturation}
The field $\phi$ is composed of noninteracting trivial-holonomy calorons;
the caloron itself is an energy- and pressure-free, BPS saturated configuration
and no interaction whatsoever has been included.
Hence the composite object must again be energy- and pressure-free and BPS saturated;
that means it does not only obey a second order differential equation (the equation of motion), 
but also a first order differential equation (the BPS equation).
Thus we need to find first-order equations whose solutions solve the second order equation 
\begin{equation}                                                            				\label{2.order}
\d_\tau^2 
\phi + \left( \frac{2\pi}{\beta} \right)^2 
\phi = 0
\end{equation}
as well.
There are two such equations\footnote{
The solutions to the equations
$\d_\tau \phi = \pm \frac{2\pi i}{\beta} \phi$ 
as well solve \eq{2.order};
in spite of this they are not allowed here because they are not adjoint fields.}, 
namely
\begin{equation}  											\label{linequ}
\d_\tau 
\phi = \pm \frac{2\pi i}{\beta} \lam_3 
\phi
\,.
\end{equation}
Choosing any (normalized) linear combination of Pauli matrices instead of $\lam_3$ 
would be possible equally well: 
\eq{linequ} is subject to a global gauge ambiguity.
The solutions to \eqref{linequ} are given as
\begin{equation}  											\label{solution}
\phi = C \lam_1 \exp \left( \pm \frac{2\pi i}{\beta} \lam_3 (\tau - \tau_0) \right)
\,,
\end{equation}
where $C$ and $\tau_0$ are real integration constants.
This solution is a circularly polarized oscillation.
It winds along an $S_1$ in the group manifold $S_3$ of SU(2).

Thus the demand for BPS saturation forces an elliptical polarization in \eq{result} into a circular polarization.
The undetermined and formerly independent quantities are now subject to the relations
\begin{align}
\Xi_C &= \Xi_A \,, & 
\tau_C - \tau_A &= \pm \frac \pi2 \,, &
\al_C - \al_A &= \pm \frac \pi2 \,.
\end{align}
The modulus of the oscillation and its phase-shift are still undetermined constants of integration.
The former will be considered in the following section, the latter is of no physical significance.
In addition, a global gauge ambiguity, i.\,e. the plane in which the oscillation takes place, is still present.

\subsection{Obtaining $\phi$'s modulus}  								\label{abschnitt modulus}
Let us now assume the existence of an externally given scale $\Lam$ which determines $\phi$'s modulus. 
We allow for explicit dependence of $\phi$ on the scale $\Lam$, 
the temperature $\beta$ and on Euclidean time through $\frac\tau\beta$, 
\begin{equation}
\phi = \phi \left( \beta, \Lam, \frac\tau\beta \right)
\,.
\end{equation}
As the phase found in \eq{solution} shall be preserved even in case of the presence of a scale $\Lam$, 
the right-hand side of the BPS equation
\begin{equation}  											\label{BPS}
\d_\tau \phi = V^{(1/2)}
\end{equation}
may only depend linearly on $\phi$. Besides that, we demand an analytical dependence of $V$ on $\phi$.
The potential $V$ (and its 'square-root' $V^{(1/2)}$) may depend on the temperature through the periodicity of $\phi$.
An explicit dependence on $\beta$ is not possible
because no explicit $\beta$-dependence occurs in the average over the caloron moduli space.

These conditions leave only the two possibilities
\begin{equation}                               								\label{moegl1}
\d_\tau \phi = \pm i \Lam \lam_3 \phi
\end{equation}
and\footnote{
Note that
$\frac{\phi}{|\phi|^2} = \phi^{-1} = \phi_0^{-1} \sum_{n=0}^{\infty} (-1)^n \phi_0^{-n} (\phi - \phi_0)^n$
is indeed an analytical function of $\phi$.}
\begin{equation}                                   							\label{moegl2}
\d_\tau \phi = \pm i \Lam^3 \lam_3 \frac{\phi}{|\phi|^2}
\,.
\end{equation}
Using \eqs{decomp1} and \eqref{solution}, we write 
\begin{equation}                                       							\label{decomp2}
\phi = | \phi(\beta,\Lam) | \lam_1 \exp \left( \pm \frac{ 2\pi i }{\beta} \lam_3 \tau 
\right)
\,.
\end{equation}
The modulus of $\phi$ is gauge invariant and hence a physical quantity describing a thermodynamical ground state,
so it has to be homogenous in space and time.
Inserting the decomposition \eqref{decomp2} into the first BPS equation \eqref{moegl1}, we get 
\begin{equation}                                    							\label{unmoeglich}
\Lam = \frac{2\pi}{\beta}
\,.
\end{equation}
This obviously can not be satisfied, since the scale $\Lam$ is a constant and $\beta$ is the inverse temperature. 
From the second possibility \eq{moegl2}, we get
\begin{equation}                                      							\label{mod}
|\phi| (\beta, \Lam) 
= \sqrt{ \frac{\beta \Lam^3}{2\pi} }
= \sqrt{ \frac{\Lam^3}{2\pi T} }
\,,
\end{equation}
which is no contradiction.
So, \eq{moegl2} is the only acceptable BPS equation, and $\phi$ has the form
\begin{equation}
\phi = \sqrt{ \frac{\beta \Lam^3}{2\pi} } \, \lam_1 \exp \left( \pm \frac{ 2\pi i }{\beta} \lam_3 \tau \right)
\,.
\end{equation}
\eq{mod} shows that the field $\phi$ is power suppressed in $T$, 
and hence all topologically nontrivial effects die off at high temperature.
The right-hand side of the BPS equation defines the 'square-root' of $\phi$'s potential,
\begin{equation}
V(\phi) = \tr ( V^{(1/2)} )^\dagger V^{(1/2)} = \Lam^6 \tr \phi^{-2}
\,.
\end{equation}
Under the above assumptions, the potential is unique. 
The Lagrangian for the field $\phi$ is
\begin{equation}
\mathcal L = \tr (\d_\tau \phi)^2 + V(\phi)
\,.
\end{equation}
The consequences of minimally coupling $\phi$ to the topologically trivial sector 
have been investigated in \cite{Hofmann}; 
some of the results are briefly reviewed in Sec.~3.1.

\chapter{Thermodynamical Pressure in the Electric Phase} 						\label{pressure}
In this chapter the two-loop corrections to the thermodynamical pressure 
in the electric phase of SU(2) Yang-Mills theory are computed.
In view of the evolution of the effective gauge coupling in the electric phase (cf.~\fig{coupling}),
one can constrain oneself to the case $e>\frac12$.
This will simplify the calculation.
For a numerical evaluation, the plateau value $e=5.1$ is used.
We will set up the prerequisites for the calculation in Sec.~\ref{prelims}. 
The calculation is performed in Sec.~\ref{Calculation}. 
In Sec.~\ref{results}, the two-loop contributions are compared to the one-loop result for the pressure.

\section{Feynman rules and other prerequisites}  							\label{prelims}

\subsection{One-loop pressure}
The thermodynamical pressure $P$ is defined as the derivative of the partition function $Z$ with respect to the volume of the system,
\begin{equation}
P = T \, \frac{ \d \ln Z }{ \d V }
\,.
\end{equation}
In a field theory, $P$ can be calculated order by order in a loop expansion with the help of diagrammatic techniques 
as presented in \cite{Kapusta, Landsman van Weert}.

\begin{figure}[t]
\centering
\includegraphics[width=35mm]{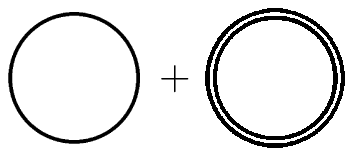}
\caption{     												\label{one loop}
One-loop contributions to the pressure.
Double lines correspond to tree-level heavy (TLH) modes, and single lines correspond to the tree-level massless (TLM) mode.}
\end{figure}

The pressure has been calculated on the one-loop level for SU(2) and SU(3) Yang-Mills theory in \cite{Hofmann}; 
we only consider the SU(2) case here.
The pressure contains a temperature dependent ground-state contribution arising from caloron 'condensation',
\begin{equation}                                							\label{g.s.}
P_{\text{g.s.}} (\lambda) = - 2 \lambda \Lambda^4 
\,,
\end{equation}
and a contribution associated with the one-loop diagrams shown in \fig{one loop},
\begin{equation}   											\label{onelooppressure}
P_{\text{one-loop}} (\lambda) 
= -\Lambda^4 \cdot \frac{ 2 \lambda^4 }{ (2\pi)^6 } 
\left( - \frac{ 2 \pi^4 }{45} + 6 \bar P \left(4 \pi e \lambda^{-3/2} \right)
\right) 
\,.
\end{equation}
The fixed mass scale $\Lambda$ is connected to the modulus of the composite Higgs field $\abs{ \phi }$ by 
\begin{equation} 											\label{Lambda}
\abs{ \phi }^2 = \frac{ \Lambda^3 }{ 2 \pi T }
\,,
\end{equation}
the dimensionless temperature $\lambda$ is defined as
\begin{equation} 											\label{lambda}
\lambda = \frac{2 \pi T}{\Lambda}
\,,
\end{equation}
and the function $\bar P$ is given as
\begin{equation}
\bar  P(a) = \int_0^{\infty} dx \, x^2 \log \left[ 1 - \exp \left( - \sqrt{ x^2 + a^2 } \right) \right]
\,.
\end{equation}

\noindent
There is also a 'nonthermal' contribution $-\Delta V$.
It is estimated as
\begin{equation}
\Delta V < \frac{|\phi|^4}{16\pi^2} 
\,.
\end{equation}
Because
\begin{equation}
\abs{ \frac{ \Delta V }{ V } } < \frac{\lam^{-3}}{32\pi^2} < 2 \cdot 10^{-6}
\,,
\end{equation}
it can be neglected in the electric phase.
(Recall that $\lam>11.65$ in the electric phase, cf. Sec.~3.1.)

\subsection{Feynman rules}  										\label{feynman rules}
For the calculation of the two-loop correction $\Delta P$ to the pressure of SU(2) being in its electric phase, 
we have the equation (see \cite{Hofmann})
\begin{equation}  											\label{feyn}
\parbox{130mm}{\includegraphics[width=130mm]{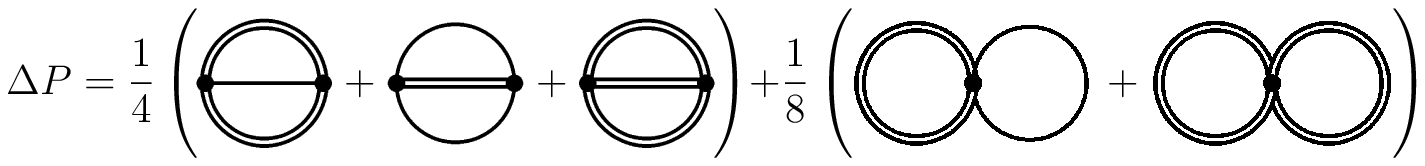}}
\,,
\end{equation}
where double lines represent TLH modes and single lines stand for TLM modes.
TLM modes will carry a color index 3, while the color indices 1 and 2 correspond to TLH modes. 
\eq{feyn} is valid for SU(N). 
In the case $\rm N=2$ considered here, the second and third diagram do not occur.

The calculation of the thermodynamical pressure is performed in unitary gauge, 
where $\phi$ is diagonal and the background is $a_\mu^{bg} = 0$.
The remaining gauge freedom is used to gauge the TLM mode to transversality,
$\d_i \, \delta a_i^{TLM} = 0$ (Coulomb gauge).
Unitary-Coulomb gauge is a completely fixed gauge, 
thus no Faddeev-Popov determinant needs to be considered 
and no ghost fields need to be introduced.

The computation is performed in the real-time formulation of thermal field theory.
Spacetime is Minkowskian with signature $(+,-,-,-)$.
Matsubara sums, which originate when working in compactified imaginary time,
are replaced with integrals over real Minkowskian momenta by means of contour integrals and analytic continuation.
For our purposes, the real-time formalism is preferable
because the implementation of constraints for the momenta of the participating fluctuations (\eqs{cutoff1}, \eqref{cutoff2})
is rather inconvenient in the imaginary-time formalism.
Moreover, the quantum and thermal part of a given fluctuation can be clearly discerned in the real-time formalism.

In the real-time formulation and in unitary-Coulomb gauge,
the Feynman rules employed in calculating the diagrams in \eq{feyn} are as follows:
The propagator for a free TLM-mode \cite{Hofmann, Kapusta, Landsman van Weert}
\begin{equation}
D^{TLM}_{\mu\nu,ab} (k,T) = - \delta_{ab} \, P^T_{\mu\nu}(k) \left( \frac{i}{k^2 + i \eps} 
+ 2 \pi \, \delta (k^2) \, n_B(|k_0|/T) \right)  
\,,
\end{equation}
where\footnote{
Static electric fields of long wavelength are completely screened due to the existence of an infinite real part
in the Debye screening mass $m_D = [\Pi_{00}(k_0=0,\vec k \to 0)]^{1/2}$; cf. App.~B} 
\begin{equation} 											\label{PT}
\begin{split}
P^T_{00}(k) &= P^T_{0i}(k) = P^T_{i0}(k) = 0   \,,  \\
P^T_{ij}(k) &= \delta_{ij} - \frac{k_i k_j}{\vec k^2}
\,,
\end{split}
\end{equation}
and $n_B$ denotes the Bose-Einstein-distribution, $ n_B(x) = \frac{1}{e^x-1} $.
The propagator for TLH-modes reads
\begin{equation}
D^{TLH}_{\mu\nu,ab} (k,T) = - \delta_{ab} \left( g_{\mu\nu} - \frac{k_{\mu} k_{\nu}}{m^2} \right) 
\left( \frac{i}{k^2 - m^2 + i \eps} + 2 \pi \, \delta (k^2-m^2) \, n_B(|k_0|/T) \right)   
\,.
\end{equation}

\noindent
The vertices are the usual ones (see \cite{Landsman van Weert, Itzykson Zuber}):  
The four-boson-vertex is 
\vspace{4mm}
\begin{multline}
\parbox[t][5mm][c]{20mm}{\includegraphics[width=29mm]{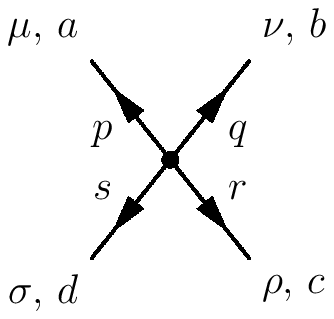}}
\quad\;\;
=
- i e^2 \, (2\pi)^4 \, \delta^4(p+q+r+s)
\left[
  \eps_{fab} \eps_{fcd} ( g^{\mu\rho} g^{\nu\sigma} - g^{\mu\sigma} g^{\nu\rho} )
\right. \\ \left.
+ \eps_{fac} \eps_{fdb} ( g^{\mu\sigma} g^{\rho\nu} - g^{\mu\nu} g^{\rho\sigma} )
+ \eps_{fad} \eps_{fbc} ( g^{\mu\nu} g^{\sigma\rho} - g^{\mu\rho} g^{\sigma\nu} )
\right]   
\,,
\end{multline}
\vspace{1mm} \par \noindent
and the three-boson-vertex is
\begin{equation}  											\label{3vertex}
\parbox{31mm}{\includegraphics[width=31mm]{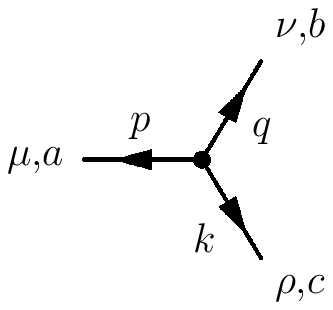}}
=
e \, (2 \pi)^4 \, \delta^4(p+q+k) \eps_{abc} 
[ g^{\mu\nu} (q-p)^{\rho} + g^{\nu\rho} (k-q)^{\mu} + g^{\rho\mu} (p-k)^{\nu} ] 
\,.
\end{equation}
According to Landsman and van Weert \cite{Landsman van Weert}, 
one has to divide every diagram by $i$ and by the number of its vertices.
Besides that, we demand that the momenta of quantum fluctuations are to be cut off at 
\begin{subequations} 											\label{cutoff1}
\begin{align}  
\abs{ p^2 - m^2 } \le \abs{ \phi }^2    								\label{cutoff1 M}   
\\ \intertext{in Minkowskian or}
p_E^2 + m^2 \le \abs{ \phi }^2       									\label{cutoff1 E}
\end{align} 
\end{subequations}
in Euclidean signature because of the existence of the compositeness scale $\abs{ \phi }$, cf. Sec.~3.1. 
The compositeness scale imposes a similar constraint on the center-of-mass energy flowing into the four-vertex, 
\begin{equation} 	 										\label{cutoff2}
\abs{ \left( p_1 + p_2 \right)^2 } \le \abs{ \phi }^2
\,,
\end{equation}
where $p_i$ are the ingoing momenta.

\subsection{Contributing diagrams}
The various contributions to the pressure on the two-loop level will be denoted as follows:
The first (last) diagram in \eq{feyn} shall be referred to as $\Delta P^{\rm MHH}$ ($\Delta P^{\rm HH}$), 
since there are two TLH and one TLM particle (two TLH particles) involved.
If we want to specify, for example, that in $\Delta P^{\rm MH}$ the TLH fluctuation 
is a vacuum fluctuation and the TLM fluctuation a thermal fluctuation, 
we write $\Delta P^{\rm MH}_{\rm TV}$ etc.
The statistical factors of \eq{feyn} are included in these definitions.

Which diagrams in \eq{feyn} do actually contribute for SU(2)?
Because of the structure of the vertex (\ref{3vertex}), 
the second and the third diagram in \eq{feyn} vanish.
Moreover, $\Delta P^{\rm MHH}_{\rm TTT}$ vanishes (for any $\rm N$ and $e$), 
since the on-shell conditions $p^2=k^2=m^2$, $q^2=0$
and energy-momentum-conservation $p+k+q=0$ cannot be satisfied simultaneously. 

As the mass of TLH fluctuations is connected to the gauge coupling via $m^2 = 4e^2 \abs{ \phi }^2$, 
the constraint \eqref{cutoff1} reads
\begin{equation} 											\label{constraint1}
\abs{ p^2 - 4e^2 |\phi|^2 } \le \abs{ \phi }^2  \qquad\text{or}\qquad p_E^2 \le \left( 1-4e^2 \right) \abs{ \phi }^2 
\,.
\end{equation}
This condition will eliminate all contributions containing quantum fluctuations of TLH modes, 
if the gauge coupling $e$ is larger than $1/2$. 
In this case, we only have the following nonvanishing contributions:
\begin{equation}
\begin{split}
\Delta P^{\rm HH}  &= \Delta P^{\rm HH}_{\rm TT} \\
\Delta P^{\rm MH}  &= \Delta P^{\rm MH}_{\rm TT} + \Delta P^{\rm MH}_{\rm VT} \\  
\Delta P^{\rm MHH} &= \Delta P^{\rm MHH}_{\rm VTT} 
\,.
\end{split}
\end{equation}
So, using the above notation, \eq{feyn} for $e>1/2$ simplifies to
\begin{equation}
\Delta P = \Delta P^{\rm HH}_{\rm TT} + \Delta P^{\rm MH}_{\rm TT} 
+ \Delta P^{\rm MH}_{\rm VT} + \Delta P^{\rm MHH}_{\rm VTT} 
\,.
\end{equation}

\section{Calculation}											\label{Calculation}
\subsection{Calculation of $\Delta P^{\rm HH}_{\rm TT}$}        					\label{calculation hhtt}
\begin{figure}[t]
\centering
\includegraphics[width=78mm]{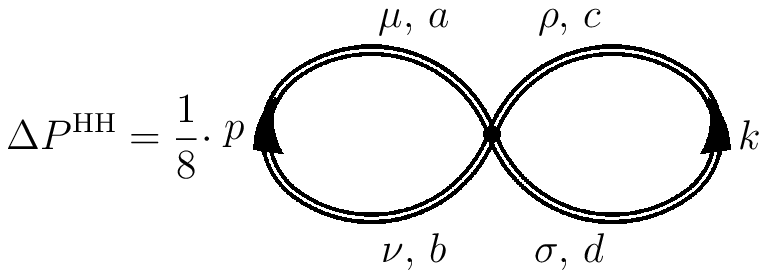}
\caption{    												\label{hh}
A local bubble diagram containing two TLH-modes} 
\end{figure}
We first write down the expression for $\Delta P^{\rm HH}_{\rm TT}$ without imposing 
the kinematical constraints \eqs{cutoff1}, \eqref{cutoff2} and
perform the constrained integrations afterwards still using the notation $\Delta P^{\rm HH}_{\rm TT}$.
According to the Feynman-rules given in section \ref{feynman rules}, 
the diagram shown in \fig{hh} with both fluctuations being thermal reads as:
\begin{equation}
\begin{split}
\Delta P^{\rm HH}_{\rm TT} & = 
\frac18
\int \frac{d^4 k}{(2\pi)^4} \int \frac{d^4 p}{(2\pi)^4} \, (-e^2)
\left[
  \eps_{adf} \eps_{fbc} \left( g^{\mu\nu} g^{\rho\sigma} - g^{\mu\rho} g^{\nu\sigma} \right)
\right. \\ & \quad \left.
+ \eps_{abf} \eps_{fdc} \left( g^{\mu\sigma} g^{\nu\rho} - g^{\mu\rho} g^{\nu\sigma} \right) 
+ \eps_{acf} \eps_{fdb} \left( g^{\mu\sigma} g^{\nu\rho} - g^{\mu\nu} g^{\sigma\rho} \right)
\right]
\\ & \quad \cdot
\left( - \delta_{ab} \right) \left( g_{\mu\nu} - \frac{k_{\mu}k_{\nu}}{m^2} \right) 
2\pi \, \delta(k^2-m^2) \, n_B (|k_0|/T) 
\\ & \quad \cdot
\left( - \delta_{cd} \right) \left( g_{\rho\sigma} - \frac{p_{\rho}p_{\sigma}}{m^2} \right) 
2\pi \, \delta(p^2-m^2) \, n_B (|p_0|/T) 
\\ & = - \frac{ e^2 }{ 2 } \int \frac{d^4 k}{(2\pi)^4} \int \frac{d^4 p}{(2\pi)^4}
\left( 12 - 3 \frac{p^2}{m^2} - 3 \frac{k^2}{m^2} +  \frac{p^2 k^2}{m^4} - \frac{(p k)^2}{m^4} \right)
\\ & \quad
\cdot 2 \pi \, \delta (k^2-m^2) \, n_B (|k_0|/T) \cdot 2 \pi \, \delta (p^2-m^2) \, n_B (|p_0|/T)
\,.
\end{split}
\end{equation}
(The contraction of Lorentz and color indices is deferred to Appendix~\ref{indices}.)
In evaluating this integral, we will first perform the integrations
over the zero-components of the momenta and thus eliminate the $\delta$-functions.
Therefore we rewrite the $\delta$-functions as follows:
\begin{equation}
\begin{split}
\Delta P^{\rm HH}_{\rm TT} 
& = 
- \frac{ e^2 }{ 2 } \int \frac{d^3 k}{(2\pi)^3} \int \frac{d^3 p}{(2\pi)^3} \int dk_0 \int dp_0
\left( 7 - \frac{ p_0^2 k_0^2 - 2 p_0 k_0 \vec p \vec k + ( \vec p \vec k )^2 }{m^4} \right)
\\ & \quad
\cdot \frac1{ 2 \sqrt{ \vec k^2 + m^2 }}
\left[ \delta \left( k_0 - \sqrt{ \vec k^2 + m^2 } \right) + \delta \left( k_0 + \sqrt{ \vec k^2 + m^2 } \right) \right] 
\cdot n_B (|k_0|/T)
\\ & \quad
\cdot \frac1{ 2 \sqrt{ \vec p^2 + m^2 }}
\left[ \delta \left( p_0 - \sqrt{ \vec p^2 + m^2 } \right) + \delta \left( p_0 + \sqrt{ \vec p^2 + m^2 } \right) \right] 
\cdot n_B (|p_0|/T)
\,.
\end{split}
\end{equation}
Now we notice that the two terms containing
\begin{equation}
\delta \left( k_0 - \sqrt{ \vec k^2 + m^2 } \right) 
\delta \left( p_0 - \sqrt{ \vec p^2 + m^2 } \right) 
\end{equation}
or
\begin{equation}
\delta \left( k_0 + \sqrt{ \vec k^2 + m^2 } \right) 
\delta \left( p_0 + \sqrt{ \vec p^2 + m^2 } \right) 
\,,
\end{equation}
respectively, yield the same result
since the rest of the integrand is invariant under simultaneous reflections $p_0 \to - p_0$, $k_0 \to - k_0$.
The same holds true for the remaining two summands.
So the expression is split into two contributions as
\begin{equation}
\begin{split}
\Delta P^{\rm HH}_{\rm TT} & = - \frac{ e^2 }{ 4 } \sum_{\pm} \int \frac{d^3 k}{(2\pi)^3} \int \frac{d^3 p}{(2\pi)^3} 
\, \frac1{ \sqrt{ \vec k^2 + m^2 } \sqrt{ \vec p^2 + m^2 }}
\\ & \quad
\cdot 
\left( 7 - \frac
{ (\vec p^2 + m^2) (\vec k^2 + m^2) \mp 2 \sqrt{ \vec p^2 + m^2 } \sqrt{ \vec k^2 + m^2 } \vec p \vec k 
+ ( \vec p \vec k )^2 }
{m^4} 
\right)
\\ & \quad
\cdot n_B \left( \sqrt{ \vec p^2 + m^2 } / T \right) \cdot n_B \left( \sqrt{ \vec k^2 + m^2 } / T \right) 
\\
& = - \frac{ e^2 }{ 4 } \sum_{\pm} \int \frac{d^3 k}{(2\pi)^3} \int \frac{d^3 p}{(2\pi)^3} 
\, \frac1{ \sqrt{ \vec k^2 + m^2 } \sqrt{ \vec p^2 + m^2 }}
\\ & \quad
\cdot 
\left( 6 
- \frac{ \vec p^2 }{ m^2 } - \frac{ \vec k^2 }{ m^2 } 
- \frac{ \vec p^2 \vec k^2 + ( \vec p \vec k )^2 }{m^4} 
\pm 2 \vec p \vec k \frac{ \sqrt{ \vec p^2 + m^2 } \sqrt{ \vec k^2 + m^2 } }{m^4} 
\right)
\\ & \quad
\cdot n_B \left( \sqrt{ \vec p^2 + m^2 } / T \right) \cdot n_B \left( \sqrt{ \vec k^2 + m^2 } / T \right) 
\,.
\end{split}
\end{equation}

\noindent
The tree-level mass of TLH-modes in SU(2) is given as 
\begin{equation}  											\label{masse}
m = 2 e \, |\phi|
\,,
\end{equation}
so the introduction of dimensionless momenta $\vec y$ and $\vec x$ according to
\begin{equation} 											\label{scaling}
\vec k = \vec y \cdot |\phi|   
\qquad\text{and}\qquad
\vec p = \vec x \cdot |\phi|   
\end{equation}
is suggested.
Moreover, temperature $T$ and compositeness scale $\abs{ \phi }$ can be expressed
in terms of the dimensionless temperature $\lambda$ and the mass scale $\Lambda$ 
(see \eqs{Lambda} and \eqref{lambda}) as
\begin{equation} 											\label{dimlos}
\begin{split}
|\phi| &= \Lambda \lambda^{-1/2}  
\,,
\\
|\phi| / T &= 2 \pi \lambda^{-3/2}
\,.
\end{split}
\end{equation}
Finally, we introduce three-dimensional polar coordinates for the scaled momenta.
The angle between $\vec x$ and $\vec y$ can be chosen to be the polar angle $\theta$.
$\Delta P^{\rm HH}_{\rm TT}$ is then expressed as
\begin{equation}   											\label{hhtt noch mit summe}
\begin{split} 
\Delta P^{\rm HH}_{\rm TT} 
& = 
- \frac{ e^2 \Lambda^4 \lambda^{-2} }{ 2 (2\pi)^4 } 
\sum_{\pm} 
\int_0^{\infty} dx \int_0^{\infty} dy \int_{-1}^1 d \cos\theta \,
\frac{ x^2 y^2 }{ \sqrt{ x^2 + 4e^2 } \sqrt{ y^2 + 4e^2 } }  
\\ & \quad \cdot
\left( 6 - \frac{ x^2 }{ 4e^2 } - \frac{ y^2 }{ 4e^2 } - \frac{ x^2 y^2 }{ 16e^4 } \left( 1 + \cos^2\theta \right) 
\pm 2 x y \cos \theta \, \frac{ \sqrt{ x^2 + 4e^2 } \sqrt{ y^2 + 4^2 }}{ 16e^4 } 
\right) 
\\ & \quad 
\cdot n_B \left( 2 \pi \lambda^{-3/2} \sqrt{ x^2 + 4e^2 } \right) 
\cdot n_B \left( 2 \pi \lambda^{-3/2} \sqrt{ y^2 + 4e^2 } \right) 
\,.
\end{split}  
\end{equation} 

\noindent
Now we have to implement the constraint for the center-of-mass energy in the vertex, 
\begin{equation}  											\label{exactlythesame}
\abs{(p+k)^2 } \le \abs{\phi}^2
\,.
\end{equation}
Of course, this condition has to undergo the same manipulations as the integrand itself: 
$p^2$ and $k^2$ have to be replaced with $m^2$,
$p_0 k_0$ with $\pm \sqrt{ \vec p^2 + m^2 } \sqrt{ \vec k^2 + m^2 }$,
\begin{equation}
\abs{ 2 m^2 \pm 2 \sqrt{ \vec p^2 + m^2 } \sqrt{ \vec k^2 + m^2 } - 2 \vec p \vec k } \le |\phi|^2 
\,,
\end{equation}
the momenta have to be rescaled as \eq{scaling}, the expression for the mass from \eq{masse} has to be inserted, 
and a change to polar coordinates has to be done.
We then have
\begin{equation} 											\label{grenze hhtt}
\abs{ 4e^2 \pm \sqrt{ x^2 + 4e^2 } \sqrt{ y^2 + 4e^2 } - x y \cos \theta }
\le \frac{1}{2}  
\,.
\end{equation}  

\begin{figure}
\centering
\includegraphics[width=5cm]{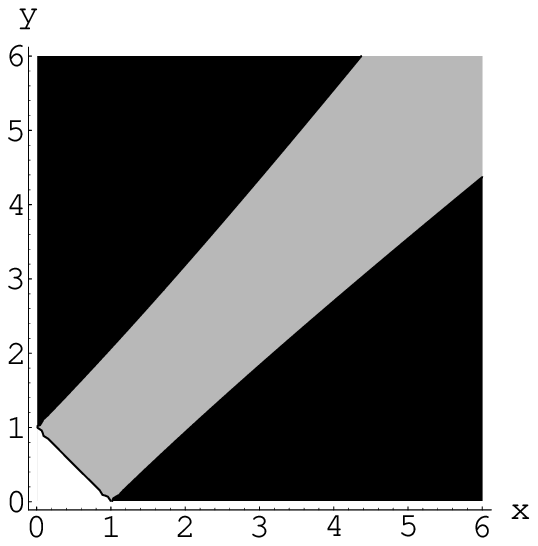}
\hspace{3cm}
\includegraphics[width=5cm]{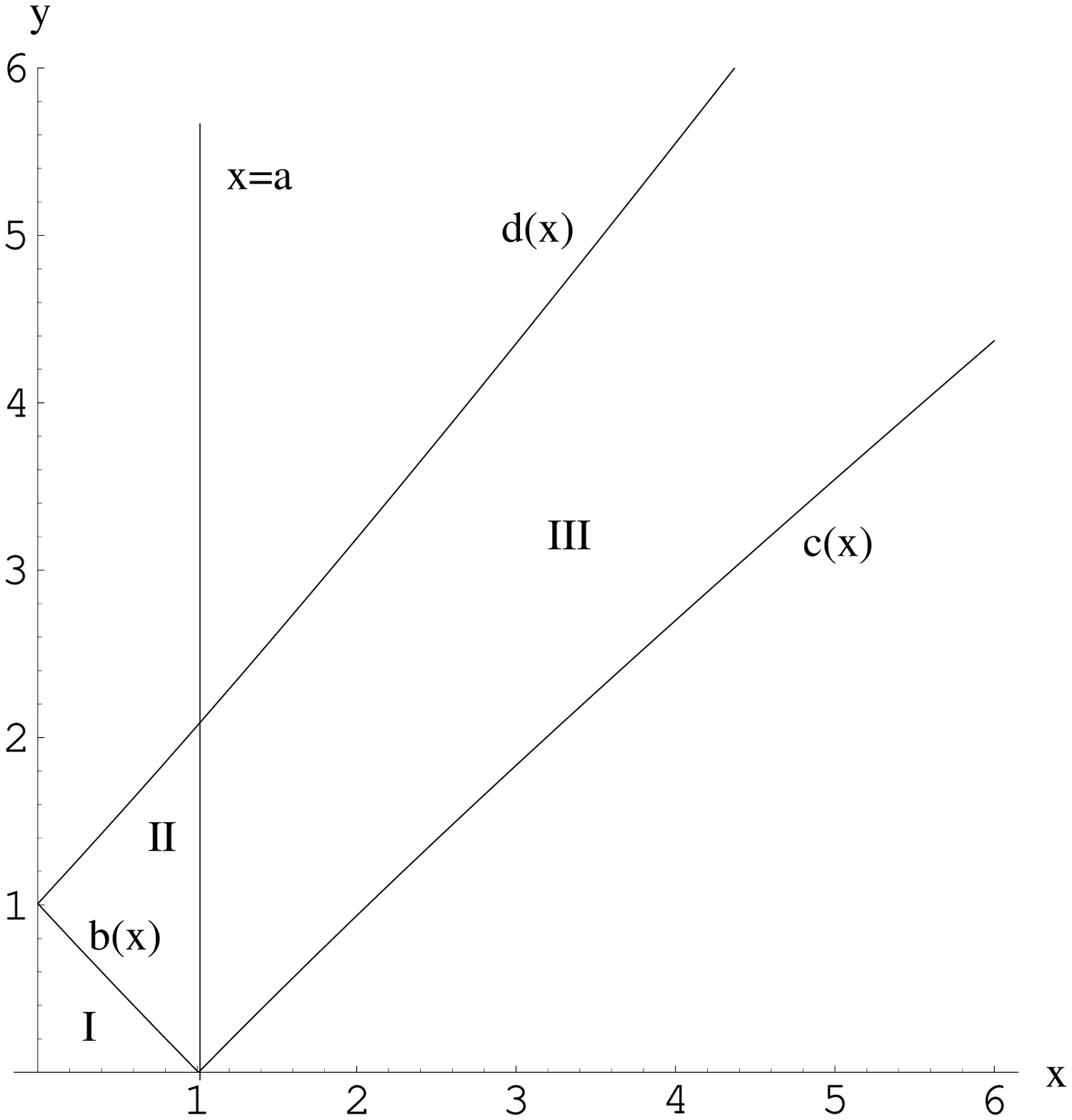}
\caption{  												\label{hhttgrenzen}
For determination of integration limits for $\cos\theta$ depending on $x$ and $y$. 
Left-hand side:
In the white region, the upper limit of integration is 1;
in the grey region, it is $g(x,y)$ as defined in \eq{def g}.
From the black region no contribution arises.
Right-hand side:
The integration is split into two bounded parts (I and II) and one unbounded part (III) as indicated.} 
\end{figure}

\noindent
For $e$ larger than $\frac{1}{2\sqrt2}$, 
condition (\ref{grenze hhtt}) with the positive sign cannot be fulfilled, 
and the corresponding integral in \eq{hhtt noch mit summe} is zero.
For the negative sign, condition (\ref{grenze hhtt}) can be fulfilled,
and we will implement it as an additional constraint on the region of integration for $\cos \theta$.
The condition
\begin{equation} 
\abs{ 4e^2 - \sqrt{ x^2 + 4e^2 } \sqrt{ y^2 + 4e^2 } - x y \cos \theta }
=
- 4e^2 + \sqrt{ x^2 + 4e^2 } \sqrt{ y^2 + 4e^2 } + x y \cos \theta 
\le \frac{1}{2}  
\end{equation}  
is equivalent to
\begin{equation} 											\label{def g}
\cos \theta 
\le 
\frac1{xy} \left( \frac12 + 4 e^2 - \sqrt{ x^2 + 4e^2 } \sqrt{ y^2 + 4e^2 } \right)
\equiv
g(x,y) 
\,,
\end{equation}
and thus $\cos \theta$ is to be integrated in the range $ [-1,1] \cap (-\infty,g] $. 
This can be done by replacing the upper limit of integration for $\cos \theta$ with
\begin{equation}
\max \left\{ -1 , \min \left\{ 1,g \right\} \right\}
\,.
\end{equation}
So, we have in total
\begin{equation}   											\label{integrand hhtt}
\begin{split} 
\Delta P^{\rm HH}_{\rm TT} 
& = 
- \frac{ e^2 \Lambda^4 \lambda^{-2} }{ 2 (2\pi)^4 } 
\int_0^{\infty} dx 
\int_0^{\infty} dy 
\int\limits_{-1}^{ \max \left\{ -1 , \min \left\{ 1,g \right\} \right\} } d \cos\theta \,
\frac{ x^2 y^2 }{ \sqrt{ x^2 + 4e^2 } \sqrt{ y^2 + 4e^2 } }  
\\ & \quad \cdot
\left( 6 - \frac{ x^2 }{ 4e^2 } - \frac{ y^2 }{ 4e^2 } - \frac{ x^2 y^2 }{ 16e^4 } \left( 1 + \cos^2\theta \right) 
- 2 x y \cos \theta \, \frac{ \sqrt{ x^2 + 4e^2 } \sqrt{ y^2 + 4^2 }}{ 16e^4 } 
\right) 
\\ & \quad 
\cdot n_B \left( 2 \pi \lambda^{-3/2} \sqrt{ x^2 + 4e^2 } \right) 
\cdot n_B \left( 2 \pi \lambda^{-3/2} \sqrt{ y^2 + 4e^2 } \right) 
\,.
\end{split}  
\end{equation} 
In this form the integrals can be performed numerically.
Instead of doing this, we will have a closer look at the function $g$ 
in order to be able to give the boundaries of integration explicitly. 
For the following compare with \fig{hhttgrenzen}.

If $g$ is greater than 1, \eq{def g} is fulfilled automatically and $\cos\theta$ has to be integrated
from -1 to 1; this is the case in the white area in \fig{hhttgrenzen}, 
bounded by the curve
\begin{equation}  											\label{integrale hhtt anfang}
b(x) = \frac{ - x - 8 e^2 x + \sqrt{1+16e^2} \sqrt{x^2+4e^2}}{8e^2}
\,.  
\end{equation}
This curve intersects with the $x$-axis at
\begin{equation}
a = \sqrt{ 1 + \frac1{16 e^2}}  
\,.
\end{equation}
If $-1<g<1$, the upper limit for the integration of $\cos\theta$ is $g$. 
The region where this is true is shaded in grey in \fig{hhttgrenzen};
it is enclosed by the lines 
\begin{equation}
\begin{split}
c(x) &= \frac{ x + 8 e^2 x - \sqrt{1+16e^2} \sqrt{x^2+4e^2}}{8e^2}
\,,  \\
d(x) &= \frac{ x + 8 e^2 x + \sqrt{1+16e^2} \sqrt{x^2+4e^2}}{8e^2}  
\,,
\end{split}
\end{equation}
and $b(x)$ as given above.
For all other values of $x$ and $y$ (shaded in black), 
we have $g<-1$ and thus no contribution.
This can be summarized as follows:
In \eq{integrand hhtt} the integral
$\int dx   \int dy   \int\limits d \cos\theta$
is decomposed as
\begin{equation} 											\label{integrale hhtt ende}
\begin{split}
\int_0^a dx \int_0^b dy \int_{-1}^1 d \cos\theta 
+ \int_0^a dx \int_b^d dy \int_{-1}^g d \cos\theta 
+ \int_a^{\infty} dx \int_c^d dy \int_{-1}^g d \cos\theta 
\end{split}
\end{equation}
where integrand and prefactor are as in \eq{integrand hhtt}.
The integrations over $\cos\theta$ can be performed analytically;
further evaluation needs to be done numerically.
For the result see \fig{ergebnis hhtt}. 

\fig{hhttgrenzen} illustrates that the integration over loop momenta is strongly restricted 
to a narrow band around $|\vec p| = |\vec k|$ due to the compositeness constraint \eq{cutoff2}.

\subsection{Calculation of $\Delta P^{\rm MH}_{\rm TT}$ and $\Delta P^{\rm MH}_{\rm VT}$}
\begin{figure}[ht]
\centering
\includegraphics[width=78mm]{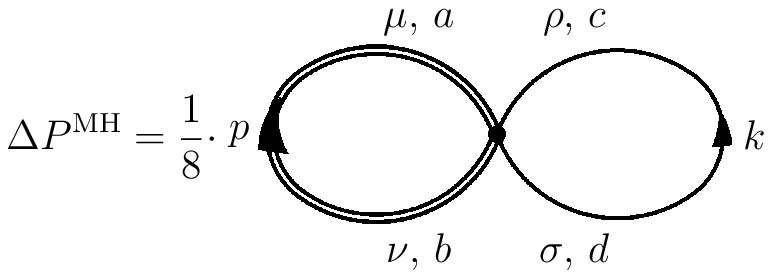}
\caption{   												\label{diagram mh}
A local bubble diagram containing one TLH-mode and one TLM-mode} 
\end{figure}
For the diagram depicted in \fig{diagram mh}, the Feynman rules yield
\begin{align} 											\label{math mh}
	\Delta P^{\rm MH} 
& = 
	\frac18 \int \frac{d^4 k}{(2\pi)^4} \int \frac{d^4 p}{(2\pi)^4} \, (-e^2)
	\left[
		\eps_{adf} \eps_{fbc} \left( g^{\mu\nu} g^{\rho\sigma} - g^{\mu\rho} g^{\nu\sigma} \right)
\right. \notag \\ & \quad \left.
		+ \eps_{abf} \eps_{fdc} \left( g^{\mu\sigma} g^{\nu\rho} - g^{\mu\rho} g^{\nu\sigma} \right) 
		+ \eps_{acf} \eps_{fdb} \left( g^{\mu\sigma} g^{\nu\rho} - g^{\mu\nu} g^{\sigma\rho} \right)
	\right]
\notag \\ & \quad \cdot
	\left( - \delta_{cd} \right) \left( g_{\rho\sigma} - \frac{p_{\rho}p_{\sigma}}{m^2} \right) 
	\left[ \frac{i}{p^2-m^2+i\eps} + 2\pi \, \delta(p^2-m^2) \, n_B (|p_0|/T) \right]
\\ & \quad \cdot
	\left( - \delta_{ab} \right) P^T_{\mu\nu} (k)
	\left[ \frac{i}{k^2+i\eps} + 2\pi \, \delta(k^2) \, n_B (|k_0|/T) \right]
\notag \\ & =
	- \frac{ e^2 }{ 2 } \int \frac{d^4 k}{(2\pi)^4} \int \frac{d^4 p}{(2\pi)^4}
	\left( - 6 + 2 \frac{p^2}{m^2} + \frac{\vec p^2}{m^2} 
	-  \frac{(\vec p \vec k)^2}{m^2 \vec k^2} \right)
\notag \\ & \quad 
	\cdot \left[ \frac{i}{p^2-m^2+i\eps} + 2\pi \, \delta(p^2-m^2) \, n_B (|p_0|/T) \right]
	\left[ \frac{i}{k^2+i\eps} + 2\pi \, \delta(k^2) \, n_B (|k_0|/T) \right]
\notag
\,.
\end{align}

\subsubsection{Calculation of $\Delta P^{\rm MH}_{\rm TT}$}
For both fluctuations being thermal, we have
\begin{equation}
\begin{split}
\Delta P^{\rm MH}_{\rm TT} 
&=
- \frac{ e^2 }{ 2 } \int \frac{d^4 k}{(2\pi)^4} \int \frac{d^4 p}{(2\pi)^4}
\left( - 6 + 2 \frac{p^2}{m^2} + \frac{\vec p^2}{m^2} 
-  \frac{(\vec p \vec k)^2}{m^2 \vec k^2} \right)
\\ & \quad
\cdot 2 \pi \, \delta (k^2) \, n_B (|k_0|/T) \cdot 2 \pi \, \delta (p^2-m^2) \, n_B (|p_0|/T)
\,.
\end{split}
\end{equation}
The evaluation of this diagram is similar to the calculation in Sec.~\ref{calculation hhtt}.
Integrating over the zero components of $p$ and $k$, we have
\begin{equation}
\begin{split}
\Delta P^{\rm MH}_{\rm TT} 
&=
- \frac{ e^2 }{ 2 } \int \frac{d^3 k}{(2\pi)^3} \int \frac{d^3 p}{(2\pi)^3} \int d k_0 \int d p_0
\left( - 4 + 2 \frac{\vec p^2}{m^2} -  \frac{(\vec p \vec k)^2}{m^2 \vec k^2} \right)
\\ & \quad \cdot
\frac{1}{ 2 \sqrt{ \vec p^2 + m^2 } }
\left[ \delta \left( p_0 - \sqrt{ \vec p^2 + m^2 } \right) + \delta \left( p_0 + \sqrt{ \vec p^2 + m^2 } \right) \right]
\cdot n_B (|p_0|/T)
\\ & \quad \cdot
\frac{1}{ 2 \sqrt{ \vec k^2 } }
\left[ \delta \left( k_0 - \sqrt{ \vec k^2 } \right) + \delta \left( k_0 + \sqrt{ \vec k^2 } \right) \right]
\cdot n_B (|k_0|/T) 
\\ &=
- \frac{ e^2 }{ 2 }\sum_{\pm} \int \frac{d^3 k}{(2\pi)^3} \int \frac{d^3 p}{(2\pi)^3}
\left( - 4 + 2 \frac{\vec p^2}{m^2} -  \frac{(\vec p \vec k)^2}{m^2 \vec k^2} \right)
\\ & \quad
\cdot 
\frac{1}{ 2 \sqrt{ \vec p^2 + m^2 } \sqrt{ \vec k^2 } }
\cdot n_B \left( \sqrt{ \vec k^2 } / T \right)   
\cdot n_B \left( \sqrt{ \vec p^2 + m^2 } / T \right)   
\,.
\end{split}
\end{equation}
Inserting the mass given in \eq{masse}, scaling momenta as in \eq{scaling},
introducing dimensionless variables as in \eq{dimlos} and polar coordinates, we obtain
\begin{equation}  											\label{obtain}
\begin{split}
\Delta P^{\rm MH}_{\rm TT} 
&= 
- \frac{ e^2 \Lambda^4 \lambda^{-2} }{ 2 (2\pi)^4 } 
\sum_{\pm} 
\int_0^{\infty} dx \int_0^{\infty} dy \int_{-1}^1 d \cos\theta
\\ & \quad \cdot
\left( - 4 + \frac{ x^2 }{ 4e^2 } - \frac{ x^2 \cos^2\theta }{ 4e^2 } \right) 
\frac{x^2 y}{ \sqrt{ x^2 + 4e^2 } }  
\\ & \quad
\cdot n_B \left( 2 \pi \lambda^{-3/2} \sqrt{ x^2 + 4e^2 } \right) 
\cdot n_B \left( 2 \pi \lambda^{-3/2} y \right) 
\,.
\end{split}  
\end{equation} 
At this stage the $\pm$ contributions in this equation are the same, but after imposing the condition 
\begin{equation} 											\label{spe}
\abs{ (p+k)^2 } \le \abs{ \phi }^2
\,,
\end{equation}
this is no longer the case.
By the above manipulations, \eq{spe} is transformed into
\begin{equation}
\abs{ m^2 \pm 2 \sqrt{ \vec p^2 + m^2 } \sqrt{ \vec k^2 } - 2 \vec p \vec k } \le \abs{ \phi }^2
\end{equation}
and finally into
\begin{equation} 											\label{grenze mhtt}
\abs{ 4e^2 \pm 2 y \sqrt{ x^2 + 4e^2 } - 2 x y \cos \theta } \le 1   
\,.
\end{equation}  
Again, condition (\ref{grenze mhtt}), which limits the range of integration for $\cos\theta$, 
can only be fulfilled for the negative sign. 
The condition 
\begin{equation} 
\abs{ 4e^2 - 2 y \sqrt{ x^2 + 4e^2 } - 2 x y \cos \theta } \le 1   
\end{equation}  
is satisfied  if and only if
\begin{equation}
\frac{ - 1 + 4 e^2 - 2 y \sqrt{ x^2 + 4 e^2 } }{ 2 x y } 
\equiv \tau_2 
\le \cos \theta 
\le \tau_1   
\equiv \frac{ + 1 + 4 e^2 - 2 y \sqrt{ x^2 + 4 e^2 } }{ 2 x y }  
\,,
\end{equation}
and thus $\cos \theta$ is to be integrated over the interval $[-1,1] \cap [\tau_2,\tau_1]$.
So the upper and lower limits of integration in \eq{obtain} are
\begin{equation}
\max \left\{ -1 , \min \left\{ 1 , \tau_1 \right\} \right\}
\qquad\text{and}\qquad
\min \left\{ 1 , \max \left\{ -1 , \tau_2 \right\} \right\}
\end{equation}
respectively.
Hence the result is
\begin{equation}  											\label{integrand mhtt}
\begin{split}
\Delta P^{\rm MH}_{\rm TT} 
&= 
- \frac{ e^2 \Lambda^4 \lambda^{-2} }{ 2 (2\pi)^4 } 
\int_0^{\infty} dx \int_0^{\infty} dy 
\int\limits_{ \min \left\{ 1 , \max \left\{ -1 , \tau_2 \right\} \right\} }^{ \max \left\{ -1 , \min \left\{ 1 , \tau_1 \right\} \right\} } 
d \cos\theta \,
\frac{x^2 y}{ \sqrt{ x^2 + 4e^2 } }  
\\ & \quad \cdot
\left( - 4 + \frac{ x^2 }{ 4e^2 } - \frac{ x^2 \cos^2\theta }{ 4e^2 } \right) 
\cdot n_B \left( 2 \pi \lambda^{-3/2} \sqrt{ x^2 + 4e^2 } \right) 
\cdot n_B \left( 2 \pi \lambda^{-3/2} y \right) 
\,.
\end{split}  
\end{equation} 

\noindent
Again, we want to give the limits of integration explicitly.
To do this, the points with $\tau_{1/2}=\pm1$ need to be determined. 
They are
\begin{equation}
\begin{split}
\tau_1 = \pm 1 
\qquad&\LRa\qquad
y = y_{1\pm} \equiv \frac{ 4 e^2 + 1 }{ 2 \left( \pm x + \sqrt{ x^2 + 4 e^2 } \right) }   \\
\tau_2 = \pm 1
\qquad&\LRa\qquad
y = y_{2\pm} \equiv \frac{ 4 e^2 - 1 }{ 2 \left( \pm x + \sqrt{ x^2 + 4 e^2 } \right) }  
\,.
\end{split}
\end{equation}
The following inequalities are satisfied:
\begin{equation}
\begin{split}
y_{2+} \le y_{2-} \le y_{1+} \le y_{1-} \qquad&\LRa\qquad x \le \eta \equiv \frac{ 2 e }{ \sqrt{ 16 e^2 -1 } }   
\\
y_{2+} \le y_{1+} \le y_{2-} \le y_{1-} \qquad&\LRa\qquad x \ge \eta 
\,. 
\end{split}
\end{equation}
Taking into account all possible combinations of $x$ and $y$ and the corresponding limits of integration,
the integral in \eq{integrand mhtt} splits into six parts, namely
\begin{equation}
\begin{split}
& \quad
  \int_0^{\eta} dx \int_{y_{2+}}^{y_{2-}} dy \int_{\tau_2}^1 d t   
+ \int_0^{\eta} dx \int_{y_{2-}}^{y_{1+}} dy \int_{-1}^1 d t  
+ \int_0^{\eta} dx \int_{y_{1+}}^{y_{1-}} dy \int_{-1}^{\tau_1} d t  
\\ & + 
  \int_{\eta}^{\infty} dx \int_{y_{2+}}^{y_{1+}} dy \int_{\tau_2}^1 d t  
+ \int_{\eta}^{\infty} dx \int_{y_{1+}}^{y_{2-}} dy \int_{\tau_2}^{\tau_1} d t  
+ \int_{\eta}^{\infty} dx \int_{y_{2-}}^{y_{1-}} dy \int_{-1}^{\tau_1} d t  
.
\end{split}
\end{equation}
The result of the numerical evaluation is shown in \fig{ergebnis mhtt}.

\subsubsection{Calculation of $\Delta P^{\rm MH}_{\rm VT}$}
For this case, we have from \eq{math mh} the expression
\begin{equation}
\begin{split}
\Delta P^{\rm MH}_{\rm VT} 
& =
- \frac{ e^2 }{ 2 }
\int \frac{d^4k}{(2\pi)^4} \int \frac{d^4p}{(2\pi)^4} \,
2 \pi \, \delta \left( p^2-m^2 \right) \, n_B \left( \abs{ p_0 } / T \right)
\\ & \quad \cdot
\left( - 6 + 2 \frac{p^2}{m^2} + \frac{ \vec p^2 }{m^2} - \frac{ ( \vec p \vec k )^2 }{m^2 \vec k^2} \right)
\frac{ i }{ k^2 + i \eps }
\,.
\end{split}
\end{equation}
As there is one thermal fluctuation ($p$) and one vacuum fluctuation ($k$) involved, 
only $p_0$ can be integrated by eliminating the $\delta$-function,
\begin{equation}
\begin{split}
\Delta P^{\rm MH}_{\rm VT} 
&= 
- \frac{ e^2 }{ 2 }
\int \frac{d^4k}{(2\pi)^4} \int \frac{d^3p}{(2\pi)^3} \int d p_0 \,
n_B \left( \sqrt{ \vec p^2 + m^2 } / T \right)
\frac{ i }{ k^2 + i \eps }
\\ & \quad \cdot
\left( - 4 + \frac{ \vec p^2}{m^2} - \frac{ ( \vec p \vec k)^2 }{m^2 \vec k^2} \right)
\frac{ \delta \left( p_0 - \sqrt{ \vec p^2 + m^2 } \right) + \delta \left( p_0 + \sqrt{ \vec p^2 + m^2 } \right) }
{2 \sqrt{ \vec p^2 + m^2 } }
\\ & = 
- \frac{ e^2 }{ 2 }
\sum_{\pm} \int \frac{d^4k}{(2\pi)^4} \int \frac{d^3p}{(2\pi)^3} \,
n_B \left( \sqrt{ \vec p^2 + m^2 } / T \right)
\frac{ i }{ k^2 + i \eps }
\\ & \quad \cdot
\left( - 4 + \frac{\vec p^2}{m^2} - \frac{ ( \vec p \vec k )^2 }{ m^2 \vec k^2 } \right)
\frac{1}{ 2 \sqrt{ \vec p^2 + m^2 } }
\,.
\end{split}
\end{equation}

\noindent
The implementation of the compositeness constraints
\begin{equation} 											\label{ignore}
\abs{ (p+k)^2 } \le \abs{ \phi }^2 
\end{equation}
and 
\begin{equation} 											\label{dontignore}
\abs{ k^2 } \le \abs{ \phi }^2
\end{equation}
is more difficult as in the previous calculations.
We will therefore ignore \eq{ignore} and give an estimate for $\Delta P^{\rm MH}_{\rm VT}$
by only taking into account \eqref{dontignore}.
The two contributions $\pm$ are equal, $k$ is analytically continued to Euclidean momenta. 
This yields the upper bound
\begin{equation}
\begin{split}
\abs{ \Delta P^{\rm MH}_{\rm VT} }
& \le 
\frac{e^2}{2} \cdot \!\!\!
\int\limits_{k_E^2 \le \abs{ \phi }^2} \!\!\!
\frac{d^4k_E}{(2\pi)^4} \int \frac{d^3p}{(2\pi)^3} \,
\frac{ n_B \left( \sqrt{ \vec p^2 + m^2 } / T \right) }{ \sqrt{ \vec p^2 + m^2 } }
\cdot \frac{ 1 }{ k_E^2 }
\cdot \abs{ - 4 + \frac{\vec p^2}{m^2} - \frac{ ( \vec p \vec k )^2 }{ m^2 \vec k^2 } }
\,.
\end{split}
\end{equation}

\noindent
As usual (inserting the expression for the mass, scaling the momenta 
and polar coordinates for the 3-vector $\vec p$ and the 4-vector $k_E$), 
we obtain
\begin{equation}
\begin{split}
\abs{ \Delta P^{\rm MH}_{\rm VT} }
& \le 
\frac{ e^2 \Lambda^4 \lambda^{-2} }{ 8 (2\pi)^4 }
\int_0^{\infty} dx \int_{-1}^1 d \cos\theta \,
\frac{x^2}{ \sqrt{ x^2 + 4e^2 } }
\\ & \quad
\cdot n_B \left( 2 \pi \lambda^{-3/2} \sqrt{ x^2 + 4e^2 } \right) 
\cdot \abs{ -4 + \frac{x^2}{4e^2} ( 1 - \cos^2 \theta ) }
\,.
\end{split}
\end{equation}
The numerical evaluation of this upper bound is shown in \fig{ergebnis mhvt}.

\subsection{Calculation of $\Delta P^{\rm MHH}_{\rm VTT}$}
\begin{figure}[t]
\centering
\includegraphics[width=76mm]{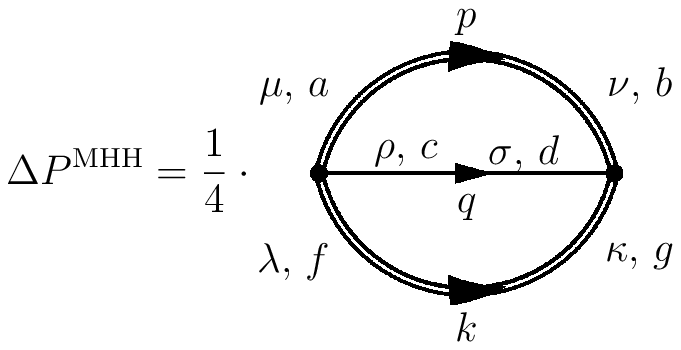}
\caption{  												\label{fd mhh}
A nonlocal diagram containing two TLH and one TLM mode} 
\end{figure}
The contribution $\Delta P^{\rm MHH}_{\rm VTT}$ shown in \fig{fd mhh} corresponds to the expression
\begin{equation}
\begin{split}
	\Delta P^{\rm MHH}_{\rm VTT}
& =
	\frac14 \cdot \frac{e^2}{2i} \cdot
	\int \frac{d^4 k}{(2 \pi)^4} \int \frac{d^4 p}{(2 \pi)^4} \int \frac{d^4 q}{(2 \pi)^4} \,
	(2\pi)^4 \, \delta(p+q+k)
\\ & \quad 
	\cdot \eps_{acf} \left[ g^{\mu\rho} (p-k)^{\lambda} + g^{\rho\lambda} (k-q)^{\mu} + g^{\lambda\mu} (q-p)^{\rho} \right]
\\ &\quad 
	\cdot \eps_{bdg} \left[ g^{\nu\sigma} (k-p)^{\kappa} + g^{\sigma\kappa} (q-k)^{\nu} + g^{\kappa\nu} (p-q)^{\sigma} \right]
\\ & \quad 
	\cdot (-\delta_{ab})   \left( g_{\mu\nu} - \frac{ p_{\mu} p_{\nu} }{ m^2 } \right)   2\pi \, \delta (p^2-m^2) \, n_B ( |p_0|/T ) 
\\ & \quad
	\cdot (-\delta_{cd})   \left( g_{\rho\sigma} - \frac{ k_{\rho} k_{\sigma} }{ m^2 } \right)   2\pi \, \delta (k^2-m^2) \, n_B ( |k_0|/T ) 
	\cdot (-\delta_{fg}) \, P^T_{\lambda\kappa} (q) \, \frac{i}{q^2+i\eps} 
\\ & = 
	- \frac{ e^2 }{ 4 } \cdot
	\int \frac{d^4p}{(2\pi)^4} \int \frac{d^4k}{(2\pi)^4} \,
	\frac{1}{(k+p)^2+i\eps}
\\ & \quad \cdot 
	\left[
		16 \left( m^2 - \frac{ (kp)^2 }{m^2} \right) 
		- \frac{ \vec k^2 \vec p^2 - (\vec k \vec p)^2 }{(\vec p + \vec k)^2} 
		\left( 8 + 4 \frac{(kp)^2}{m^4} \right)
	\right]
\\ & \quad
	\cdot 2\pi \, \delta \left( p^2-m^2 \right) \, n_B \left( \abs{p_0} / T \right)
	\cdot 2\pi \, \delta \left( k^2-m^2 \right) \, n_B \left( \abs{k_0} / T \right)
\,.
\end{split}
\end{equation}
The vacuum propagator $[ (k+p)^2+i\eps ]^{-1}$ has a singularity at
$(p+k)^2=2(m^2+kp)=0$, that is at $kp = -m^2$.
In spite of this, the term containing 
\begin{equation}
16 \Big( m^2 - \frac{ (kp)^2 }{m^2} \Big)
\end{equation}
is regular everywhere even for $\eps=0$ because
\begin{equation}
\frac{16}{(k+p)^2} \frac{m^4 - (kp)^2}{m^2} 
= \frac{16}{m^2} \frac{m^4 - (kp)^2}{2(m^2+kp)} 
= \frac{8}{m^2} (m^2 - kp)
= 8 \left( 1 - \frac{kp}{m^2} \right)
\,.
\end{equation}
So $i\eps$ can be dropped for this term.
This is, however, not the case for the second part which is proportional to 
\begin{equation}
\frac{ \vec k^2 \vec p^2 - (\vec k \vec p)^2 }{(\vec p + \vec k)^2} \left( 8 + 4 \frac{(kp)^2}{m^4} \right)
\,.
\end{equation}

\noindent
In this diagram we again have to take care of constraining the momenta of vacuum fluctuations,
\begin{equation} 											\label{grenze mhhvtt}
\abs{ q^2 } = \abs { (p+k)^2 } = 2 \abs{ m^2 +pk } \le \abs{ \phi }^2
\,.
\end{equation}
This is exactly the same condition as has been found in the calculation of
$\Delta P^{\rm HH}_{\rm TT}$ (compare \eq{exactlythesame} and the following), 
so we can take the integration limits from Sec.~\ref{calculation hhtt}.

We once again reformulate the $\delta$-functions, 
\begin{equation}
\begin{split}
\Delta P^{\rm MHH}_{\rm VTT}
& =
- \frac{ e^2 }{ 4 } \cdot
\int \frac{d^3p}{(2\pi)^3} \int \frac{d^3k}{(2\pi)^3} \int d p_0 \int d k_0 \,
\frac{1}{2(m^2+kp)+i\eps}
\\ & \quad
\cdot \frac1{ 2 \sqrt{ \vec p^2 + m^2 }}
\left[ \delta \left( p_0 - \sqrt{ \vec p^2 + m^2 } \right) + \delta \left( p_0 + \sqrt{ \vec p^2 + m^2 } \right) \right]
n_B \left( \abs{p_0} / T \right)
\\ & \quad
\cdot \frac1{ 2 \sqrt{ \vec k^2 + m^2 }}
\left[ \delta \left( k_0 - \sqrt{ \vec k^2 + m^2 } \right) + \delta \left( k_0 + \sqrt{ \vec k^2 + m^2 } \right) \right]
n_B \left( \abs{k_0} / T \right)
\\ & \quad
\cdot
\left[
16 \left( m^2 - \frac{ (kp)^2 }{m^2} \right) 
- \frac{ \vec k^2 \vec p^2 - (\vec k \vec p)^2 }{(\vec p + \vec k)^2} 
\left( 8 + 4 \frac{(kp)^2}{m^4} \right)
\right]
\end{split}
\end{equation}
and perform the integrations over $p_0$ and $k_0$, 
\begin{equation}
\begin{split}
\Delta P^{\rm MHH}_{\rm VTT}
& =
- \frac{e^2}{16} \cdot
\int \frac{d^3p}{(2\pi)^3} \int \frac{d^3k}{(2\pi)^3} \,
\frac{1}{ m^2 -  \sqrt{ \vec p^2 + m^2 } \sqrt{ \vec k^2 + m^2 } +  \vec p \vec k + i \eps }
\\ & \quad
\cdot \frac1{ \sqrt{ \vec k^2 + m^2 } \sqrt{ \vec p^2 + m^2 }} \,
n_B \left( \sqrt{ \vec p^2 + m^2 } / T \right) \,
n_B \left( \sqrt{ \vec k^2 + m^2 } / T \right)
\\ & \quad
\cdot
\left[
16 \left( m^2 - \frac{ \left( \sqrt{ \vec k^2 + m^2 } \sqrt{ \vec p^2 + m^2 } + \vec k \vec p \right)^2 }{m^2} \right) 
\right. \\ & \qquad \left.
- \frac{ \vec k^2 \vec p^2 - (\vec k \vec p)^2 }{(\vec p + \vec k)^2} 
\left( 8 + 4 \frac{ \left( \sqrt{ \vec k^2 + m^2 } \sqrt{ \vec p^2 + m^2 } + \vec k \vec p \right)^2 }{m^4} \right)
\right]
\,,
\end{split}
\end{equation}
rescale momenta by setting
\begin{equation}
\vec p = \vec x \cdot \abs{ \phi } 
\qquad \text{and} \qquad
\vec k = \vec y \cdot \abs{ \phi }
\,,
\end{equation}
go to polar coordinates, and introduce dimensionless variables $\lambda$ and $\Lambda$ as in \eq{dimlos}.
Finally we arrive at
\begin{equation}   											\label{ganzlang}
\begin{split}
\Delta P^{\rm MHH}_{\rm VTT}
&=
- \frac{ e^2 \Lambda^4 \lambda^{-2} }{ 8 (2\pi)^4 }
\int dx \int dy \int dt \,
\frac{ x^2 y^2 }{ \sqrt{ x^2 + 4e^2 } \sqrt{ y^2 + 4e^2 } }
\\ & \quad \cdot
n_B \left( 2 \pi \lambda^{-3/2} \sqrt{ x^2 + 4e^2 } \right) \,
n_B \left( 2 \pi \lambda^{-3/2} \sqrt{ y^2 + 4e^2 } \right)
\\ & \quad \cdot
\frac{1}
{ 4e^2 - \sqrt{ x^2 + 4e^2 } \sqrt{ y^2 + 4e^2 } - x y t + i \eps }
\cdot
\left[
\frac{ x^2 y^2 ( t^2 - 1 ) }{ x^2 + y^2 + 2xyt }
\right. \\ & \quad \left. \cdot
\left(
12 + \frac{x^2}{e^2} + \frac{y^2}{e^2} + \frac{x^2 y^2}{4e^4} ( 1 + t^2 )
+ \frac{xyt}{2e^4} \sqrt{x^2+4e^2} \sqrt{y^2+4e^2}
\right)
\right. \\ & \quad \left.
- 16
\left( 
x^2 + y^2 + \frac{x^2 y^2}{4e^2} ( 1 + t^2 ) + \frac{xyt}{2e^2} \sqrt{x^2+4e^2} \sqrt{y^2+4e^2} 
\right)
\right]
\,.
\end{split}
\end{equation}
The condition in \eq{grenze mhhvtt} is implemented exactly as in the case of $\Delta P^{\rm HH}_{\rm TT}$, 
see \eq{integrale hhtt anfang} to \eqref{integrale hhtt ende}; 
$\cos\theta$ has been abbreviated by $t$ in \eq{ganzlang}.

As already indicated above, the term proportional to $\frac{x^2 y^2 (t^2-1)}{x^2 + y^2 + 2xyt}$ has a singularity.
We evaluate the integral numerically by prescribing a small value for $\eps$.
In Tab.~\ref{epsilon}, one can see that for decreasing $\eps$ 
the real part converges to a finite value, while the imaginary part converges to zero.
For our purposes, working with $\eps = 10^{-8}$ is sufficient for determining the real part; the imaginary  part is ignored.
The second term is regular, and the evaluation is straightforward by setting $\eps=0$.
The result of numerical evaluation is shown in \fig{ergebnis mhhvtt}.
\begin{table}[t]
\centering
\begin{tabular}{l||l|l||l|l||}
        & \multicolumn{2}{c||}{$\lam = 20$}  &  \multicolumn{2}{c||}{$\lam = 100$}   \\
\hline
$\eps$  &  real part  &  imag. part  & real part  & imag. part    \\
\hline
$10^{ -2}$ & $-1.46989 \cdot 10^{-3}$ & $-2.39 \cdot 10^{ -4}$ & $-6.22191 \cdot 10^{-4}$ & $-1.01 \cdot 10^{ -4}$   \\
$10^{ -3}$ & $-1.65682 \cdot 10^{-3}$ & $-8.31 \cdot 10^{ -5}$ & $-7.01275 \cdot 10^{-4}$ & $-3.51 \cdot 10^{ -5}$   \\
$10^{ -4}$ & $-1.71600 \cdot 10^{-3}$ & $-2.70 \cdot 10^{ -5}$ & $-7.26311 \cdot 10^{-4}$ & $-1.14 \cdot 10^{ -5}$   \\
$10^{ -5}$ & $-1.73472 \cdot 10^{-3}$ & $-8.62 \cdot 10^{ -6}$ & $-7.34229 \cdot 10^{-4}$ & $-3.65 \cdot 10^{ -6}$   \\
$10^{ -6}$ & $-1.74064 \cdot 10^{-3}$ & $-2.73 \cdot 10^{ -6}$ & $-7.36732 \cdot 10^{-4}$ & $-1.16 \cdot 10^{ -6}$   \\
$10^{ -7}$ & $-1.74251 \cdot 10^{-3}$ & $-8.60 \cdot 10^{ -7}$ & $-7.37524 \cdot 10^{-4}$ & $-3.63 \cdot 10^{ -7}$   \\
$10^{ -8}$ & $-1.74308 \cdot 10^{-3}$ & $-2.44 \cdot 10^{ -7}$ & $-7.37760 \cdot 10^{-4}$ & $-9.63 \cdot 10^{ -8}$   \\
$10^{ -9}$ & $-1.74321 \cdot 10^{-3}$ & $-5.68 \cdot 10^{ -8}$ & $-7.37799 \cdot 10^{-4}$ & $-1.62 \cdot 10^{ -8}$   \\
$10^{-10}$ & $-1.74324 \cdot 10^{-3}$ & $-1.32 \cdot 10^{ -8}$ & $-7.37807 \cdot 10^{-4}$ & $-2.74 \cdot 10^{ -9}$   \\
$10^{-11}$ & $-1.74324 \cdot 10^{-3}$ & $-1.92 \cdot 10^{ -9}$ & $-7.37777 \cdot 10^{-4}$ & $-1.09 \cdot 10^{ -9}$   \\
$10^{-12}$ & $-1.74324 \cdot 10^{-3}$ & $-2.02 \cdot 10^{-10}$ & $-7.37778 \cdot 10^{-4}$ & $-2.04 \cdot 10^{-10}$  
\end{tabular}
\caption{												\label{epsilon}
Results of numerical evaluation of the singular part of
$\Delta P^{\rm MHH}_{\rm VTT} / P_{\text{one-loop}}$ 
for $\lam=20$ and $\lam=100$. 
In the limit $\eps\to0$, the real part converges to a finite value; 
with $\eps=10^{-8}$, already four digits are stable.
The imaginary part goes to zero for vanishing $\eps$.}
\end{table}

\section{Results}  								\label{results}
\begin{figure}[p]
\centering
\includegraphics[width=7cm]{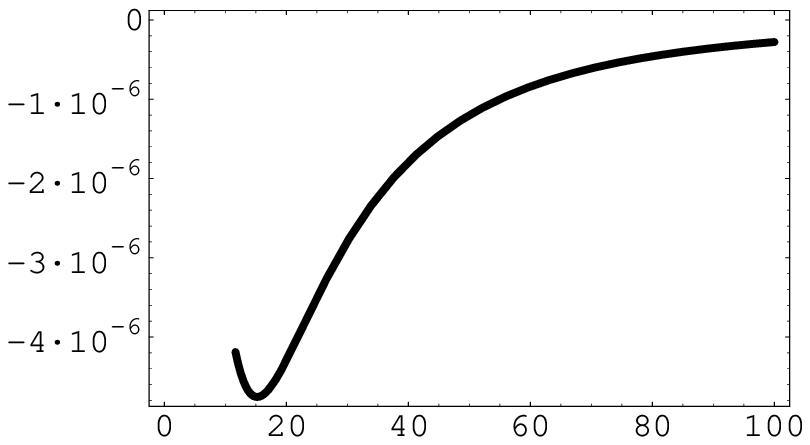}
\hfill
\includegraphics[width=7cm]{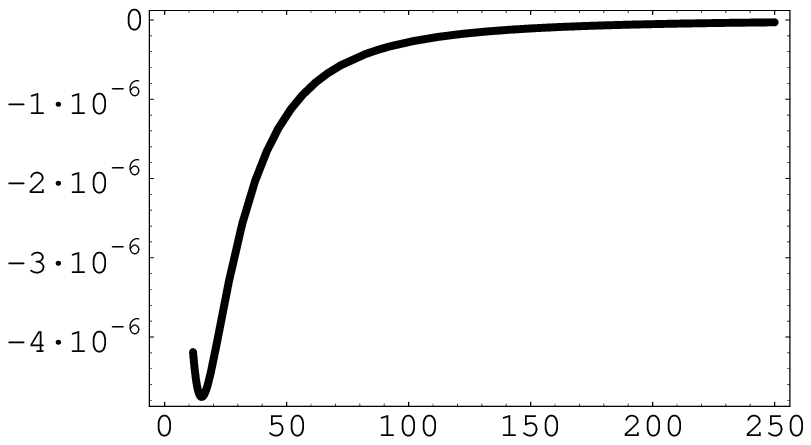}
\caption{   												\label{ergebnis hhtt}
The ratio of $\Delta P^{\rm HH}_{\rm TT}$ and $P_{\text{one-loop}}$ plotted for $11.65<\lambda<100$ and $11.65<\lambda<250$.} 
\end{figure}

\begin{figure}[p]
\centering
\includegraphics[width=7cm]{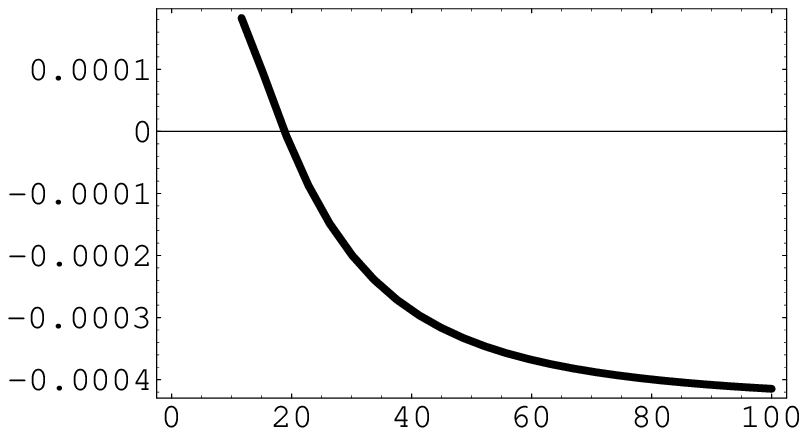}
\hfill
\includegraphics[width=7cm]{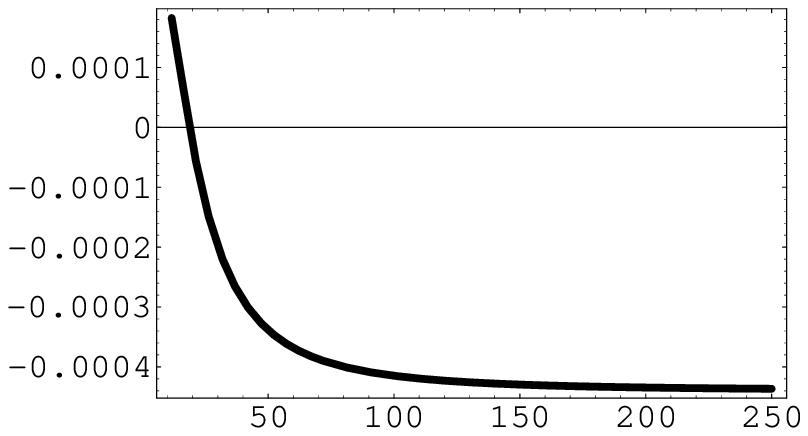}
\caption{ 												\label{ergebnis mhtt}
The ratio of $\Delta P^{MH}_{TT}$ and $P_{\text{one-loop}}$ plotted for $11.65<\lambda<100$ and $11.65<\lambda<250$.} 
\end{figure}

\begin{figure}[p]
\centering
\includegraphics[width=7cm]{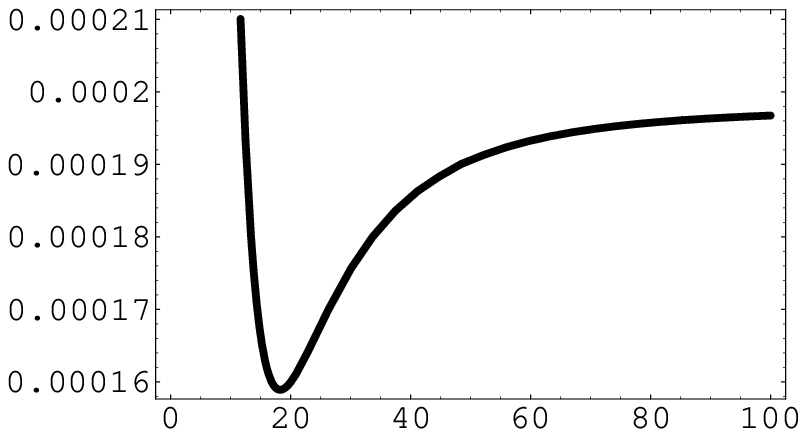}
\hfill
\includegraphics[width=7cm]{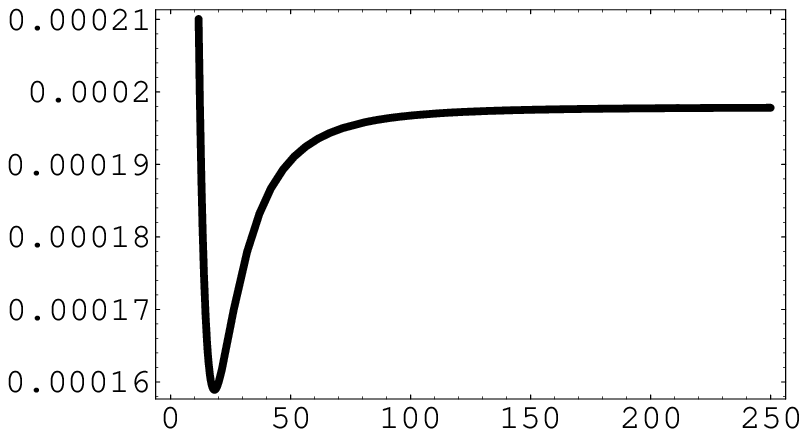}
\caption{  												\label{ergebnis mhvt}
An upper bound for the ratio of $\Delta P^{\rm MH}_{\rm VT}$ and $P_{\text{one-loop}}$ 
plotted for $11.65<\lambda<100$ and $11.65<\lambda<250$.} 
\end{figure}

\begin{figure}[p]
\centering
\includegraphics[width=7cm]{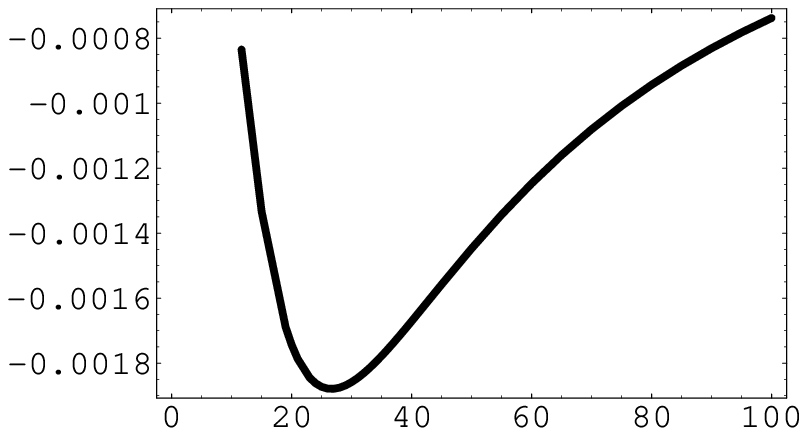}
\hfill
\includegraphics[width=7cm]{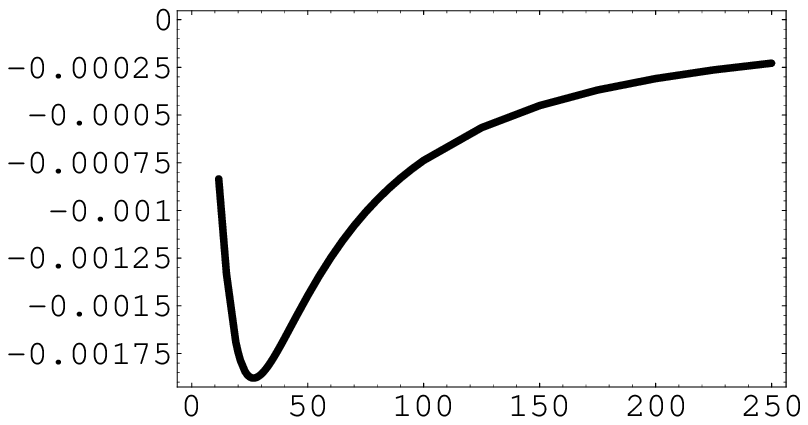}
\caption{ 												\label{ergebnis mhhvtt}
The ratio of $\Delta P^{\rm MHH}_{\rm VTT}$ and $P_{\text{one-loop}}$ plotted for $11.65<\lambda<100$ and $11.65<\lambda<250$.} 
\end{figure}

To compare the corrections arising from the two-loop diagrams to the one-loop pressure, 
we plot the ratio of each of the contributions to $\Delta P$ and $P_{\text{one-loop}}$ from \eq{onelooppressure}
as a function of the dimensionless temperature $\lam_c = 11.65 \le \lam \le 250$. 
The results are shown in \fig{ergebnis hhtt} through \fig{ergebnis mhhvtt}.
The one-loop pressure does not include the ground-state contribution \eq{g.s.}.
For the effective gauge coupling, the plateau value $e=5.1$ is used for all temperatures $\lam$.
Throughout most of the electric phase, this is admissible, 
but the logarithmic pole of $e$ at the critical temperature $\lam_c$ is ignored.
The nonlocal diagram is the dominating contribution for $\lam < 100$.
Throughout the electric phase, the corrections arising from two-loop diagrams are tiny;
the ratio of two-loop to one-loop contribution is $2 \cdot 10^{-3}$ at most.
With rising temperature, the two-loop contributions decrease.
The contributions $\Delta P^{\rm MHH}_{\rm VTT}$ and $\Delta P^{\rm HH}_{\rm TT}$ approach zero for large temperatures.
In contrast to that, $\Delta P^{\rm MH}_{\rm TT}$ becomes constant.
For $\Delta P^{\rm MH}_{\rm VT}$ we obtained only a rough estimate. 
For temperatures close to the phase transition, there is a dip in the two-loop corrections to the pressure.
The minimum of the dominating contribution is at $\lam \sim 27$.

The microscopic interpretation is as follows: 
Close to the phase transition at $\lam_c$, the monopole mass $M_{\text{monopole}} \propto \frac Te$ decreases sizeably. 
This increases the scattering of TLM modes off magnetic monopoles,
with the consequence of a dip in the dominating two-loop contribution.
With rising temperature the monopoles become massive and dilute, and scattering processes are suppressed.
But as even for asymptotically high temperatures massive and dilute scattering centers are present,
the contribution $\Delta P^{\rm MH}_{\rm TT}$ remains finite.

\clearpage

\chapter{Summary and Outlook}

In this thesis, we computed the phase and the modulus of a composite, adjoint Higgs field $\phi$
relevant for the thermodynamical description of an SU(2) Yang-Mills theory.
The phase was defined by giving a spatial and moduli space average over an 
adjointly transforming two-point function and demanding BPS-saturation.
The modulus could be inferred after assuming the existence of an externally given scale.
It was seen that the field $\phi$ exploits the instantaneous long range correlations in the classical caloron solution.
The nontrivial temporal winding was obtained only after averaging over the entire admissible part of the moduli space.
From the BPS equation, $\phi$'s potential was uniquely deduced. 

The two-loop contribution to the pressure of SU(2) Yang-Mills theory was computed.
There are four contributing Feynman diagrams, among them one nonlocal. This one is seen to be dominant.
The two-loop corrections are smaller than $2 \cdot 10^{-3}$ as compared to the one-loop result.
Therefore the underlying picture of only very weakly interacting thermal quasiparticles is confirmed.
The importance of the compositeness constraints in this process was pointed out.
A microscopic interpretation in terms of TLM modes scattering off magnetic monopoles was given.

We saw that the compositeness constraints rule out (at least in the large coupling regime) a number of
diagrams which one would naively expect to contribute.
It would be interesting to consider this for higher loop-diagrams.
Possibly, from some loop order upward only a few classes of diagrams will survive.
The effects of the two-loop corrections for the pressure on the evolution of the gauge coupling 
are currently being worked on \cite{Hofmann Rohrer}.
A review on the discussed approach to SU(2) and SU(3) Yang-Mills thermodynamics
(containing also the material presented here) and especially
indicating implications on particle physics and cosmology is available in \cite{Hofmann}.

\begin{appendix}
{\small

\chapter{Appendix to Chapter 3} 
\section{Notation and conventions}									\label{not&conv}
In Chapter~3, the following conventions are used:  
SU(2) gauge field and field strength are written in matrix notation as
\begin{equation}
A_\mu = A^a_\mu \, \frac{\lam^a}{2}
\qquad\text{and}\qquad
F_{\mu\nu} = F^a_{\mu\nu} \, \frac{\lam^a}{2}  
\,,
\end{equation}
where $\lam^a$ are the Pauli matrices.
The covariant derivative is defined as
\begin{equation}
D_{\mu} = \d_{\mu} - i A_{\mu}  
\,,
\end{equation}
where the coupling constant is absorbed in the gauge field.
The field strength is in matrix notation
\begin{equation}  
F_{\mu\nu} 
= \d_{\mu} A_{\nu} - \d_{\nu} A_{\mu} - i \left[ A_{\mu},A_{\nu} \right] 
\end{equation}
and in components
\begin{equation}
F^a_{\mu\nu} = \d_\mu A^a_\nu - \d_\nu A^a_\mu + \eps^{abc} A^b_\mu A^c_\nu
\,.
\end{equation}
We work in a Euclidean spacetime with $x = ( \tau , \vec x )$.
Latin indices run from 1 to 3, Greek indices from 1 to 4.
Upper and lower indices are not distinguished. 
The summation convention is implied if not specified otherwise. 
The totally antisymmetric symbol $\eps_{\mu\nu\rho\sigma}$ has $\eps_{1234}=\eps^{1234}=+1$.
The four-gradient is 
\begin{equation}
\d_\mu = \frac{\d}{\d x_\mu} 
\,,
\end{equation}
derivatives with respect to time and space coordinates are written as
\begin{equation}
\d_4 = \frac{\d}{\d\tau}  
\qquad \text{and} \qquad 
\d_i = \frac{\d}{\d x_i} 
\end{equation}
respectively. 
The derivative with respect to the radial coordinate $r=|\vec x|$ is written as
\begin{equation}
\d_r = \frac{\d}{\d r}
\,.
\end{equation}
The Pauli matrices are denoted by $\lambda^a$, they have the explicit form
\begin{equation}
\lambda^1 = \begin{pmatrix}0&1\\1&0\end{pmatrix} 
\,,\qquad
\lambda^2 = \begin{pmatrix}0&-i\\i&0\end{pmatrix}
\,,\qquad
\lambda^3 = \begin{pmatrix}1&0\\0&-1\end{pmatrix}
\,.
\end{equation}
The Pauli matrices fulfill
\begin{equation}
\begin{split}
\left[ \lambda^a , \lambda^b \right] &= 2 i \eps_{abc} \lambda^c  \\
\lambda^a \lambda^b &= \delta_{ab} + i \eps_{abc} \lambda^c  \\
\lambda^a \lambda^b \lambda^c 
&= \delta_{ab} \lambda^c - \delta_{ac} \lambda^b + \delta_{bc} \lambda^a + i \eps_{abc}  
\,.
\end{split}
\end{equation}
The traces of products of Pauli-matrices are
\begin{equation}  											\label{paulispur}
\begin{split}
\tr \lambda^a &= 0  \\
\tr \lambda^a \lambda^b &= 2 \delta_{ab}  \\
\tr \lambda^a \lambda^b \lambda^c &= 2 i \eps_{abc}  \\
\tr \lambda^a \lambda^b \lambda^c \lambda^d 
&= 2 \left( \delta_{ab} \delta_{cd} + \delta_{ad} \delta_{bc}- \delta_{ac} \delta_{bd} \right)  \\
\tr \lambda^a \lambda^b \lambda^c \lambda^d \lambda^e
&= 2 i \left( \delta_{ab} \eps_{cde} + \delta_{cd} \eps_{abe} 
   - \delta_{cd} \eps_{abd} + \delta_{de} \eps_{abc} \right)
\,.
\end{split}
\end{equation}
The 't~Hooft symbols $\eta$ and $\bar\eta$ are defined as
\begin{equation}
\begin{split}
\et a\mu\nu & = \eps_{a\mu\nu} + \delta_{a\mu} \delta_{\nu4} - \delta_{a\nu} \delta_{\mu4}   \\
\etb a\mu\nu & = \eps_{a\mu\nu} - \delta_{a\mu} \delta_{\nu4} + \delta_{a\nu} \delta_{\mu4} 
\,. 
\end{split}
\end{equation}
The symbols $\eta$ ($\bar\eta$) are (anti) self-dual and antisymmetric in the vector indices,
\begin{equation}
\begin{split}
\et a\mu\nu &= \frac12 \eps_{\mu\nu\alpha\beta} \, \et a\alpha\beta  \\
\etb a\mu\nu &= - \frac12 \eps_{\mu\nu\alpha\beta} \, \etb a\alpha\beta  \\
\et a\mu\nu &= - \et a\nu\mu  \\
\etb a\mu\nu &= - \etb a\nu\mu
\,.
\end{split}
\end{equation}
They fulfill a number of useful relations:
\begin{equation}     											\label{relations}
\begin{split}
\et a\mu\nu \et b\mu\nu &= 4 \delta_{ab}  \\
\et a\lambda\mu \et a\lambda\nu &= 3 \delta_{\mu\nu}  \\
\et a\mu\nu \et a\mu\nu &= 12  \\
\et a\lambda\mu \et b\lambda\nu &= \delta_{ab} \delta_{\mu\nu} + \eps_{abc} \et c\mu\nu  \\
\eps_{abc} \et b\mu\nu \et c\kappa\lambda &= 
    \delta_{\mu\kappa} \et a\nu\lambda + \delta_{\nu\lambda} \et a\mu\kappa
  - \delta_{\mu\lambda} \et a\nu\kappa - \delta_{\nu\kappa} \et a\mu\lambda  \\
\eps_{abc} \et b\mu\nu \et c\mu\lambda &= 2 \et a\nu\lambda  \\
\et a\mu\nu \et b\nu\lambda \et c\lambda\mu &= 4 \eps_{abc}
\,. 
\end{split}
\end{equation}
The relations \eqref{relations} hold for $\bar\eta$ as well.
Besides that,
\begin{equation}
\begin{split}
\et a\mu\nu \et a\kappa\lambda
&= \delta_{\mu\kappa} \delta_{\nu\lambda} - \delta_{\mu\lambda} \delta_{\nu\kappa}
  + \eps_{\mu\nu\kappa\lambda}  \\
\etb a\mu\nu \etb a\kappa\lambda
&= \delta_{\mu\kappa} \delta_{\nu\lambda} - \delta_{\mu\lambda} \delta_{\nu\kappa}
  - \eps_{\mu\nu\kappa\lambda}  \\
\eps_{\lambda\mu\nu\sigma} \et a\rho\sigma
&=
\delta_{\rho\lambda} \et a\mu\nu + \delta_{\rho\nu} \et a\lam\mu + \delta_{\rho\mu} \et a\nu\kappa  \\
- \eps_{\lambda\mu\nu\sigma} \etb a\rho\sigma
&=
\delta_{\rho\lambda} \etb a\mu\nu + \delta_{\rho\nu} \etb a\lam\mu + \delta_{\rho\mu} \etb a\nu\kappa  \\
\et a\mu\nu \etb b\mu\nu &= 0
\,.
\end{split}
\end{equation}

\section{Caloron field strength} 									\label{P&Pmunu}
The SU(2) field strength in the 'nonperturbative' convention is
\begin{equation}
F^a_{\mu\nu} = \d_\mu A^a_\nu - \d_\nu A^a_\mu + \eps^{abc} A^b_\mu A^c_\nu
\,.
\end{equation}
Inserting the singular gauge instanton field
\begin{equation}
A_\mu^a = - \etb a\mu\kappa \d_\kappa \ln \Pi 
= - \etb a\mu\kappa \frac{ \d_\kappa \Pi }{ \Pi }
\,, 
\end{equation}
we get
\begin{align}
F_{\mu\nu}^a 
& = \notag
     - \etb a\nu\kappa \d_\mu \frac{\d_\kappa\Pi}{\Pi}
     + \etb a\mu\kappa \d_\nu \frac{\d_\kappa\Pi}{\Pi}
     + \eps^{abc} \etb b\mu\kappa \etb c\nu\lambda \frac{(\d_\kappa\Pi)(\d_\lam\Pi)}{\Pi^2}  
\\ &= \notag
     - \etb a\nu\kappa \d_\mu \frac{\d_\kappa\Pi}{\Pi}
     + \etb a\mu\kappa \d_\nu \frac{\d_\kappa\Pi}{\Pi}
     + \left( \etb a\mu\nu \delta_{\kappa\lambda} + \etb a\kappa\lambda \delta_{\mu\nu} 
         - \etb a\mu\lambda \delta_{\kappa\nu} - \etb a\kappa\nu \delta_{\mu\lambda} \right) \frac{ (\d_\kappa \Pi)(\d_\lambda \Pi) }{\Pi^2}
\\ &= 
     - \etb a\nu\kappa \frac{\Pi ( \d_\mu \d_\kappa \Pi) - (\d_\kappa \Pi)(\d_\mu \Pi)}{\Pi^2}  
     + \etb a\mu\kappa \frac{\Pi ( \d_\nu \d_\kappa \Pi) - (\d_\kappa \Pi)(\d_\nu \Pi)}{\Pi^2}  
\\ & \quad \notag
     + \left( \etb a\mu\nu \frac{ (\d_\kappa \Pi)(\d_\kappa \Pi) }{\Pi^2} 
	 - \etb a\mu\lambda \frac{ (\d_\nu \Pi)(\d_\lambda \Pi) }{\Pi^2}
	 - \etb a\kappa\nu \frac{ (\d_\kappa \Pi)(\d_\mu \Pi) }{\Pi^2} \right)
\\ &= \notag
  \etb a\mu\nu \frac{ (\d_\kappa \Pi)(\d_\kappa \Pi) }{\Pi^2} 
- \etb a\nu\kappa \frac{\Pi ( \d_\mu \d_\kappa \Pi) - 2 (\d_\kappa \Pi)(\d_\mu \Pi)}{\Pi^2}  
+ \etb a\mu\kappa \frac{\Pi ( \d_\nu \d_\kappa \Pi) - 2 (\d_\kappa \Pi)(\d_\nu \Pi)}{\Pi^2}  
\end{align}
For convenience, we define the quantities
\begin{equation}   											\label{P}
P = \frac{ (\d_\kappa \Pi)(\d_\kappa \Pi) }{\Pi^2}  
\qquad\text{and}\qquad
P_{\mu\nu} = \frac{ \Pi (\d_\mu \d_\nu \Pi ) - 2 (\d_\mu\Pi)(\d_\nu\Pi) }{\Pi^2}
\,,
\end{equation}
such that the instanton field strength reads
\begin{equation}
F_{\mu\nu}^a = \etb a\mu\nu P - \etb a\nu\kappa P_{\mu\kappa} + \etb a\mu\kappa P_{\nu\kappa}
\,.
\end{equation}
The symbols \eqref{P} have the properties
\
\begin{equation}
P_{\mu\nu} = P_{\nu\mu}  
\qquad\text{and}\qquad
P_{\mu\mu} = -2P 
\,. 
\end{equation}

\noindent
The pre-potential $\Pi$ for a single caloron is given in \eq{prepotential}.  
Using the fact that it depends on the radial coordinate $r=|\vec x|$ only (and on Euclidean time),
$\Pi(\tau,\vec x) = \Pi(\tau,r)$,
$P$ and $P_{\mu\nu}$ can be expressed in terms of the derivatives of $\Pi$ with respect to $\tau$ and $r$.
In particular, the Laplace equation
\begin{equation}
\Pi^{-1} \d_\mu\d_\mu \Pi = 0
\,,
\end{equation}
which results from the demand for self-duality, translates into
\begin{equation}
\frac{\d_4^2 \Pi}{\Pi} + \frac{\d_r^2 \Pi}{\Pi} + 2 \frac{\d_r \Pi}{r\,\Pi} = 0
\,.
\end{equation}

\noindent
At an arbitrary point $x=(\tau,\vec x)$, the components of $P_{\mu\nu}$ are explicitly given as
\begin{equation}
\begin{split}
P_{44} (x) 
& = 
\frac{\d_4^2 \Pi(x)}{\Pi(x)} - 2 \frac{[\d_4\Pi(x)]^2}{\Pi^2(x)}  \\
P_{4i} (x)
& =
\frac{x_i}{r} \left( \frac{\d_4\d_r\Pi(x)}{\Pi(x)} - 2 \frac{[\d_4\Pi(x)] [\d_r\Pi(x)]}{\Pi^2(x)} \right)  \\
P_{ij} (x)
& =
\delta_{ij} \frac{\d_r\Pi(x)}{r\,\Pi(x)}
+ \frac{x_i x_j}{r^2} \left(
\frac{\d_r^2\Pi(x)}{\Pi(x)} - \frac{\d_r\Pi(x)}{r\,\Pi(x)} - 2 \frac{[\d_r\Pi(x)]^2}{\Pi^2(x)} \right)
\,,
\end{split}
\end{equation}
and
\begin{equation}
P(x) = \frac{[\d_4\Pi(x)]^2}{\Pi^2(x)} + \frac{[\d_r\Pi(x)]^2}{\Pi^2(x)}
\,.
\end{equation}
At $x=(\tau,0)$, the tensor $P_{\mu\nu}(\tau,0)$ is diagonal and its components are
\begin{equation}
\begin{split}
P_{4i}(\tau,0) &= P_{i4}(\tau,0) = 0  \\
P_{44}(\tau,0) &= \frac{\d_4^2 \Pi(\tau,0)}{\Pi(\tau,0)} - 2 \frac{[\d_4\Pi(\tau,0)]^2}{\Pi^2(\tau,0)}  \\
P_{ij}(\tau,0) &= - \frac13 \delta_{ij} \frac{\d_4^2\Pi(\tau,0)}{\Pi(\tau,0)}
\,,
\end{split}
\end{equation}
and
\begin{equation}
P(\tau,0) = \frac{[\d_4\Pi(\tau,0)]^2}{\Pi^2(\tau,0)}
\,.
\end{equation}

\section{Details to section \ref{abschnitt calculation}} 						\label{details}
All spacetime dependent objects are to be evaluated at the same time $\tau$,
so that we may, for simplicity of notation, suppress the time dependence
and write $F_{\mu\nu}(\vec x)$ instead of $F_{\mu\nu}(\tau, \vec x)$ 
and $\{0,\vec x\}$ instead of $\{(\tau,0),(\tau,\vec x)\}$ etc.
Besides that, we abbreviate the time derivative with a dot and the radial derivative with a prime.

In the following some expressions containing $P_{\mu\nu}$ and $P$,
which we will need later, are calculated:
\begin{equation}											\label{NRanfang}
\begin{split}
\etb b\mu\kappa \etb b\nu\rho P_{\nu\kappa}(0) \, P_{\mu\rho}(\vec x) 
&= \big( \delta_{\mu\nu} \delta_{\kappa\rho} - \delta_{\mu\rho} \delta_{\nu\kappa} \big) \, P_{\nu\kappa}(0) \, P_{\mu\rho}(\vec x) 
\\ &=
P_{\mu\nu}(0) \, P_{\mu\nu}(\vec x) - P_{\nu\nu}(0) \, P_{\mu\mu}(\vec x)
\\ &=
P_{\mu\nu}(0) \, P_{\mu\nu}(\vec x) - 4 P(0) \, P(\vec x)
\end{split}
\end{equation}
\begin{equation}
\begin{split}
\etb a\mu\kappa \etb c\nu\rho P_{\nu\kappa}(0) P_{\mu\rho}(\vec x)
   & =
       \etb a\mu4 \etb c4\rho P_{44}(0) P_{\mu\rho}(\vec x) + \etb a\mu k \etb cn\rho P_{nk}(0) P_{\mu\rho} (\vec x)
\\ & =
       - P_{44}(0) P_{ac}(\vec x) - \frac13 \etb a\mu n \etb cn\rho \frac{\ddot\Pi(0)}{\Pi(0)} P_{\mu\rho}(\vec x)
\\ & =
       - P_{44}(0) P_{ac}(\vec x) 
       - \frac13   \big( \eps_{a\mu n} + \delta_{\mu4} \delta_{an} \big)   \big( \eps_{cn\rho} - \delta_{cn} \delta_{\rho4} \big)
         \frac{\ddot\Pi(0)}{\Pi(0)} P_{\mu\rho}(\vec x)
\\ & =
       - P_{44}(0) P_{ac}(\vec x) 
       - \frac13 \frac{\ddot\Pi(0)}{\Pi(0)} 
         \big( P_{ac}(\vec x) - \delta_{ac} P_{ii}(\vec x) - \delta_{ac} P_{44}(\vec x) \big)
\\ & =
       - P_{44}(0) P_{ac}(\vec x) 
       - \frac13 \frac{\ddot\Pi(0)}{\Pi(0)}   \big( P_{ac}(\vec x) - \delta_{ac} P_{\mu\mu}(\vec x) \big)
\\ & =
       - P_{44}(0) P_{ac}(\vec x) 
       - \frac13 \big( P_{44}(0) + 2 P(0) \big)   \big( P_{ac}(\vec x) + 2 \delta_{ac} P(\vec x) \big)
\\ & =
       - \frac23   \big( 2 P_{44}(0) + P(0) \big)   P_{ac}(\vec x)
       - \frac23   \big( P_{44}(0) + 2 P(0) \big)   \delta_{ac} P(\vec x)
\end{split}
\end{equation}
\begin{equation}
P_{ii}(\vec x) = - \frac{\ddot\Pi(r)}{\Pi(r)} - 2 \frac{\Pi'^2(r)}{\Pi^2(r)}
\end{equation}
\begin{equation}
\begin{split}
P_{\mu\nu}(0) \, P_{\mu\nu}(\vec x) 
&=
P_{44}(0) \, P_{44}(\vec x) + P_{mn}(0) \, P_{mn}(\vec x)  \\
&= 
P_{44}(0) \, P_{44}(\vec x) - \frac13 \frac{\ddot\Pi(0)}{\Pi(0)} \, P_{ii}(\vec x)  \\
&= 
\left( \frac{\ddot\Pi(0)}{\Pi(0)} - 2 \frac{\dot\Pi^2(0)}{\Pi^2(0)} \right)
\left( \frac{\ddot\Pi(r)}{\Pi(r)} - 2 \frac{\dot\Pi^2(r)}{\Pi^2(r)} \right)
+ \frac13 \frac{\ddot\Pi(0)}{\Pi(0)} 
\left( \frac{\ddot\Pi(r)}{\Pi(r)} + 2 \frac{\Pi'^2(r)}{\Pi^2(r)} \right)
\end{split}
\end{equation}
\begin{equation}
P_{\mu\nu}(0) P_{\mu\nu}(\vec x) - P(0) P(\vec x)
=
\left( \frac{\dot\Pi^2(0)}{\Pi^2(0)} - \frac23 \frac{\ddot\Pi(0)}{\Pi(0)} \right)
\left( 3 \frac{\dot\Pi^2(r)}{\Pi^2(r)}
- 2 \frac{\ddot\Pi(r)}{\Pi(r)} - \frac{\Pi'^2(r)}{\Pi^2(r)} \right)
\end{equation}
\begin{equation}
\begin{split}
\etb a\kappa\rho \, P_{\mu\kappa}(0) \, P_{\mu\rho}(\vec x) 
&=
\etb a4\rho \, P_{44}(0) \, P_{4\rho}(\vec x) + \etb ak\rho \, P_{mk}(0) \, P_{m\rho}(\vec x)  
\\ & =
P_{44}(0) \, P_{4a}(\vec x) - \frac 13 \frac{\ddot\Pi(0)}{\Pi(0)} \, \etb am\rho \, P_{m\rho}(\vec x)
\\ & =
P_{44}(0) \, P_{4a}(\vec x) + \frac 13 \frac{\ddot\Pi(0)}{\Pi(0)} \, P_{4a}(\vec x)
\\ & =
\left( P_{44}(0) + \frac 13 \frac{\ddot\Pi(0)}{\Pi(0)} \right) P_{4a}(\vec x)
\\ & = 
2 \frac{x^a}r \left( \frac{\dot\Pi'(r)}{\Pi(r)} - 2 \frac{\dot\Pi(r) \, \Pi'(r)}{\Pi^2(r)} \right)
\left( \frac23 \frac{\ddot\Pi(0)}{\Pi(0)} - 2 \frac{\dot\Pi^2(0)}{\Pi^2(0)} \right)
\end{split}
\end{equation}
\begin{equation}											\label{NRende}
\begin{split}
\frac{x^a x^f}{r^2} \, \etb f\kappa\rho \, P_{\mu\kappa}(0) \, P_{\mu\rho}(\vec x) 
&=
2 \frac{x^a}r \left( \frac{\dot\Pi'(r)}{\Pi(r)} - 2 \frac{\dot\Pi(r) \, \Pi'(r)}{\Pi^2(r)} \right)
\left( \frac23 \frac{\ddot\Pi(0)}{\Pi(0)} - 2 \frac{\dot\Pi^2(0)}{\Pi^2(0)} \right)
\end{split}
\end{equation}

\absatz{Integrand of \eqref{twopoint}}
Inserting the result for the Wilson lines from \eq{wline} and writing the field strength in components, 
we have \eq{4terme 1}, which in simplified notation reads
\begin{align} 												\label{4terme}
& 
\notag
	\tr \lambda^a \, F_{\mu\nu} (0) \, \left\{ 0,\vec x \right\} \,
	F_{\mu\nu} (\vec x) \, \left\{ \vec x,0 \right\}  
\\ & = \notag
	\frac12 \tr 
	\Bigg[
	\lam^a \lam^b \left( \cos g(r) + i \lam^c \frac{x^c}{r} \sin g(r) \right)
	\lam^d \left( \cos g(r) - i \lam^e \frac{x^e}{r} \sin g(r) \right)  
	\Bigg]
	F_{\mu\nu}^b (0) F_{\mu\nu}^d (\vec x) 
\\ & = \notag
	\frac 12  \cos^2 \! g(r)
	\cdot \tr [\lam^a \lam^b \lam^d] 
	\cdot F_{\mu\nu}^b (0) \, F_{\mu\nu}^d (\vec x) 
\\ & \quad
	- \frac i2  \sin g(r) \, \cos g(r) 
	\cdot \tr [\lam^a \lam^b \lam^d \lam^e] 
	\cdot \frac{x^e}{r}
	\cdot F_{\mu\nu}^b (0) \, F_{\mu\nu}^d (\vec x) 
\\ & \quad \notag
	+ \frac i2 \sin g(r) \, \cos g(r) 
	\cdot \tr [\lam^a \lam^b \lam^c \lam^d] 
	\cdot \frac{x^c}{r} 
	\cdot F_{\mu\nu}^b (0) \, F_{\mu\nu}^d (\vec x) 
\\ & \quad \notag
	+ \frac 12 \sin^2 \! g(r)
	\cdot \tr [\lam^a \lam^b \lam^c \lam^d \lam^e]
	\cdot \frac{x^c x^e}{r^2} 
	\cdot F_{\mu\nu}^b (0) \, F_{\mu\nu}^d (\vec x) 
\,.
\end{align}
We will calculate each of the four summands separately.
Therefore we write the Lorentz-trace of the field components $F_{\mu\nu}^b (0) F_{\mu\nu}^d (\vec x)$  
in terms of the symbols $P$ and $P_{\mu\nu}$ introduced in \eqs{P},
\begin{align}
& F_{\mu\nu}^b (0) F_{\mu\nu}^d (\vec x) \notag
\\ & =
	\big[ \etb b\mu\nu P(0) + \etb b\mu\kappa P_{\nu\kappa}(0) - \etb b\nu\kappa P_{\mu\kappa}(0) \big]
	\big[ \etb d\mu\nu P(\vec x) + \etb d\mu\rho P_{\nu\rho}(\vec x) - \etb d\nu\rho P_{\mu\rho}(\vec x) \big]
\\ & = \notag
	2 \bigg\{
	\delta_{bd} \big[ P_{\mu\nu}(0) \, P_{\mu\nu}(\vec x) - 2 P(0) \, P(\vec x) \big]
	+ \eps_{bdf} \etb f\kappa\rho P_{\mu\kappa}(0) \, P_{\mu\rho}(\vec x)
	- \etb b\mu\kappa \etb d\nu\rho P_{\nu\kappa}(0) \, P_{\mu\rho}(\vec x)
	\bigg\}  
\,.
\end{align}
Furthermore, we need the traces of products of Pauli matrices \eqs{paulispur},
some of the relations for 't~Hooft symbols from Sec.~\ref{not&conv},
and the expressions prepared in \eqs{NRanfang} through \eqref{NRende}.
The first of the four terms in \eq{4terme} is
\begin{align} 												\label{1.term}
& 
	\frac12 \cos^2 \! g(r) \cdot \tr [\lam^a \lam^b \lam^d] 
	\cdot F_{\mu\nu}^b (0) \, F_{\mu\nu}^d (\vec x) 
\notag \\ & =
	i \cos^2 \! g(r) \cdot \eps^{abd} \, 
	F_{\mu\nu}^b (0) \, F_{\mu\nu}^d (\vec x) 
\notag \\ & =
	2 i \cos^2 \! g(r) \cdot 
	\etb a\kappa\rho \, P_{\mu\kappa}(0) \, P_{\mu\rho}(\vec x)
\notag \\ & =
	- 4 i \cos^2 \! g(r) \, \frac{x^a}{r} 
	\left( \frac{\dot\Pi^2(0)}{\Pi^2(0)} - \frac23 \frac{\ddot\Pi(0)}{\Pi(0)} \right)
	\left( \frac{\dot\Pi'(r)}{\Pi(r)} - 2 \frac{\dot\Pi(r) \, \Pi'(r)}{\Pi^2(r)} \right)
\,.
\end{align}
In the third line, we used that
\begin{equation*}
\eps^{abd} \etb b\mu\kappa \etb d\nu\rho P_{\nu\kappa}(0) P_{\mu\rho}(\vec x)
=
\eps^{adb} \etb d\rho\nu \etb b\kappa\mu P_{\kappa\nu}(0) P_{\rho\mu}(\vec x)
=
- \eps^{abd} \etb b\mu\kappa \etb d\nu\rho P_{\nu\kappa}(0) P_{\mu\rho}(\vec x)
\end{equation*}
vanishes.
The second summand is
\begin{align} 												\label{2.term}
& 
	- \frac i2 \cos g(r) \, \sin g(r) 
	\cdot \tr [\lam^a \lam^b \lam^d \lam^e]
	\cdot \frac{x^e}{r} 
	\cdot F_{\mu\nu}^b (0) \, F_{\mu\nu}^d (\vec x) 
\notag \\ & =
	- i \cos g(r) \, \sin g(r) 
	\cdot \big( \delta^{ab} \delta^{de} - \delta^{ac} \delta^{be} + \delta^{bd} \delta^{ae} \big)
	\cdot \frac{x^e}{r} 
	\cdot F_{\mu\nu}^b (0) \, F_{\mu\nu}^d (\vec x) 
\notag \\ & =
	- i \cos g(r) \, \sin g(r) 
	\cdot \frac{x^e}{r} \cdot
	\Big\{
		3 \delta_{ae} \big[ P_{\mu\nu}(0) \, P_{\mu\nu}(\vec x) - 2 P(0) \, P(\vec x) \big]
		+ 2 \eps_{aef} \etb f\kappa\rho \, P_{\mu\kappa}(0) \, P_{\mu\rho}(\vec x)
\notag \\ & \quad
		- \big( \etb a\mu\kappa \etb e\nu\rho - \etb e\mu\kappa \etb a\nu\rho \big) P_{\nu\kappa}(0) \, P_{\mu\rho}(\vec x)
		- \delta_{ae} \etb b\mu\kappa \etb b\nu\rho \, P_{\nu\kappa}(0) \, P_{\mu\rho}(\vec x)
	\Big\}
\notag \\ & =
	- i \cos g(r) \, \sin g(r) 
	\cdot \frac{x^a}{r} \cdot
	\Big\{
		3 \big[ P_{\mu\nu}(0) \, P_{\mu\nu}(\vec x) - 2 P(0) \, P(\vec x) \big]
		- \big[ P_{\mu\nu}(0) \, P_{\mu\nu}(\vec x) - 4 P(0) \, P(\vec x) \big]
	\Big\}
\notag \\ & =
	- 2 i \cos g(r) \, \sin g(r)
	\cdot \frac{x^a}{r} \cdot
	\Big\{ P_{\mu\nu}(0) P_{\mu\nu}(\vec x) - P(0) P(\vec x) \Big\}
\notag \\ & =
	- 2 i \cos g(r) \, \sin g(r) 
	\cdot \frac{x^a}{r} \cdot
	\left( \frac{\dot\Pi^2(0)}{\Pi^2(0)} - \frac23 \frac{\ddot\Pi(0)}{\Pi(0)} \right)
	\left( 3 \frac{\dot\Pi^2(r)}{\Pi^2(r)} - 2 \frac{\ddot\Pi(r)}{\Pi(r)} - \frac{\Pi'^2(r)}{\Pi^2(r)} \right)
\,.
\end{align}
The third summand is
\begin{align} 												\label{3.term}
& 
	\frac i2 \cos g(r) \, \sin g(r) 
	\cdot \tr [\lam^a \lam^b \lam^c \lam^d]
	\cdot \frac{x^c}{r} \cdot
	F_{\mu\nu}^b (0) \, F_{\mu\nu}^d (\vec x) 
\notag \\ & =
	i \cos g(r) \, \sin g(r) 
	\cdot \big( \delta^{ab} \delta^{cd} - \delta^{ac} \delta^{bd} + \delta^{bc} \delta^{ad} \big)
	\cdot \frac{x^c}{r} \cdot
	F_{\mu\nu}^b (0) \, F_{\mu\nu}^d (\vec x) 
\notag \\ & =
	i \cos g(r) \, \sin g(r) 
	\cdot \frac{x^c}{r} \cdot
	\Big\{
	- \delta_{ac} \big[ P_{\mu\nu}(0) \, P_{\mu\nu}(\vec x) - 2 P(0) \, P(\vec x) \big]
\notag \\ & \quad
	- \big( \etb a\mu\kappa \etb c\nu\rho + \etb c\mu\kappa \etb a\nu\rho \big) \, P_{\nu\kappa}(0) \, P_{\mu\rho}(\vec x)
	+ \delta_{ac} \etb b\mu\kappa \etb b\nu\rho \, P_{\nu\kappa}(0) \, P_{\mu\rho}(\vec x)
	\Big\}
\notag \\ & =
	- 2 i \cos g(r) \, \sin g(r) \cdot
	\left\{ \frac{x^a}{r} P(0) \, P(\vec x) + \frac{x^c}{r} \etb a\mu\kappa \etb c\nu\rho \, P_{\nu\kappa}(0) \, P_{\mu\rho}(\vec x) \right\}
\notag \\ & =
	2 i \cos g(r) \, \sin g(r) 
	\cdot \frac{x^a}{r} \cdot
	\left( \frac{\dot\Pi^2(0)}{\Pi^2(0)} - \frac23 \frac{\ddot\Pi(0)}{\Pi(0)} \right)
	\left( \frac{\dot\Pi^2(r)}{\Pi^2(r)} - 3 \frac{\Pi'^2(r)}{\Pi^2(r)} + 2 \frac{\Pi''(r)}{\Pi(r)} \right)
\,.
\end{align}
The fourth and last one is
\begin{align} 												\label{4.term}
& 
	\frac12 \sin^2 \! g(r)
	\cdot \tr [\lam^a \lam^b \lam^c \lam^d \lam^e]
	\cdot \frac{x^c x^e}{r^2} \cdot
	F_{\mu\nu}^b (0) \, F_{\mu\nu}^d (\vec x) 
\notag \\ & =
	i \sin^2 \! g(r)
	\cdot \big( \delta_{ab} \eps_{cde} + \delta_{cd} \eps_{abe} - \delta_{ce} \eps_{abd} + \delta_{de} \eps_{abc} \big)
	\cdot \frac{x^c x^e}{r^2} 
	\cdot F_{\mu\nu}^b (0) \, F_{\mu\nu}^d (\vec x) 
\notag \\ & =
	i \sin^2 \! g(r)
	\cdot \left( 2 \eps_{abc} \frac{x^c x^d}{r^2} - \eps_{abd} \right)
	\cdot F_{\mu\nu}^b (0) \, F_{\mu\nu}^d (\vec x) 
\notag \\ & =
	i \sin^2 \! g(r)
	\cdot
	\bigg\{
		\left( 2 \eps_{abc} \eps_{bdf} \frac{x^c x^d}{r^2} - 2 \delta_{af} \right) 
		\etb f\kappa\rho \, P_{\mu\kappa}(0) \, P_{\mu\rho}(\vec x)
		- 2 \eps_{abc} \frac{x^c x^d}{r^2} \etb b\mu\kappa \etb d\nu\rho \, P_{\nu\kappa}(0) \, P_{\mu\rho}(\vec x)
	\bigg\}
\notag \\ & =
	- 2 i \sin^2 \! g(r) 
	\cdot \etb a\kappa\rho \, P_{\mu\kappa}(0) \, P_{\mu\rho}(\vec x)  
\notag \\ & =
	- 4 i \sin^2 \! g(r) 
	\cdot \frac{x^a}{r} \cdot
	\left( \frac{\dot\Pi^2(0)}{\Pi^2(0)} - \frac23 \frac{\ddot\Pi(0)}{\Pi(0)} \right)
	\left( \frac{\dot\Pi'(r)}{\Pi(r)} - 2 \frac{\dot\Pi(r) \, \Pi'(r)}{\Pi^2(r)} \right)
\,.
\end{align}
Summing \eqs{1.term} through \eqref{4.term} 
and using
\begin{equation}
\cos^2 \al - \sin^2 \al = \cos 2\al
\qquad\text{and}\qquad
\cos \al \, \sin \al = \frac12 \sin 2 \al
\end{equation}
yields (with the $\tau$'s reinserted)
\begin{align}                              
& 
	\tr \lambda^a \, F_{\mu\nu} (\tau,0) \, \left\{ (\tau,0),(\tau,\vec x) \right\} \,
	F_{\mu\nu} (\tau,\vec x) \, \left\{ (\tau,\vec x),(\tau,0) \right\}  
\notag \\ & =
	2 i \, \frac{x^a}{r}
	\left( \frac{\dot\Pi^2(\tau,0)}{\Pi^2(\tau,0)} - \frac23 \frac{\ddot\Pi(\tau,0)}{\Pi(\tau,0)} \right)
	\cdot
	\Bigg\{
		2 \cos (2 g(\tau,r)) 
		\left( 2 \frac{\dot\Pi(\tau,r) \, \Pi'(\tau,r)}{\Pi^2(\tau,r)} - \frac{\dot\Pi'(\tau,r)}{\Pi(\tau,r)} \right)
\notag \\ & \quad 
		+ \sin (2 g(\tau,r))
		\left( 2 \frac{\Pi'^2(\tau,r)}{\Pi^2(\tau,r)} - 2 \frac{\dot\Pi^2(\tau,r)}{\Pi^2(\tau,r)} 
		+ \frac{\ddot\Pi(\tau,r)}{\Pi(\tau,r)} - \frac{\Pi''(\tau,r)}{\Pi(\tau,r)} \right)
	\Bigg\}
\,.
\end{align}

\chapter{Appendix to Chapter 4}

\section{Details to section \ref{Calculation}}									\label{indices}
In this appendix, the contraction of Lorentz and color indices appearing in 
the expressions for the Feynman diagrams Figs.~\ref{hh}, \ref{diagram mh}, \ref{fd mhh} is performed.
For diagrams containing TLM fluctuations and hence $P^T_{\mu\nu}$ as defined in \eq{PT}, the following formulae will be useful:
\begin{equation}											\label{rules}
\begin{split}
P^T_{\mu\nu}(q) \, q^\nu &= 0  \\
P^T_{\mu\nu}(q) \, g^{\mu\nu} &= -2  \\
P^T_{\mu\nu}(q) \, p^\mu k^\nu &= \vec p \vec k - \frac{(\vec q \vec p)(\vec q \vec k)}{\vec q^2}  \\
P^T_{\mu\nu}(q) \, p^\mu p^\nu &= \vec p^2 - \frac{(\vec q \vec p)^2}{\vec q^2}  
\,.
\end{split}
\end{equation}

\absatz{Concerning $\Delta P^{\rm HH}$}
The color indices $a$, $b$, $c$ and $d$ take the values 1,2 only 
because we chose the particles propagating in the loops to be TLH-modes.
Index $f$ is summed over 1,2,3.
\begin{align} 
& 
\left[
\eps_{adf} \eps_{fbc} \left( g^{\mu\nu} g^{\rho\sigma} - g^{\mu\rho} g^{\nu\sigma} \right)
+ \eps_{abf} \eps_{fdc} \left( g^{\mu\sigma} g^{\nu\rho} - g^{\mu\rho} g^{\nu\sigma} \right) 
+ \eps_{acf} \eps_{fdb} \left( g^{\mu\sigma} g^{\nu\rho} - g^{\mu\nu} g^{\sigma\rho} \right)
\right]
\notag \\ & 
\cdot
\left( - \delta_{ab} \right) \left( g_{\mu\nu} - \frac{k_{\mu}k_{\nu}}{m^2} \right) 
\cdot
\left( - \delta_{cd} \right) \left( g_{\rho\sigma} - \frac{p_{\rho}p_{\sigma}}{m^2} \right) 
\notag \\
& = 
\eps_{acf} \eps_{acf} 
\left( g^{\mu\nu} g^{\rho\sigma} - g^{\mu\rho} g^{\nu\sigma} 
- g^{\mu\sigma} g^{\nu\rho} + g^{\mu\nu} g^{\rho\sigma} \right) 
\cdot
\left( g_{\mu\nu} - \frac{k_{\mu}k_{\nu}}{m^2} \right) 
\cdot
\left( g_{\rho\sigma} - \frac{p_{\rho}p_{\sigma}}{m^2} \right) 
\\ & = 
4 \left( g^{\mu\nu} g^{\rho\sigma} - g^{\mu\rho} g^{\nu\sigma} \right) 
\cdot
\left( g_{\mu\nu} - \frac{k_{\mu}k_{\nu}}{m^2} \right) 
\cdot
\left( g_{\rho\sigma} - \frac{p_{\rho}p_{\sigma}}{m^2} \right) 
\notag \\
& = 
4 \left( 12 - 3 \frac{p^2}{m^2} - 3 \frac{k^2}{m^2} +  \frac{p^2 k^2}{m^4} - \frac{(p k)^2}{m^4} \right)
\notag
\end{align}

\absatz{Concerning $\Delta P^{\rm MH}$}
Since $k$ and $p$ are associated with a TLM-mode and a TLH-mode respectively, 
we have to set $a=b=3$ and sum over $c,d=1,2$. \eqs{rules} have to be used.
\begin{align} 												\label{mh tensor}
& 
\left[
  \eps_{adf} \eps_{fbc} \left( g^{\mu\nu} g^{\rho\sigma} - g^{\mu\rho} g^{\nu\sigma} \right)
+ \eps_{abf} \eps_{fdc} \left( g^{\mu\sigma} g^{\nu\rho} - g^{\mu\rho} g^{\nu\sigma} \right) 
+ \eps_{acf} \eps_{fdb} \left( g^{\mu\sigma} g^{\nu\rho} - g^{\mu\nu} g^{\sigma\rho} \right)
\right]
\notag \\ & 
\cdot
\left( - \delta_{cd} \right) \left( g_{\rho\sigma} - \frac{p_{\rho}p_{\sigma}}{m^2} \right) 
\cdot
\left( - \delta_{ab} \right) P^T_{\mu\nu} (k)
\notag \\
&=
\eps_{acf} \eps_{acf} 
\left( g^{\mu\nu} g^{\rho\sigma} - g^{\mu\rho} g^{\nu\sigma} 
- g^{\mu\sigma} g^{\nu\rho} + g^{\mu\nu} g^{\sigma\rho} \right)
\cdot
\left( g_{\rho\sigma} - \frac{p_{\rho}p_{\sigma}}{m^2} \right) 
\cdot
P^T_{\mu\nu} (k)
\notag \\ &=
4 \left( g^{\mu\nu} g^{\rho\sigma} - g^{\mu\rho} g^{\nu\sigma} \right)
\cdot
\left( g_{\rho\sigma} - \frac{p_{\rho}p_{\sigma}}{m^2} \right) 
\cdot
P^T_{\mu\nu} (k)
\\ & =
4 \left( \left( 3 - \frac{p^2}{m^2} \right) g^{\mu\nu} + \frac{p^{\mu}p^{\nu}}{m^2} \right)
\cdot P_{\mu\nu}^T (k) 
\notag \\ &=
4 \cdot \left( -6 + 2 \frac{p^2}{m^2} + \frac{\vec p^2}{m^2} - \frac{(\vec p \vec k)^2}{m^2 \vec k^2} \right)
\notag
\end{align}

\absatz{Concerning $\Delta P^{\rm MHH}$}
The momentum $q$ is associated with a TLM fluctuation, and the TLH fluctuations carry momenta $p$ and $k$.
Thus we have $f,g=3$ and $a,b,c,d=1,2$, and hence
\begin{equation}
\eps_{acf} \eps_{bdg} \delta_{ab} \delta_{cd} \delta_{fg}
= \sum_{a=1}^2 \sum_{c=1}^2 \eps_{ac3} \eps_{ac3} = 2  
\,.
\end{equation}
The Lorentz structure resulting from the propagators and polarization tensors is 
\begin{align}
&
\left[ g^{\mu\rho} (k-p)^{\lambda} + g^{\rho\lambda} (q-k)^{\mu} + g^{\lambda\mu} (p-q)^{\rho} \right]
\left[ g^{\nu\sigma} (p-k)^{\kappa} + g^{\sigma\kappa} (k-q)^{\nu} + g^{\kappa\nu} (q-p)^{\sigma} \right]
\notag \\ & \cdot
\left( g_{\mu\nu} - \frac{ p_{\mu} p_{\nu} }{ m^2 } \right)
\left( g_{\rho\sigma} - \frac{ k_{\rho} k_{\sigma} }{ m^2 } \right)
P^T_{\lambda\kappa} (q)
\,.
\end{align}
Contracting the Lorentz indices and inserting $P^T_{\lambda\kappa} (q)$ by using the rules \eqref{rules} yields 
\begin{align}
&
	- 2 \left( \vec k^2 - \frac{ (\vec k \vec q)^2 }{ \vec q^2 } \right)
	- 2 \left( \vec p^2 - \frac{ (\vec p \vec q)^2 }{ \vec q^2 } \right)
	+ 6 \left( \vec k \vec p - \frac{ (\vec k \vec q) (\vec p \vec q) }{ \vec q^2 } \right)
\\ \notag &
	+ 2 k^2 - 4kq - 4pq + 2 p^2 + 4 q^2
\\ \notag &
	+ \frac{1}{m^2} 
	\bigg[
		\left( \vec p^2 - \frac{ (\vec p \vec q)^2 }{ \vec q^2 } \right)   \left( k^2 + q^2 \right)
		+ \left( \vec k^2 - \frac{ (\vec k \vec q)^2 }{ \vec q^2 } \right)   \left( p^2 + q^2 \right)
		- 4 \left( \vec k \vec p - \frac{ (\vec k \vec q)(\vec p \vec q) }{ \vec q^2 } \right) kp
\\ \notag & 
		- 4 (kp)^2 + 4 (kp)(kq) - 2 (kq)^2 + 4 (kp)(pq) - 2 (pq)^2
	\bigg]
\\ \notag &
	+ \frac{1}{m^4}
	\left[
		- \left( \vec p^2 - \frac{ (\vec p \vec q)^2 }{ \vec q^2 } \right) (kq)^2
		- \left( \vec k^2 - \frac{ (\vec k \vec q)^2 }{ \vec q^2 } \right) (pq)^2
		+ 2 \left( \vec k \vec p - \frac{ (\vec k \vec q)(\vec p \vec q) }{ \vec q^2 } \right) (kq)(pq)
	\right]
\,.
\end{align}
Now we may use energy momentum conservation and set $q=-p-k$.
Moreover, we notice that 
\begin{equation}
\vec k^2 - \frac{ (\vec k^2 + \vec k \vec p)^2 }{ (\vec k + \vec p)^2 }
= \vec p^2 - \frac{ (\vec p^2 + \vec k \vec p)^2 }{ (\vec k + \vec p)^2 }
= \frac{ (\vec k^2 + \vec k \vec p) (\vec p^2 + \vec k \vec p) }{ (\vec k + \vec p)^2 } - \vec k \vec p
= \frac{ \vec k^2 \vec p^2 - (\vec k \vec p)^2 }{ (\vec k + \vec p)^2 }
\,.
\end{equation}
This leads to the following:
\begin{align}
& 
	10 \frac{ (\vec k \vec p)^2 - \vec k^2 \vec p^2 }{ (\vec k + \vec p)^2 } + 10 k^2 + 16 kp + 10 p^2
\\ \notag &
	+ \frac{1}{m^2} 
	\left(
		- 2 k^4 - 8 k^2 kp - 16 (kp)^2 - 8 kp p^2 - 2 p^4 
		- \frac{ (\vec k \vec p)^2 - \vec k^2 \vec p^2 }{ (\vec k + \vec p)^2 } (3 k^2 + 8 kp + 3 p^2)
	\right)
\\ \notag &
	+ \frac{1}{m^4}
	\frac{ (\vec k \vec p)^2 - \vec k^2 \vec p^2 }{ (\vec k + \vec p)^2 }
	\left( k^4 + 4 k^2 kp + 4 (kp)^2 + 2 k^2 p^2 + 4 kp p^2 + p^4 \right)
\,.
\end{align}
We are interested in only one contribution from this diagram, namely $\Delta P^{\rm MHH}_{\rm VTT}$.
Fortunately, its calculation includes the two on mass shell conditions $p^2=m^2$ and $k^2=m^2$.
They reduce the former expression to
\begin{equation}
16 \left( m^2 - \frac{ (kp)^2 }{m^2} \right) 
- \frac{ \vec k^2 \vec p^2 - (\vec k \vec p)^2 }{(\vec p + \vec k)^2} 
\left( 8 + 4 \frac{(kp)^2}{m^4} \right)
\,.
\end{equation}

\section{Electric screening mass for TLM modes}
There are two nonvanishing contributions to the polarization tensor $\Pi_{\mu\nu}(p)$ for TLM modes on one-loop level.
They are depicted in Figs.~\ref{tadpole} and \ref{nonlocal}.
We are especially interested in the component $\Pi_{00}$ for vanishing external momentum, 
more precisely $\Pi_{00} (p_0 = 0 , |\vec p| \to 0)$.

\subsection{Tadpole contribution}
\begin{figure}[t]
\centering
\includegraphics[width=70mm]{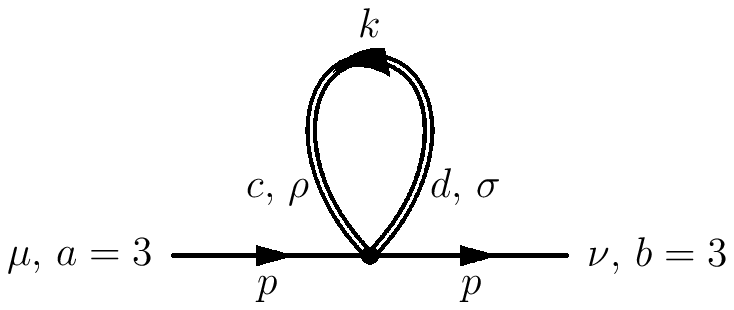}
\caption{  												\label{tadpole}
The tadpole contribution to the TLM mode polarization tensor.}
\end{figure}
According to the Feynman rules (see Sec.~\ref{prelims}), the tadpole diagram \fig{tadpole} corresponds to the expression
\begin{align}
\Pi^{\mu\nu}_{\text{tadp.}} (p)
& = 
	\frac1i \cdot \int \frac{d^4k}{(2\pi)^4} \,
	(-\delta_{cd}) \left( g_{\rho\sigma} - \frac{ k_{\rho} k_{\sigma} }{ m^2 } \right)
	\left[ \frac{i}{k^2-m^2} + 2 \pi \delta(k^2-m^2) \, n_B \left( \abs{ k_0 } / T \right) \right]
\notag \\ & \quad
	\cdot (-ie^2) \cdot
	\left[ \eps_{eab} \eps_{ecd} \left( g^{\mu\rho} g^{\nu\sigma} - g^{\mu\sigma} g^{\nu\rho} \right)
	+ \eps_{eac} \eps_{edb} \left( g^{\mu\sigma} g^{\rho\nu} - g^{\mu\nu} g^{\rho\sigma} \right)
\right. \\ & \left. \quad
	+ \eps_{ead} \eps_{ebc} \left( g^{\mu\nu} g^{\sigma\rho} - g^{\mu\rho} g^{\sigma\nu} \right) \right]
\notag \\ & =
	4e^2 \int \frac{d^4k}{(2\pi)^4}
	\left[ g^{\mu\nu} \left( 3 - \frac{k^2}{m^2} \right) + \frac{k^{\mu} k^{\nu}}{m^2} \right]
	\left[ \frac{i}{k^2-m^2} + 2 \pi \delta(k^2-m^2) \, n_B \left( \abs{k_0}/T \right) \right]
\notag 
\,.
\end{align}
When integrating the loop momentum $k$, the following constraints have to be taken into account:
\par
1. For both thermal and vacuum contribution the constraint on the center-of-mass energy in vertices, 
\begin{equation}  											 \label{con1}
\abs{ (p+k)^2 } \le \abs{ \phi }^2
\,.
\end{equation}
\par
2. For the vacuum contribution additionally the constraint on the offshellness of quantum fluctuations,
\begin{equation} 											\label{con2}
\abs{ k^2 - m^2 } \le \abs{ \phi }^2
\,.
\end{equation}

\noindent
As for the calculations of two-loop diagrams in Sec.~4, massive vacuum fluctuations for $e>\frac12$ are forbidden by \eq{con2},
i.\,e. the vacuum contribution vanishes for $e>\frac12$.
Moreover, for $e>\frac12$ and external momentum $p=0$, the thermal part also vanishes because in this case
\begin{equation}
\abs{(p+k)^2} = \abs{k^2} = m^2 = (2e)^2 \abs{\phi}^2 > \abs{\phi}^2
\,,
\end{equation}
and \eq{con1} cannot be satisfied.
Thus we have no contribution from the tadpole for $e > 1/2$.

\subsection{Nonlocal contribution}

\begin{figure}[ht]
\centering
\includegraphics[width=81mm]{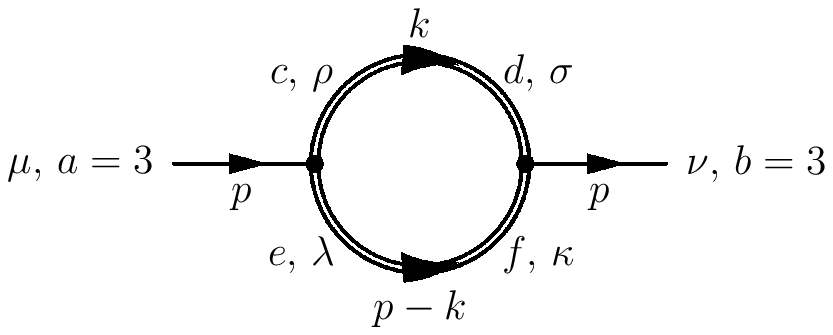}
\caption{ 												\label{nonlocal}
The nonlocal contribution to the TLM mode polarization tensor.}
\end{figure}

\setlength{\mathindent}{0pt}

The nonlocal contribution to the polarization tensor is depicted in \fig{nonlocal}. 
In formula, it reads
\begin{align}
\Pi^{\mu\nu}_{\text{nonl.}} (p) 
& = 
	\frac1{2i} \int \frac{d^4k}{(2\pi)^4} \;\;
	e^2 \cdot 
	\eps_{ace} \Big[ g^{\mu\rho} (-p-k)^{\lambda} + g^{\rho\lambda} (k-p+k)^{\mu} + g^{\lambda\mu} (p-k+p)^{\rho} \Big]
\notag \\ & \quad \cdot
	\eps_{dbf} \Big[ g^{\sigma\nu} (-k-p)^{\kappa} + g^{\nu\kappa} (p+p-k)^{\sigma} + g^{\kappa\sigma} (-p+k+k)^{\nu} \Big]
\notag \\ & \quad \cdot
	(-\delta_{cd}) \left( g_{\rho\sigma} - \frac{ k_{\rho} k_{\sigma} }{m^2} \right)
	\left[ \frac{i}{k^2-m^2} + 2\pi \, \delta(k^2-m^2) \, n_B\left(\abs{k_0}/T\right) \right]
\notag \\ & \quad \cdot
	(-\delta_{ef}) \left( g_{\lambda\kappa} - \frac{ (p-k)_{\lambda} (p-k)_{\kappa} }{m^2} \right)
\notag \\ & \quad 
	\cdot \left[ \frac{i}{(p-k)^2-m^2} + 2\pi \, \delta((p-k)^2-m^2) \, n_B\left(\abs{p_0-k_0}/T\right) \right]
\notag \\ & =
	i e^2 \int \frac{d^4k}{(2\pi)^4} \;\;
	\Bigg\{
		\left( 2k^2 - 2\frac{k^4}{m^2} - 2kp + 4 \frac{k^2 kp}{m^2} - 4\frac{(kp)^2}{m^2} 
		+ 5p^2 + 2 \frac{k^2p^2}{m^2} - \frac{p^4}{m^2} \right) g^{\mu\nu}
\notag \\ & \quad
		+ \left( 10 + 2\frac{k^2}{m^2} - 2\frac{kp}{m^2} - 3\frac{p^2}{m^2} + \frac{p^4}{m^4} \right) k^\mu k^\nu
		+ \left( - 2 - 3\frac{k^2}{m^2} + \frac{(kp)^2}{m^4} + \frac{p^2}{m^2} \right) p^\mu p^\nu
\notag \\ & \quad
		+ \left( - 5 - \frac{k^2}{m^2} + 4 \frac{kp}{m^2} - \frac{p^2 kp}{m^4} \right) \big( k^\nu p^\mu + k^\mu p^\nu \big)
\Bigg\}
\notag \\ & \quad \cdot
\left[ \frac{i}{k^2-m^2} + 2\pi \, \delta(k^2-m^2) \, n_b\left(\abs{k_0}/T\right) \right]
\notag \\ & \quad \cdot
\left[ \frac{i}{(p-k)^2-m^2} + 2\pi \, \delta((p-k)^2-m^2) \, n_B\left(\abs{p_0-k_0}/T\right) \right]
\end{align}
Again, all contributions containing a vacuum TLH fluctuation vanish for $e>1/2$.
The only nonvanishing contribution is the case of both fluctuations being thermal; it reads
\begin{align}
\Pi^{\mu\nu}_{\substack{\text{nonl.}\\ \text{therm.}}} (p) 
 & = 
	i e^2 \int \frac{d^4k}{(2\pi)^2} \;\;
	\Bigg\{
		\left( 2kp - 4\frac{(kp)^2}{m^2} + 7p^2 - \frac{p^4}{m^2} \right) g^{\mu\nu}
		+ \left( 12 - 2\frac{kp}{m^2} - 3\frac{p^2}{m^2} + \frac{p^4}{m^4} \right) k^\mu k^\nu
\notag \\ & \quad 
		+ \left( - 6 + 4 \frac{kp}{m^2} - \frac{p^2 kp}{m^4} \right) \big( k^\nu p^\mu + k^\mu p^\nu \big)
		+ \left( - 5 + \frac{(kp)^2}{m^4} + \frac{p^2}{m^2} \right) p^\mu p^\nu
	\Bigg\}
\notag \\ & \quad 
\cdot \delta(k^2-m^2) \, n_B\left(\abs{k_0}/T\right) 
\cdot \delta((p-k)^2-m^2) \, n_B\left(\abs{p_0-k_0}/T\right) 
\,.
\end{align}
\setlength{\mathindent}{29.5pt}
We now set $p_0=0$ (keeping $|\vec p| \neq 0$) and get
\begin{align}                          								\label{p0=0}
&
\Pi^{\mu\nu}_{\substack{\text{nonl.}\\ \text{therm.}}} (p_0=0,\vec p) 
\notag \\ & = 
	i e^2 \int \frac{d^4k}{(2\pi)^2} \;\;
	\Bigg\{
		\left( - 2 \vec k \vec p - 4\frac{( \vec k \vec p)^2}{m^2} - 7 \vec p^2 - \frac{\vec p^4}{m^2} \right) g^{\mu\nu}
		+ \left( 12 + 2\frac{\vec k \vec p}{m^2} + 3\frac{\vec p^2}{m^2} + \frac{\vec p^4}{m^4} \right) k^\mu k^\nu
\notag \\ & \quad 
		+ \left( - 6 - 4 \frac{\vec k \vec p}{m^2} - \frac{\vec p^2 \vec k \vec p}{m^4} \right) \big( k^\nu p^\mu + k^\mu p^\nu \big)
		+ \left( - 5 + \frac{(\vec k \vec p)^2}{m^4} - \frac{\vec p^2}{m^2} \right) p^\mu p^\nu
	\Bigg\}
\notag \\ & \quad 
	\cdot \delta(k^2-m^2) \cdot  \delta((p-k)^2-m^2) \cdot \big[ n_B\left(\abs{k_0}/T\right) \big]^2
\end{align}
The two $\delta$-functions can for $p \neq 0$ be rewritten as
\begin{align}
& 
	\delta(k^2 - m^2) \cdot \delta((p-k)^2 - m^2)
=
	\delta(k^2 - m^2) \cdot \delta(p^2 - 2pk)
\notag \\ &=
	\frac{1}{2\sqrt{\vec k^2+m^2}} 
	\bigg[
		\delta \left( k_0 - \sqrt{\vec k^2 + m^2} \right) \cdot
		\delta \left( p_0^2 - \vec p^2 - 2 p_0 \sqrt{\vec k^2 + m^2} + 2 \vec p \vec k \right)
\notag \\ & \quad 
		+ \delta \left( k_0 + \sqrt{\vec k^2 + m^2} \right) \cdot
		\delta \left( p_0^2 - \vec p^2 + 2 p_0 \sqrt{\vec k^2 + m^2} + 2 \vec p \vec k \right)
	\bigg]
\notag \\ & =
	\frac{1}{2\sqrt{|\vec k|^2+m^2}} 
	\bigg[
		\delta \left( k_0 - \sqrt{ |\vec k|^2 + m^2} \right) \cdot
		\delta \left( p_0^2 - |\vec p|^2 - 2 p_0 \sqrt{|\vec k|^2 + m^2} + 2 |\vec p| |\vec k| \cos\theta \right)
\notag \\ & \quad 
		+ \delta \left( k_0 + \sqrt{|\vec k|^2 + m^2} \right) \cdot
		\delta \left( p_0^2 - |\vec p|^2 + 2 p_0 \sqrt{|\vec k|^2 + m^2} + 2 |\vec p| |\vec k| \cos\theta \right)
	\bigg]
\,.
\end{align}
In the case $p_0=0$ (with $|\vec p| \neq 0$) the former expression reduces to
\begin{align}											\label{deltadelta}
& 
\delta(k^2 - m^2) \cdot \delta((p-k)^2 - m^2)
\notag \\ &=
	\frac{1}{2\sqrt{|\vec k|^2+m^2}} 
	\bigg[
		\delta \left( k_0 - \sqrt{|\vec k|^2 + m^2} \right) \cdot \delta \left( 2 |\vec p| |\vec k| \cos\theta - |\vec p|^2 \right)
\\ & \quad 
		+ \delta \left( k_0 + \sqrt{|\vec k|^2 + m^2} \right) \cdot \delta \left( 2 |\vec p| |\vec k| \cos\theta - |\vec p|^2 \right)
	\bigg]
\notag \\ &=
	\frac{1}{4 |\vec p| |\vec k| \sqrt{|\vec k|^2 + m^2}} \cdot
	\delta \left( \cos\theta - \frac{|\vec p|}{2|\vec k|} \right) 
	\cdot \bigg[ \delta \left( k_0 - \sqrt{ |\vec k|^2 + m^2} \right) + \delta \left( k_0 + \sqrt{ |\vec k|^2 + m^2} \right) \bigg]
\notag
\,.
\end{align}
Inserting \eq{deltadelta} into \eq{p0=0} yields
\begin{equation}
\begin{split}
&
\Pi^{\mu\nu}_{\substack{\text{nonl.}\\ \text{therm.}}} (p_0=0,\vec p)
\\ & = 
	\frac{ie^2}{(2\pi)^2} 
	\int_{-\infty}^{\infty} dk_0
	\int_{-p/2}^{\infty} k^2 dk
	\int_{S_2} d\Omega
	\left[ n_B (\sqrt{ k^2 + m^2 } / T) \right]^2
\\ & \quad
\cdot 
	\Bigg\{
		\left( - 8 p^2 - 2\frac{p^4}{m^2} \right) g^{\mu\nu}
		+ \left( 12 + 4\frac{p^2}{m^2} + \frac{p^4}{m^4} \right) k^\mu k^\nu
\\ & \quad \quad
		+ \left( - 6 - 2 \frac{p^2}{m^2} - \frac12 \frac{p^4}{m^4} \right) \big( k^\nu p^\mu + k^\mu p^\nu \big)
		+ \left( - 5 - \frac{p^2}{m^2} + \frac14 \frac{p^4}{m^4} \right) p^\mu p^\nu
	\Bigg\}
\\ & \quad
	\cdot \frac{ \delta( \cos\theta - \frac{p}{2k} ) }{ 4pk \sqrt{k^2+m^2} }
	\cdot \left[ \delta \left( k_0 - \sqrt{k^2+m^2} \right) + \delta \left( k_0 + \sqrt{k^2+m^2} \right) \right]
\end{split}
\end{equation}
Only the terms proportional to $g^{\mu\nu}$ or $k^\mu k^\nu$ contribute to $\Pi^{00}$.
The result is
\begin{equation}
\begin{split}
&
\Pi^{00}_{\substack{\text{nonl.}\\ \text{therm.}}} (p_0=0,\vec p) 
\\ & =
- \frac{i e^2}{2\pi} \left( 4 p + \frac{p^3}{m^2} \right)
\int_{p/2}^{\infty} dk \, \frac{k}{\sqrt{k^2+m^2}} \; n_B^2 \left(\sqrt{k^2+m^2}/T\right)
\\ & \quad 
+ \frac{i e^2}{4\pi} \left( \frac{12}{p} + 4 \frac{p}{m^2} + \frac{p^3}{m^4} \right)
\int_{p/2}^{\infty} dk \, k \sqrt{k^2+m^2} \; n_B^2 \left(\sqrt{k^2+m^2}/T\right)
\,.
\end{split}
\end{equation}
In the limit $|\vec p| \to 0$ this expression diverges,
\begin{equation}
\begin{split}
\Pi^{00}_{\substack{\text{nonl.}\\ \text{therm.}}} (p_0=0,\vec p)
& \xrightarrow[|\vec p|\to0]{}
\frac{3 i e^2}{\pi} \frac{1}{|\vec p|} 
\int_0^{\infty} dk \, k \sqrt{k^2+m^2} \; n_B^2 \left(\sqrt{k^2+m^2}/T\right)
\,.
\end{split}
\end{equation}

}
\end{appendix}

%
%
%
%
%
%
%


\begin{thebibliography}{99}
\addcontentsline{toc}{chapter}{Bibliography} 

\bibitem{Ginzburg Landau} 
	V. L. Ginzburg and L. D. Landau, JETP {\bf 20}, 1064 (1950)

\bibitem{Abrikosov} 
	A. A. Abrikosov, Sov. Phys. JETP {\bf 5}, 1174 (1957)

\bibitem{Bardeen Cooper Schrieffer}
	J. Bardeen, L. N. Cooper and J. R. Schrieffer, Phys. Rev. {\bf 106}, 162 (1957)   \\
	J. Bardeen, L. N. Cooper and J. R. Schrieffer, Phys. Rev. {\bf 108}, 1175 (1957)

\bibitem{Linde}
	A. D. Linde, Phys. Lett. B {\bf 96}, 289 (1980)

\bibitem{Hofmann}
	R. Hofmann, hep-th/0504064

\bibitem{Peskin Schroeder} 
	M. E. Peskin and D. V. Schroeder, An Introduction to Quantum Field Theory, Addison-Wesley, 1997

\bibitem{'t Hooft Veltman} 
	G. 't~Hooft and M. J. G. Veltman, Nucl. Phys. B {\bf 44}, 189 (1972)  \\
	G. 't~Hooft, Nucl. Phys. B {\bf 33}, 173 (1971)  \\
	G. 't~Hooft, Nucl. Phys. B {\bf 62}, 444 (1973)  \\
	G. 't~Hooft and M. J. G. Veltman, Nucl. Phys. B {\bf 50}, 318 (1972)

\bibitem{Nielsen Olesen} 
	H. B. Nielsen and P. Olesen, Nucl. Phys. B {\bf 61}, 45 (1973)
  
\bibitem{'t Hooft 1974} 
	G. 't~Hooft, Nucl. Phys. B {\bf 79}, 276 (1974)

\bibitem{Polyakov 74} 
	A. M. Polyakov, JETP Lett. {\bf 20}, 194 (1974)

\bibitem{Ryder} 
	L. H. Ryder, Quantum Field Theory, Cambridge Universtiy Press, 1996
	
\bibitem{Schaefer Shuryak} 
	T. Sch\"afer and E. V. Shuryak, Rev. Mod. Phys. {\bf 70}, 323 (1998)

\bibitem{Gross Pisarski Yaffe} 
	D. J. Gross, R. D. Pisarski and L. G. Yaffe, Rev. Mod. Phys. {\bf 53}, 43 (1981)

\bibitem{WeinbergII} 
	S. Weinberg, The Quantum Theory of Fields Vol. II, Cambridge University Press, 1996

\bibitem{Bogomolnyi}  
	E. B. Bogomolnyi, Sov. J. Nucl. Phys. {\bf 24}, 449 (1976)

\bibitem{Prasad Sommerfield} 
	M. K. Prasad and C. M. Sommerfield, Phys. Rev. Lett. {\bf 35}, 760 (1975)

\bibitem{BPST} 
	A. Belavin, A. Polyakov. A. Schwartz and Yu. Tyupkin, Phys. Lett. {\bf 59}, 85 (1975)

\bibitem{ADHM} 
	M. Atiyah, W. Drinfeld, N. Hitchin and Y. Manin, Phys. Lett. A {\bf 65}, 185 (1978)

\bibitem{Harrington Shepard} 
	B. J. Harrington and H. K. Shepard, Phys. Rev. D {\bf 17}, 2122 (1978)

\bibitem{Nahm 1984} 
	W. Nahm, Lect. notes in Physics. 201, eds. G. Denardo, e.~a. (1984), p.189  

\bibitem{Lee Lu} 
	K. Lee and C. Lu, Phys. Rev. D {\bf 58}, 025011 (1998) 

\bibitem{Kraan van Baal 1} 
	T. C. Kraan and P. van Baal, Nucl. Phys. B {\bf 533}, 627 (1998)

\bibitem{Kraan van Baal 2} 
	T. C. Kraan and P. van Baal, Phys. Lett. B {\bf 435}, 389 (1998)

\bibitem{Diakonov et. al.} 
	D. Diakonov, N. Gromov V. Petrov, and S. Slizovskiy, Phys. Rev. D {\bf 70}, 036003 (2004) 

\bibitem{Kapusta} 
	J. Kapusta, Finite Temperature Field Theory, Cambridge University Press, 1989

\bibitem{Landsman van Weert} 
	N. P. Landsman and C. G. van Weert, Phys. Rep. {\bf 145}, 141 (1987)

\bibitem{Itzykson Zuber} 
	C. Itzykson and J.-B. Zuber, Quantum Field Theory, McGraw-Hill, 1985

\bibitem{Hofmann Rohrer}
	R. Hofmann and J. Rohrer, work in progress

\end{thebibliography}
\end{document}